\documentclass[twoside,11pt]{article}

\usepackage[final]{melba}  

\usepackage{amsmath,amsfonts}

\usepackage{booktabs}
\usepackage{float}

\melbaheading{4}{https://www.melba-journal.org/article/21657}{2021}{1-48}{09/2020}{03/2021}{Adalberto Claudio Quiros, Roderick Murray-Smith, \& Ke Yuan}{Medical Imaging with Deep Learning (MIDL) 2020}{ Marleen de Bruijne, Tal Arbel, Ismail Ben Ayed, Hervé Lombaert}

\ShortHeadings{PathologyGAN: Learning deep representations of cancer tissue}{A.C.Q., R.M.S., \& K.Y.}
\firstpageno{1}

\title{PathologyGAN: \\ Learning deep representations of cancer tissue}

\author{\name Adalberto Claudio Quiros \email a.claudio-quiros.1@research.gla.ac.uk \\  
	\addr School of Computing Science, University of Glasgow, Glasgow, Scotland, UK
	\AND
	\name Roderick Murray-Smith \email roderick.murray-smith@glasgow.ac.uk \\
	\addr School of Computing Science, University of Glasgow, Glasgow, Scotland, UK
	\AND
	\name Ke Yuan \email ke.yuan@glasgow.ac.uk \\
	\addr School of Computing Science, University of Glasgow, Glasgow, Scotland, UK
}

\begin{document}

\maketitle

\begin{abstract}
    
    

    Histopathological images of tumors contain abundant information about how tumors grow and how they interact with their micro-environment. 
    Better understanding of tissue phenotypes in these images could reveal novel determinants of pathological processes underlying cancer, and in turn improve diagnosis and treatment options. Advances of Deep learning makes it ideal to achieve those goals, however, its application is limited by the cost of high quality labels from patients data. Unsupervised learning, in particular, deep generative models with representation learning properties provides an alternative path to further understand cancer tissue phenotypes, capturing tissue morphologies. 
    
    In this paper, we develop a framework which allows Generative Adversarial Networks (GANs) to capture key tissue features and uses these characteristics to give structure to its latent space. To this end, we trained our model on two different datasets, an H\&E colorectal cancer tissue from the National Center for Tumor diseases (NCT, Germany) and an H\&E breast cancer tissue from the Netherlands Cancer Institute (NKI, Netherlands) and Vancouver General Hospital (VGH, Canada). Composed of 86 slide images and 576 tissue micro-arrays (TMAs) respectively. 
	    
    We show that our model generates high quality images, with a Fr\'echet Inception Distance (FID) of 16.65 (breast cancer) and 32.05 (colorectal cancer). We further assess the quality of the images with cancer tissue characteristics (e.g. count of cancer, lymphocytes, or stromal cells), using quantitative information to calculate the FID and showing consistent performance of 9.86. Additionally, the latent space of our model shows an interpretable structure and allows semantic vector operations that translate into tissue feature transformations. Furthermore, ratings from two expert pathologists found no significant difference between our generated tissue images from real ones. 
    
    The code, generated images, and pretrained models are available at ~\url{https://github.com/AdalbertoCq/Pathology-GAN}
\end{abstract}


\section{Introduction}
    Cancer is a disease with extensive heterogeneity, where malignant cells interact with immune cells, stromal cells, surrounding tissues and blood vessels. Histological images, such as haematoxylin and eosin (H\&E) stained tissue microarrays (TMAs) or whole slide images (WSI), are a high-throughput imaging technology used to study such diversity. Despite being ubiquitous in clinical settings, analytical tools of H\&E images remain primitive, making these valuable data largely under-explored. Consequently, cellular behaviours and the tumor microenvironment recorded in H\&E images remain poorly understood. Increasing our understanding of such microenvironment interaction holds the key for improved diagnosis and treatment of cancer. 
    
    \begin{figure}[!ht]
        \centering
        \includegraphics[scale=0.15]{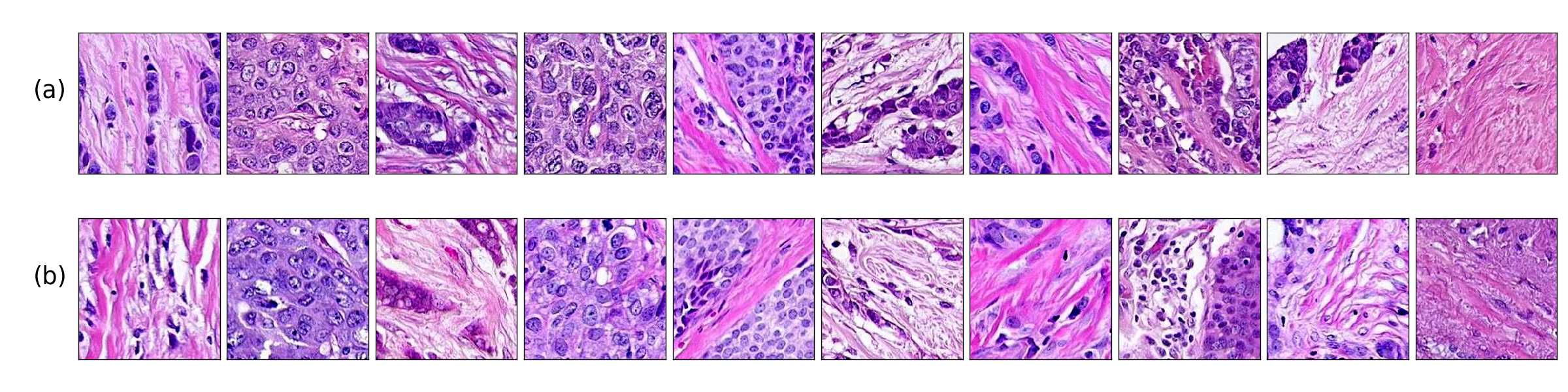}
        \includegraphics[scale=0.15]{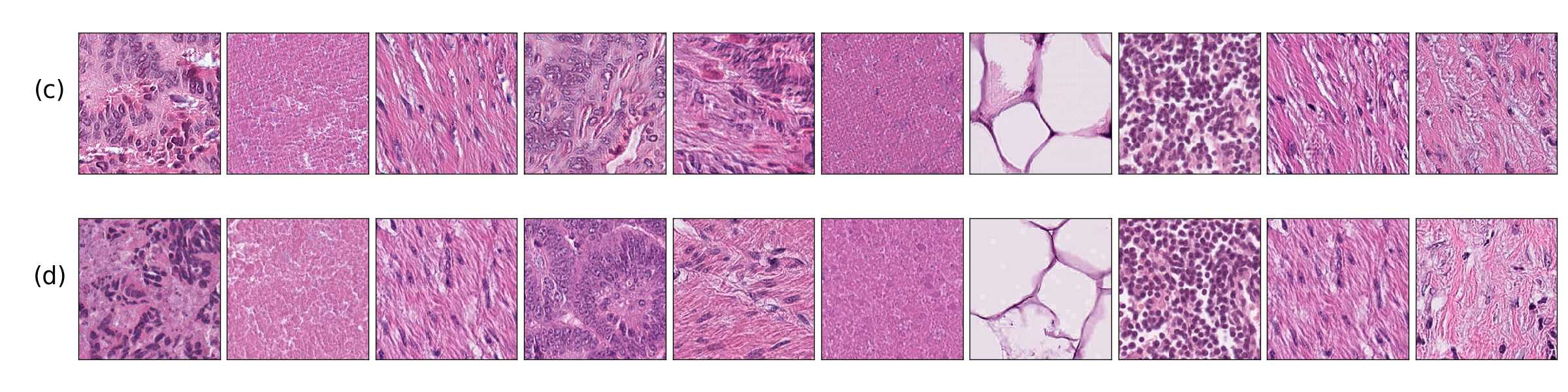}
        \caption{(a): Images ($224\times224$) from PathologyGAN trained on H\&E breast cancer tissue. (b): Real images, Inception-V1 closest neighbor to the generated above in (a). (c): Images ($224\times224$) from PathologyGAN trained on H\&E colorectal cancer tissue. (d) Real images, Inception-V1 closest neighbor to the generated above in (c).}
        \label{fig:hand_picked_samples}
    \end{figure}

    The motivation for our work is to develop methods that could lead to a better understanding of phenotype diversity between/within tumors. We hypothesize that this diversity could be substantial given the highly diverse genomic and transcriptomic landscapes observed in large scale molecular profiling of tumors across multiple cancer types \citep{Campbell2020}. We argue that representation learning with GAN-based models is the most promising to achieve our goal for the two following reasons: 
    \begin{enumerate}
        \item By being able to generate high fidelity images, a GAN could learn the most relevant descriptions of tissue phenotype.
        \item The continuous latent representation learned by GANs could help us quantify differences in tissue architectures free from supervised information. 
    \end{enumerate}
	    
    In this paper, we propose to use Generative Adversarial Networks (GANs) to learn representations of entire tissue architectures and define an interpretable latent space (e.g. colour, texture, spatial features of cancer and normal cells, and their interaction). To this end, we present the following contributions:
    \begin{enumerate}
        \item We propose PathologyGANs to generate high fidelity cancer tissue images from a structured latent space. The model combines BigGAN \citep{Brock2018}, StyleGAN \citep{Karras2019} and Relativistic Average Discriminator \citep{Jolicoeur-Martineau2018}.
        \item We assess the quality of the generated images through two different methods: convolutional Inception-V1 features and prognostic features of the cancer tissue, such as counts and densities of different cell types \citep{Beck2011, Yuan2012}. Both features are benchmarked with the Fr\'echet Inception Distance (FID). The results show that the model captures pathologically meaningful representations, and when evaluated by expert pathologists, generated tissue images are not distinct from real tissue images.
        \item We show that our model induces an ordered latent space based on tissue characteristics (e.g. cancer cell density or tissue type), this allows to perform linear vector operations that transfer into high level tissue image changes.
    \end{enumerate}

\section{Related Works}
    
    Deep learning has been widely applied in digital pathology, from these we can differentiate between supervised and unsupervised learning approaches. 
    
    Supervised applications range from mitosis and cell detection \citep{Tellez2018, Xu2019b, Zhang2019b}, nuclei and tumor segmentation \citep{Qu2019, Qaiser2019}, histological subtype classification \citep{Coudray2018, Wei2019}, to survival and prognosis modeling \citep{Katzman2018, Lee2018}. Recently, there have been developments on relating phenotype to the molecular underpining of tumors, in particular genomic characteristics \citep{Coudray2018, Schmauch2020, Woerl2020, Fu2020, Coudray2020, Kather2020}, and spatial transcriptomics \citep{Vickovic2019, He2020, Bergenstrahle2020, Schmauch2020, Wang2020, Bergenstrahle2020_2}. Furthermore, previous traditional computer vision approaches \citep{Beck2011, Yuan2012} already identified correlation between phenotype patterns and patient survival. These works highlight the importance and opportunities that building tissue phenotype representations bring, providing insight into survival or genomic information purely from tissue images such as TMA or WSIs. 
    
    Nevertheless, these methods require data labeling which is usually costly in time and effort, this is particularly the case for sequencing derived molecular labels. In addition, deep learning approaches have a lack of interpretability, which is also a major limiting factor in making a real impact in clinical practice. 
    
    Unsupervised learning applications mostly focus on nuclei \citep{Xu2015, Mahmood2018}, tissue \citep{deBel2018}, or region-of-interest segmentation \citep{Gadermayr2018, Gadermayr2019}, besides stain transformation \citep{Rana2018, Xu2019} and normalization \citep{Zanjani2018}. Within unsupervised learning, generative models have been briefly used for tissue generation \citep{Levine2020}, however this model lacks of representation learning properties. On the other hand, there has been some initial work on building cell and nuclei representations with models such as InfoGAN \citep{Hu2019} and Sparse Auto-Encoders \citep{Hou2019}, although these models focus either on small sections of images or cells instead of larger tiles of tissue. 
    
    Building phenotype representations based on tissue architecture and cellular attributes remains a field to be further explored. Generative models offer the ability to create tissue representations without expensive labels and representations not only correlated with a predicted outcome (as in discriminative models), rather creating representations based on the similarities across the characteristics of tissue samples. Take a point mutation for example, it is now understood that mutations are frequently only shared in subpopulations of cancer cells within a tumor \citep{Gerstung2020, Dentro2020}. Therefore, it's difficult to know if a point mutation is presented in the cells recorded in a image. Fundamentally, supervised approach is limited by the fact that molecular and clinical labels are often obtained from materials that are physically different from the ones in the images. The associations are therefore highly indirect and subject to many confounding factors
    
    From generative models, Generative Adversarial Networks (GANs) have become increasingly popular, applied to different domains from imaging to signal processing. GANs \citep{Goodfellow2014} are able to learn high fidelity and diverse data representations from a target distribution. This is done with a generator, $G(z)$, that maps random noise, $\boldsymbol{z} \sim p_{\boldsymbol{z}}(z)$, to samples that resemble the target data, $\boldsymbol{x} \sim p_{\text { data }}(\boldsymbol{x})$, and a discriminator, $D(x)$, whose goal is to distinguish between real and generated samples. The goal of a GAN is find the equilibrium in the min-max problem:
    \begin{equation}
        \min _{G} \max _{D} V(D, G)=\mathbb{E}_{\boldsymbol{x} \sim p_{\text { data }}(\boldsymbol{x})}[\log D(\boldsymbol{x})]+\mathbb{E}_{\boldsymbol{z} \sim p_{\boldsymbol{z}}(\boldsymbol{z})}[\log (1-D(G(\boldsymbol{z})))].
    \end{equation}
    
    Since its introduction, modeling distributions of images has become the mainstream application for GANs, firstly introduced by \cite{Radford2015}. State-of-the-art GANs such as BigGAN \citep{Brock2018} and StyleGAN \citep{Karras2019} have recently been use to generate impressive high-resolution images. Additionally, solutions like Spectral Normalization GANs \citep{Miyato2018}, Self-Attention GANs \citep{Zhang2019}, and also BigGAN have achieved high diversity images in data sets like ImageNet \citep{imagenet_cvpr09}, with 14 million images and 20 thousand different classes. 
    At the same time, evaluating these models has been a challenging task. Many different metrics such as Inception Score (IS) \citep{Salimans2016}, Fr\'echet Inception Distance (FID) \citep{Heusel2017}, Maximum Mean Discrepancy (MMD) \citep{Gretton2012}, Kernel Inception Distance (KID) \citep{Binkowski2018}, and 1-Nearest Neighbor classifier (1-NN) \citep{Lopezpaz2016} have been proposed to do so, and thorough empirical studies \citep{Huang2018, Barratt2018} have shed some light on the advantages and disadvantages of each them. However, the selection of a feature space is crucial for using these metrics.
    
    In our work we take a step towards developing a generative model that learns phenotypes through tissue architectures and cellular characteristics, introducing a GAN with representation learning properties and an interpretable latent space. We advocate that in the future these phenotype representations could give us insight about the diversity within/across cancer types and their relation to genomic, transcriptomic, and survival information; finally leading to better treatment and prognosis of the disease.
    
\section{Methods}
	\subsection{PathologyGAN}
		\begin{figure}[!t]
	        \centering
	        \includegraphics[scale=0.2]{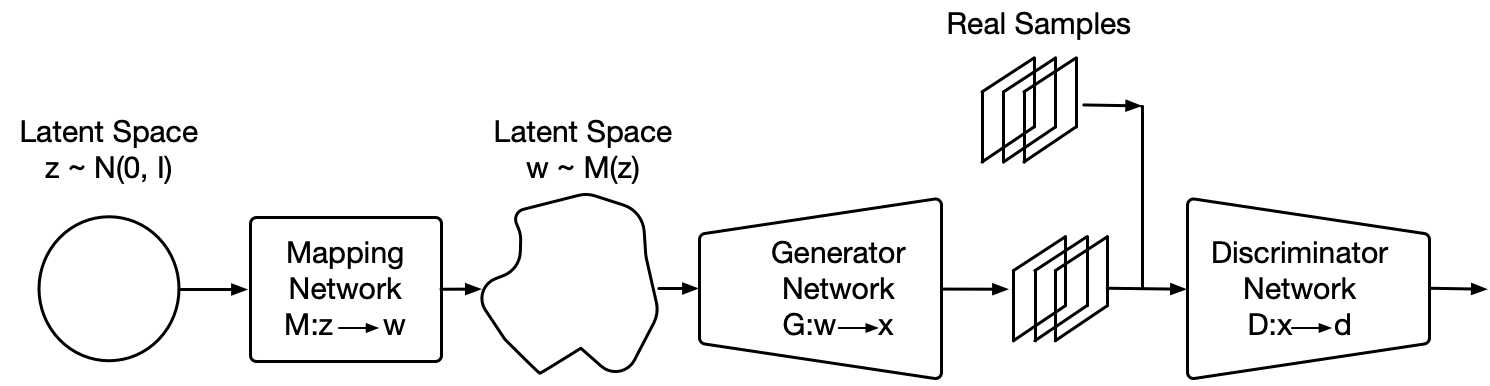}
	        \caption{High level architecture of PathologyGAN. We include details of each module's architecture in the Appendix \ref{appendix:model_arch}}
	        \label{fig:pathologygan_model}
	    \end{figure}
	    
	    
	    We used BigGAN \citep{Brock2018} as a baseline architecture and introduced changes which empirically improved the Fr\'echet Inception Distance (FID) and the structure of the latent space. 
	    
	    BigGAN has been shown to be a successful GAN in replicating datasets with a diverse number of classes and large amounts of samples, such as ImageNet with approximately $1M$ samples and $20K$ classes. For this reason, we theorize that such model will be able to learn and replicate the diverse tissue phenotypes contained in whole slide images (WSI), being able to handle the large amount of tiles/patches resulting from diving the WSIs.  
	    
	    We followed the same architecture as BigGAN, employed Spectral Normalization in both generator and discriminator, self attention layers, and we also use orthogonal initialization and regularization as mentioned in the original paper. 
	    
	    We make use of the Relativistic Average Discriminator \citep{Jolicoeur-Martineau2018}, where the discriminator's goal is to estimate the probability of the real data being more realistic than the fake. We take this approach instead of following the Hinge loss \citep{Lim2017} as the GAN objective. We find that this change makes the model convergence faster and produce higher quality images. Images using the Hinge loss did not capture the morphological structure of the tissue (we provide examples of these results in the Appendix \ref{appendix:hinge}). The discriminator, and generator loss function are formulated as in Equations 2 and 3, where $\mathbb{P}$ is the distribution of real data, $\mathbb{Q}$ is the distribution for the fake data, and $C(x)$ is the non-transformed discriminator output or critic:
	    \begin{align}
	        L_{Dis}=-\mathbb{E}_{x_{r} \sim \mathbb{P}}&\left[\log \left(\tilde{D}\left(x_{r}\right)\right)\right]-\mathbb{E}_{x_{f} \sim \mathbb{Q}}\left[\log \left(1-\tilde{D}\left(x_{f}\right)\right)\right], \\
	        L_{Gen}=-\mathbb{E}_{x_{f} \sim \mathbb{Q}}&  \left[\log\left(\tilde{D}\left(x_{f}\right)\right)\right]-\mathbb{E}_{x_{r} \sim\mathbb{P}}\left[\log \left(1-\tilde{D}\left(x_{r}\right)\right)\right], \\
	        \quad \quad \tilde{D}\left(x_{r}\right) &= \text{sigmoid}\left(C\left(x_{r}\right)-\mathbb{E}_{x_{f} \sim \mathbb{Q}} C\left(x_{f}\right)\right)),  \\ 
	        \quad \quad \tilde{D}\left(x_{f}\right) &= \text{sigmoid}\left(C\left(x_{f}\right)-\mathbb{E}_{x_{r} \sim \mathbb{P}} C\left(x_{r}\right)\right). 
	        \label{eqn:disc_loss}
	    \end{align}
	    
	    Additionally, we introduce two elements from StyleGAN \citep{Karras2019} with the purpose of allowing the generator to freely optimize the latent space and find high-level features of the cancer tissue. First, a mapping network $M$ composed by four dense ResNet layers \citep{He2016}, placed after the latent vector $z \sim \mathcal{N}(0, I)$, with the purpose of allowing the generator to find the latent space $w \sim M(z)$ that better disentangles the latent factors of variation. Secondly, style mixing regularization, where two different latent vectors $z1$ and $z2$ are run into the mapping network and fed at the same time to the generator, randomly choosing a layer in the generator and providing $w1$ and $w2$ to the different halves of the generator (e.g. on a generator of ten layers and being six the randomly selected layer, $w1$ would feed layers one to six and $w2$ seven to ten). Style mixing regularization encourages the generator to localize the high level features of the images in the latent space. We also use adaptive instance normalization (AdaIN) on our models, providing the entire latent vectors.
	    
	    We use the Adam optimizer \citep{Kingma2014} with $\beta_{1}=0.5$ and same learning rates of $0.0001$ for both generator and discriminator, the discriminator takes 5 steps for each of the generator. Each model was trained on an NVIDIA Titan RTX 24 GB for approximately 72 hours.
	
	\subsection{Datasets}
	\label{datasets}
	
	    
		To train our model, we used two different datasets, an H\&E colorectal cancer tissue from the National Center for Tumor diseases (NCT, Germany)  \citep{kather_2018} and an H\&E breast cancer tissue from the Netherlands Cancer Institute (NKI, Netherlands) and Vancouver General Hospital (VGH, Canada) \citep{Beck2011}. 
	    
	    The H\&E breast cancer dataset was built from the Netherlands Cancer Institute (NKI) cohort and the Vancouver General Hospital (VGH) cohort with 248 and 328 patients, respectively. Each of them include TMA images, along with clinical patient data such as survival time, and estrogen-receptor (ER) status. The original TMA images all have a resolution of $1128 \times 720$ pixels, and we split each of the images into smaller patches of $224 \times 224$, and allowed them to overlap by 50\%. We also performed data augmentation on these images, a rotation of $90^{\circ}$, and $180^{\circ}$, and vertical and horizontal inversion. We filtered out images in which the tissue covers less than 70\% of the area. In total this yield to a training set of 249K images and a test set of 62K. 
	    
	    The H\&E colorectal cancer dataset provides 100K tissue images of $224\times224$ resolution, each image has an associated type of tissue label: adipose, background, debris, lymphocytes, mucus, smooth muscle, normal colon mucosa, cancer-associated stroma, and colorectal adenocarcinoma epithelium (tumor). This dataset is composed of 86 H\&E stained human cancer tissue slides. In order to check the model's flexibility and ability to work with different datasets, we decided not to apply any data augmentation and use the tiles as they are provided. 
	    
	    In both datasets, we perform the partition over the total tissue patches, not according to patients, since our goal is to verify the ability to learn tissue representations.  We trained our model on the VGH/NKI and NCT datasets for $45$ and $130$ epochs, respectively.
	    
	    In Appendix \ref{appendix:datasetsize} we study the model's capacity in capturing representations with small size datasets (5K, 10K, 20K) verifying its converge and ability to generalize. 
	    
	\subsection{Evaluation metric on PathologyGAN}	
		
	    The Fr\'echet Inception Distance (FID) \citep{Heusel2017} is a common metric used to measure GANs performance, and it quantifies the GAN's ability to learn and reproduce the original data distribution. The goal of the FID score is to measure the similarity in quality and diversity between real $p_{data}(x)$ and generated data $p_{g}(x)$. 
	    
	    Instead to measuring the distance between the real and generated distributions in the pixel space, it uses a pretrained ImageNet Inception Network \citep{Szegedy2016} to extract features of each image, reducing the dimensionality of the samples and obtaining vision-relevant features. Feature samples are fitted into a multivariate Gaussian distribution obtaining real $\mathcal{N}(\mu_{data}, \Sigma_{data})$ and generated $\mathcal{N}(\mu_{g}, \Sigma_{g})$ feature distributions. Finally, it uses the Fr\'echet distance \citep{frechet1957} to measure the difference between the two distributions:
	    $$FID = d^{2}((\mu_{data}, \Sigma_{data}),(\mu_{g}, \Sigma_{g})) = ||\mu_{data} - \mu_{g}||^{2}_{2} + Tr(\Sigma_{data}+\Sigma_{g}-2(\Sigma_{data}\Sigma_{g})^{1/2})$$
	    
	    We evaluate our model by calculating the FID score from $10K$ generated images and randomly sampling $10K$ real images.
	    
	    We focus on using FID as it is a common GAN evaluation method \citep{Brock2018, Karras2019, Miyato2018, Zhang2019} that reliably captures differences between the real and generated distributions. Additionally, we provide more details in Appendix \ref{GANevalDP} comparing FID to other metrics such as Kernel Inception Distance (KID) or 1-Nearest Neighbor (1-NN) in the context of digital pathology.
	    
    \subsection{Quantification of cancer cells in generated images - Breast cancer tissue}
	\label{quantification_images}
		
		   
		In our results, we use the counts of cancer cells and other cellular information as a mean to measure the image quality and representation learning properties of our model. The motivation behind this approach is to ensure that our model capture meaningful and faithful representations of the tissue.
		
		We use this information in two different ways, first as an alternative feature space for FID as each image is translated into a vector with cellular information in the tissue, and secondly to label each generated tissue image according to the cancer cell density, allowing us to visualize the representation learning properties of our model's latent space. 
	    
\begin{figure}[!t]
		        \centering
		        \includegraphics[scale=0.1]{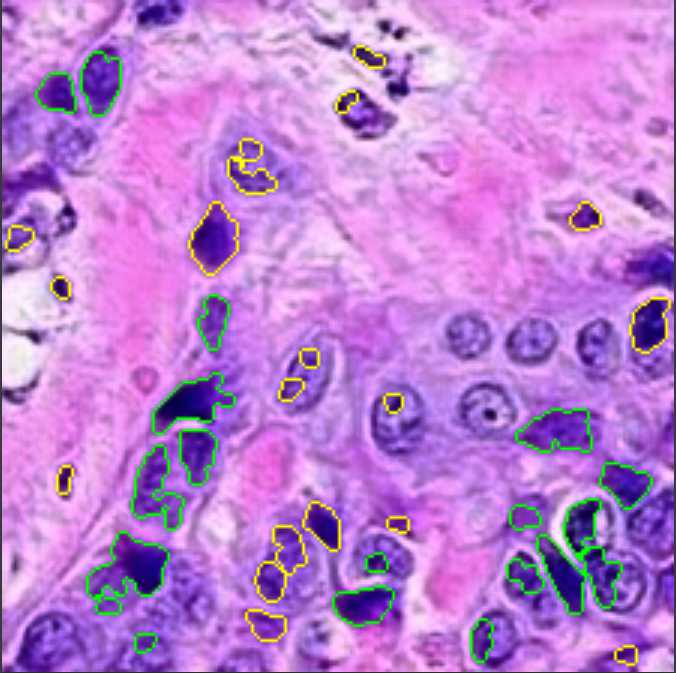}
		        \includegraphics[scale=0.1]{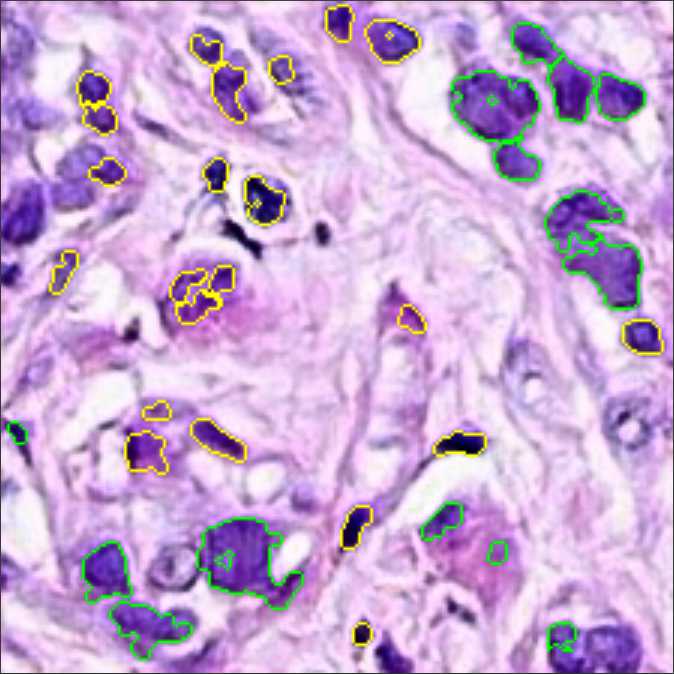}
		        \caption{CRImage identifies different cell types in our generated images. Cancer cells are highlighted with a green color, while lymphocytes and stromal cells are highlighted in yellow.}
		        \label{fig:CRImage_SVM_example}
		 \end{figure}

	    The CRImage tool \citep{Yuan2012} uses an SVM classifier to provide quantitative information about tumor cellular characteristics in tissue. This approach allows us to gather pathological information in the images, namely the number of cancer cells, the number of other types of cells (such as stromal or lymphocytes), and the ratio of tumorous cells per area. We limit the use of this information to breast cancer tissue, since the tool was developed for this specific case. Figure~\ref{fig:CRImage_SVM_example} displays an example of how the CRImage captures the different cells in the generated images, such as cancer cells, stromal cells, and lymphocytes. 
	    
	    Finally, we created 8 different classes that account for counts of cancer cells in the tissue image, and consecutively we label each generated image with the corresponding class, allowing us to more clearly visualize the relation between cancer cell density in generated images and the model's latent space.
		
	\subsection{Tissue type assignation on generated images - Colorectal cancer tissue}
	\label{tissue_images}
	    
	    
		The NCT colorectal cancer dataset provides a label along each $224\times224$ tissue sample and we make use of this information to label each generated image with a type of tissue, assigning a label to each generated image according to the 10-nearest neighbors of real images in the Inception-V1 feature space. 
		
		In Figure \ref{fig:nearest_neighbors} we show generated images on the left column and the nearest real neighbors on the remaining columns. We present the different tissue types: tumor (i), stroma (j), muscle (k), lymphocytes (l), debris (m), mucus (n), adipose (o), and background (p). From this example, we can conclude that distance in feature space gives a good reference to find the tissue type of generated images. 
		
\section{Results}
	\subsection{Image quality analysis}
	\label{gan_metrics}
    	
	    
	    
	    
	    We study the fidelity of the generated images and their distribution in relation to the original data in two different ways, through measures of FID metrics and by visualizing the closest neighbors between generated and real images. 
	    
	    We calculate Fr\'echet Inception Distance (FID) with two different approaches, with the usual convolutional features of a Inception-V1 network and with cellular information extracted from the CRImage cell classifier, as explained in Section \ref{quantification_images}. We restrict using CRImage to breast cancer tissue only since it was developed for that particular purpose.
	    
	    Table~\ref{GAN_results-table} shows that our model is able to achieve an accurate characterization of the cancer tissue. Using the Inception feature space, FID shows a stable representation for all models with values similar to ImageNet models of BigGAN \citep{Brock2018} and SAGAN \citep{Zhang2019}, with FIDs of 7.4 and 18.65, respectively or StyleGAN \citep{Karras2019} trained on FFHQ with FID of 4.40. Using the CRImage cellular information as feature space, FID shows again close representations to real tissue.  
	   
\begin{figure}[H]
			\centering
			\minipage{0.5\textwidth}
			  \includegraphics[scale=0.22, trim=250 0 250 140, clip]{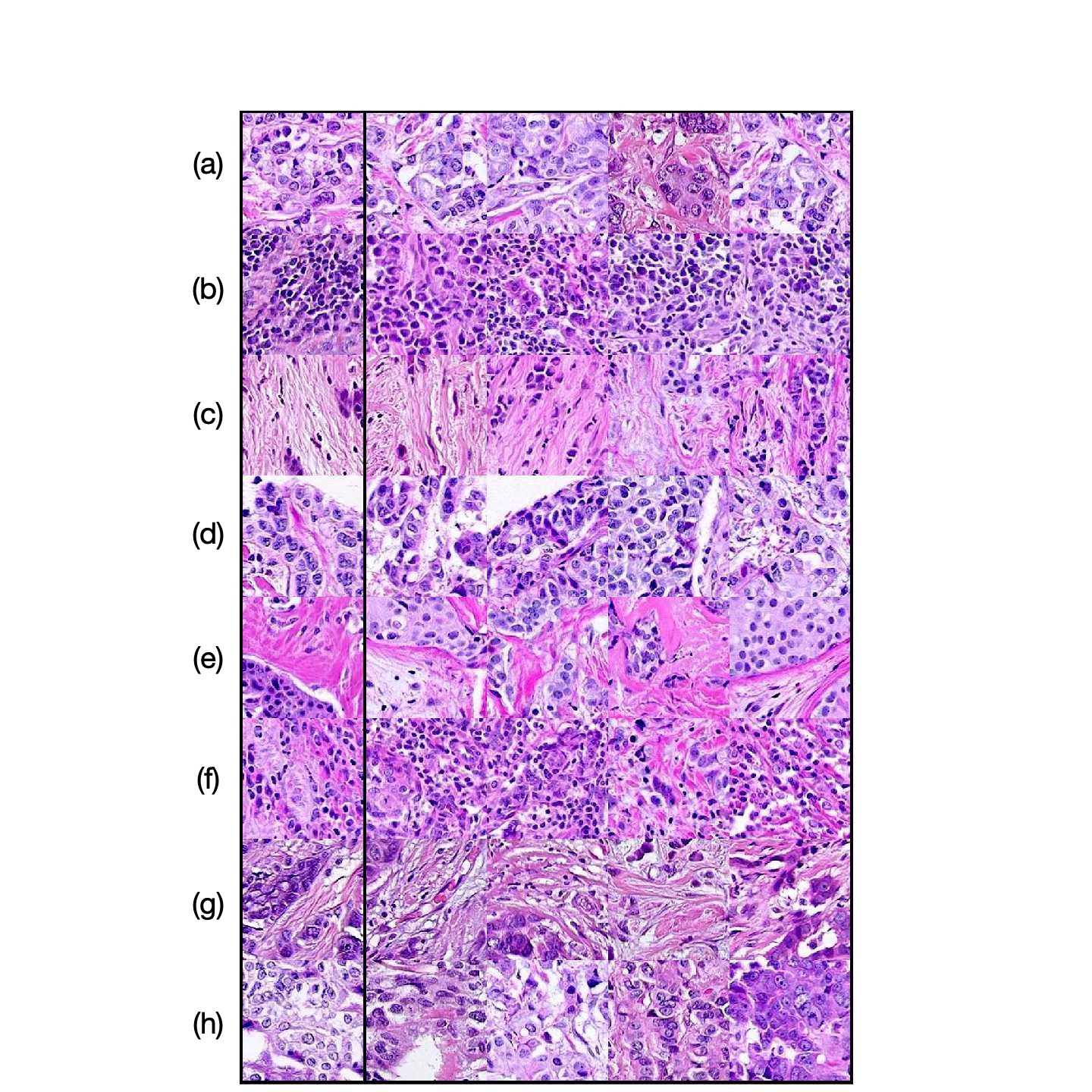}
			\endminipage\hfill
			\minipage{0.5\textwidth}
			  \includegraphics[scale=0.22, trim=225 0 250 140, clip]{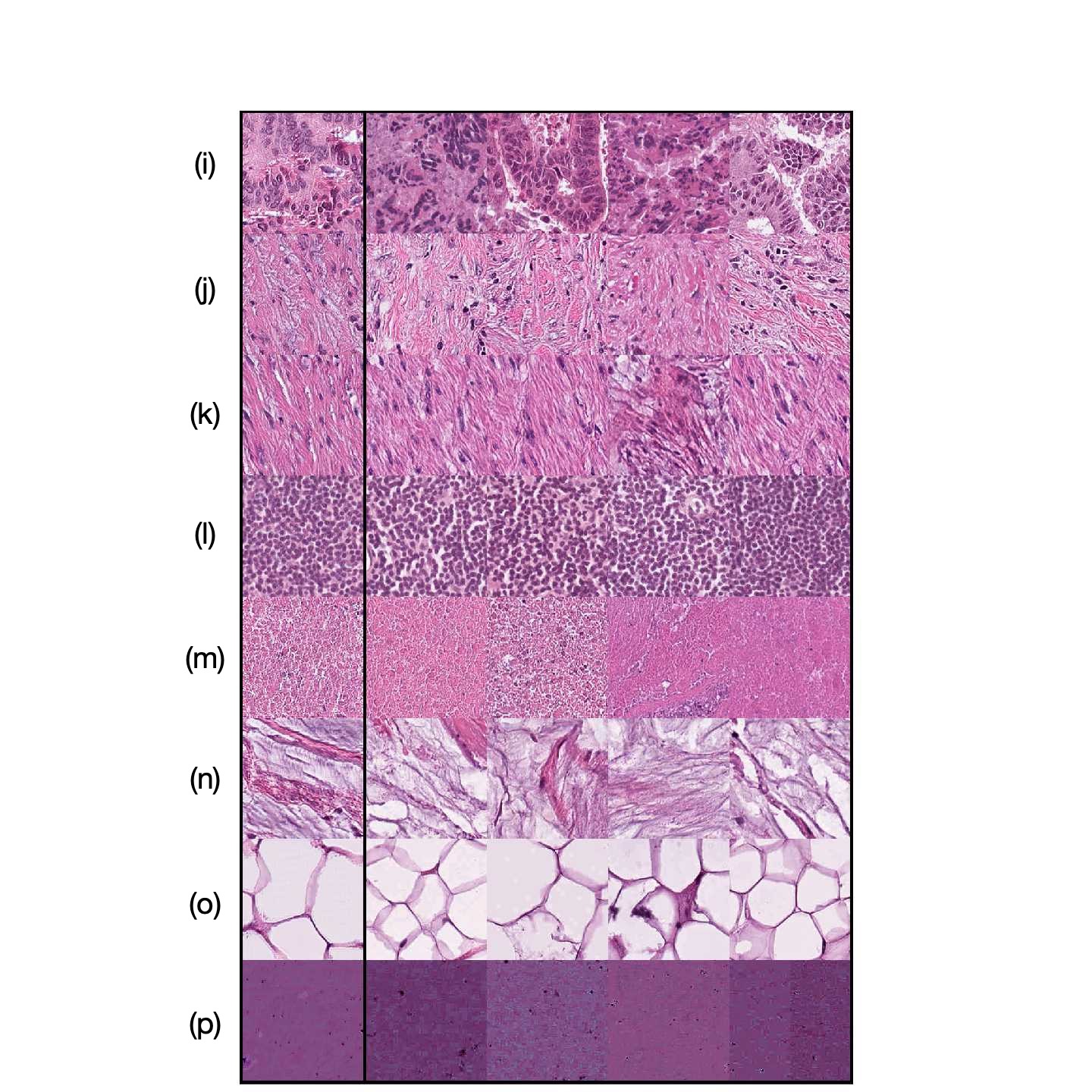}
			\endminipage
			 \caption{Nearest neighbors in Inception-V1 feature space for breast cancer (a-h) and colorectal cancer (i-p). For each row, first column images corresponds to a generated tissue samples from PathologyGAN, the remaining columns are the closest real images in feature space.}
			 \label{fig:nearest_neighbors}
		\end{figure}

		Additionally, in Figure \ref{fig:nearest_neighbors} we present samples of generated images (first column) and its closest real neighbors in Inception-V1 feature space, the images are paired by rows. (a-h) correspond to different random samples of breast cancer tissue. In the case of colorectal cancer, we provide examples of different types of tissue: tumor (i), stroma (j), muscle (k), lymphocytes (l), debris (m), mucus (n), adipose (o), and background (p). We can see that generated and real images hold the same morphological characteristics. 
		
		\begin{table}[H]
	        \centering
	        \begin{tabular}{l|l|l|l|l|l|l}
	        \toprule
	        Model     & Inception FID & Inception FID & CRImage FID \\
	                       & Colorectal       & Breast            & Breast            \\
	        \toprule
	        \midrule
	        PathologyGAN & 32.05$\pm$3 & 16.65$\pm$2.5 & 9.86$\pm$0.4 \\
	        \bottomrule
	        \bottomrule
	        \end{tabular}
	        \caption{Evaluation of PathologyGANs. Mean and standard deviations are computed over three different random initializations. The low FID scores in both feature space suggest consistent and accurate representations.}
	        \label{GAN_results-table}
	    \end{table}
    	
    \subsection{Analysis of latent representations}
    	
	
		\begin{figure}[!b]
    		\centering
    		\minipage{0.5\textwidth}
    		  \includegraphics[scale=0.145]{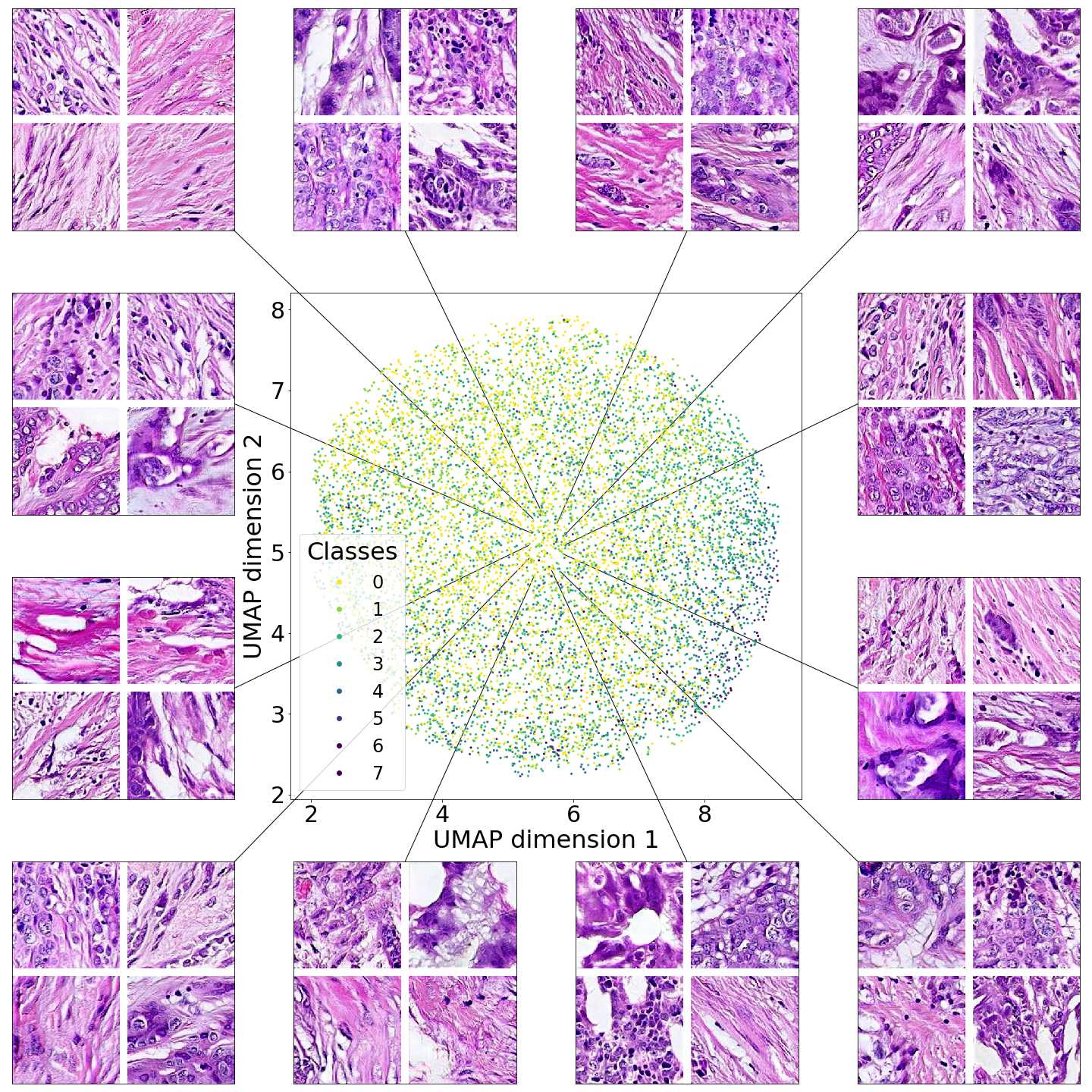}
    		\endminipage\hfill
    		\minipage{0.5\textwidth}
    		  \includegraphics[scale=0.145]{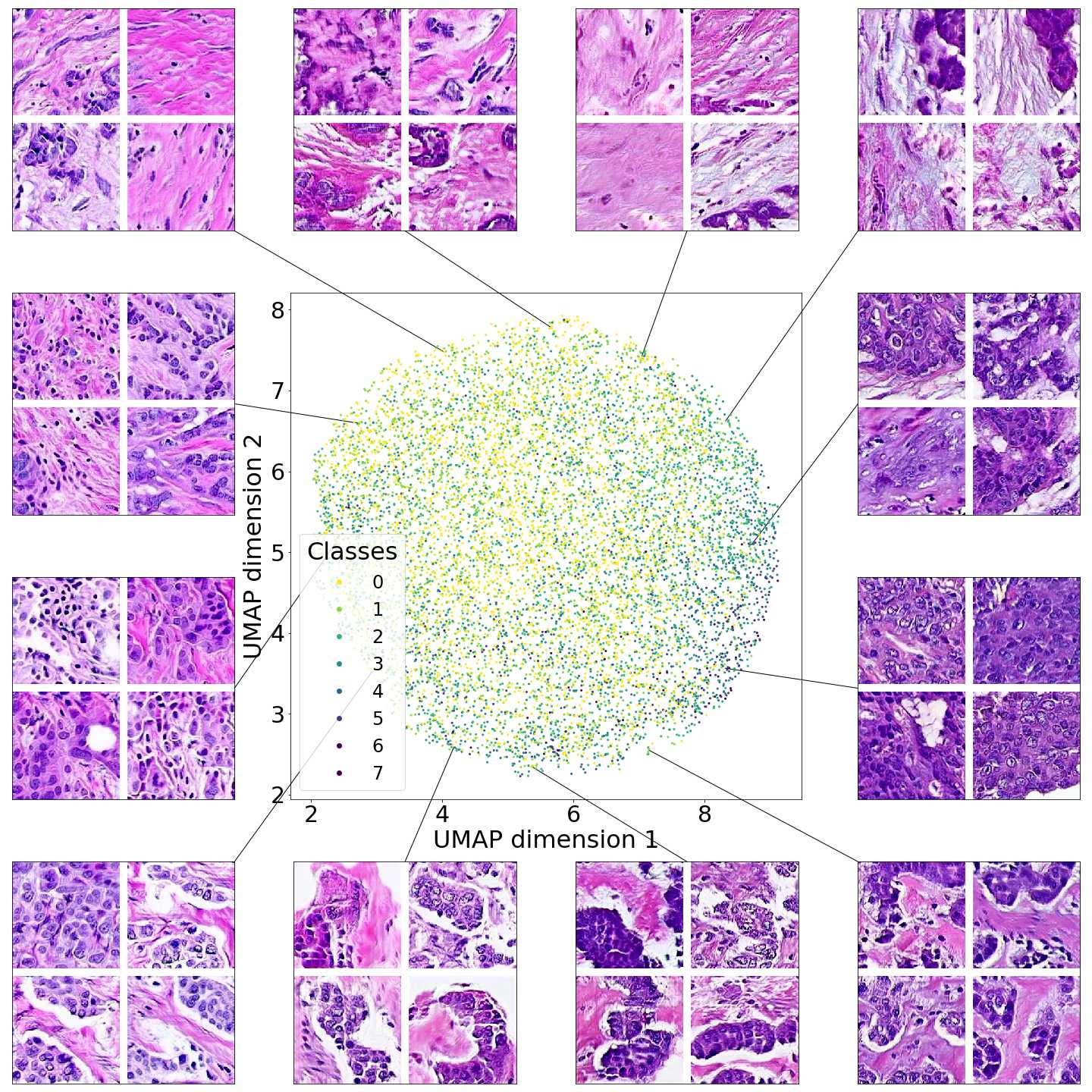}
    		\endminipage
    		 \caption{Latent space of PathologyGAN trained on breast cancer tissue from Netherlands Cancer Institute (NKI) and Vancouver General Hospital (VGH) dataset. Uniform Manifold Approximation and Projection (UMAP) representations of generated tissue samples, each generated image is labeled with the density of cancer cells, class 0 for lowest and class 8 for highest. Moving from quadrant $II$ to quadrant $IV$ in the UMAP representation corresponds to increasing the density of cancer cells in the generated tissue.}
    		 \label{fig:latent_space_vgh_nki}
    	\end{figure}   
        
    	\begin{figure}[!t]
    		\centering
    		\minipage{0.5\textwidth}
    		  \includegraphics[scale=0.145]{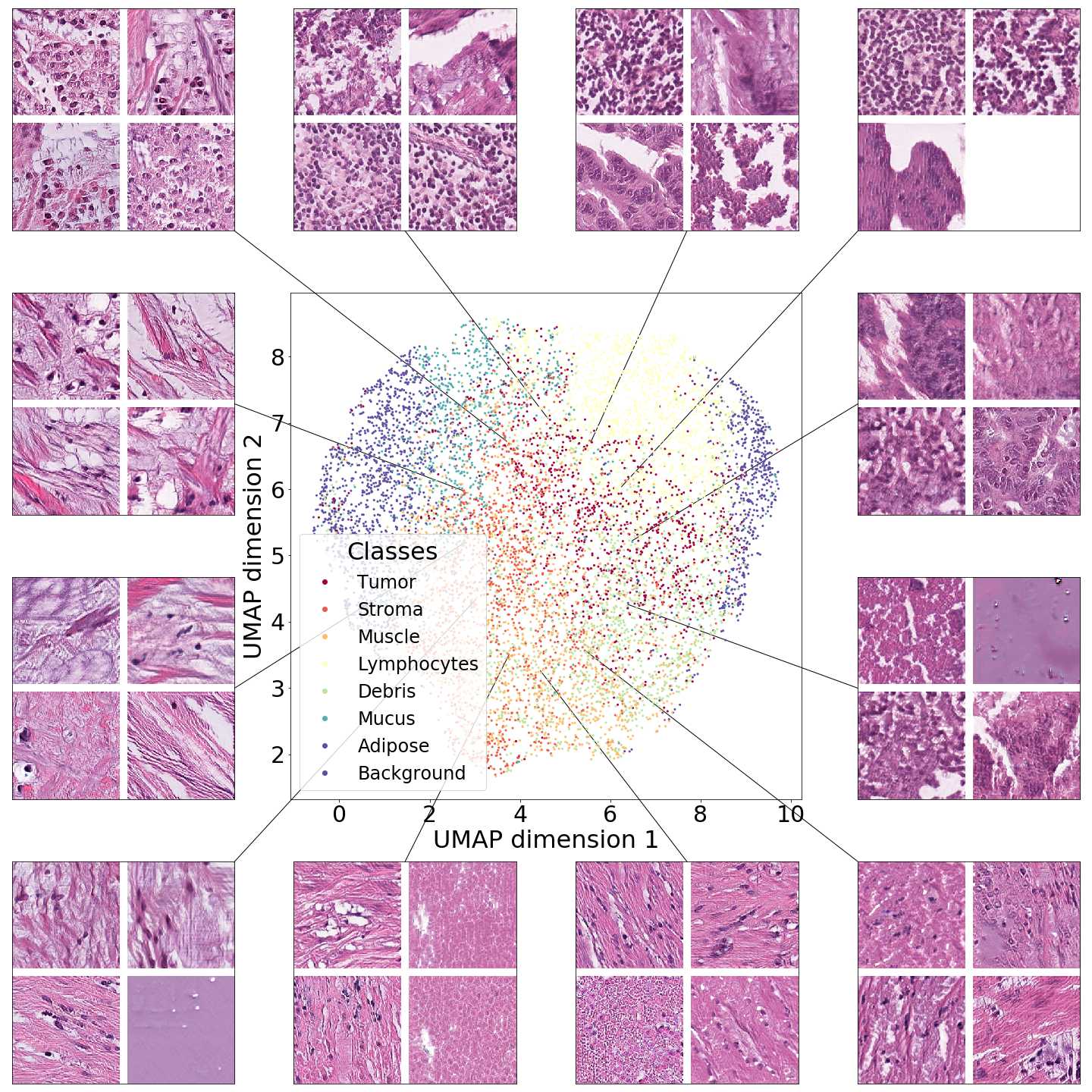}
    		\endminipage\hfill
    		\minipage{0.5\textwidth}
    		  \includegraphics[scale=0.145]{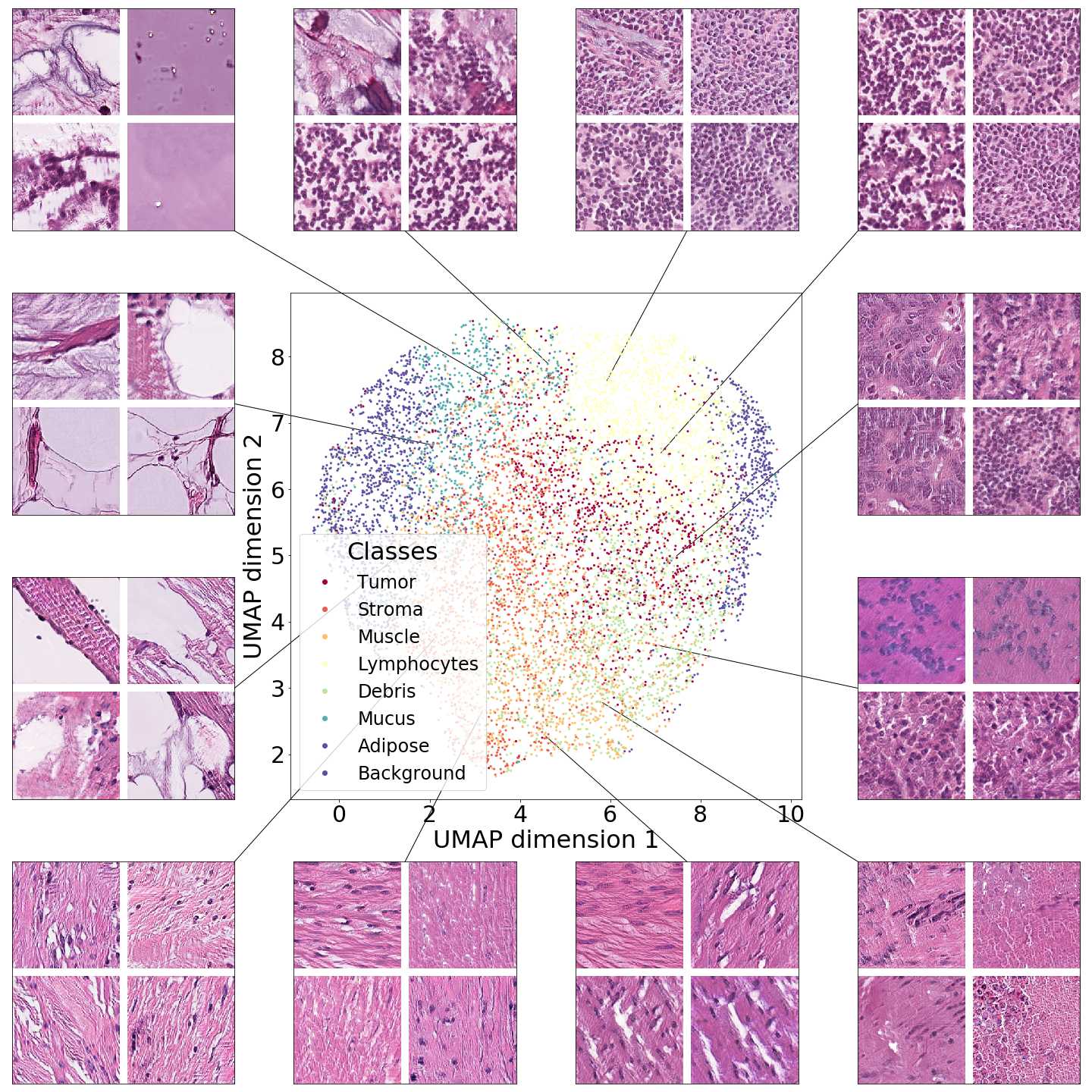}
    		\endminipage
    		 \caption{Latent space of PathologyGAN trained on colorectal cancer tissue from National Center for Tumor  (NCT) dataset. Uniform Manifold Approximation and Projection (UMAP) representations of generated tissue samples, each generated image is labeled with the type of tissue. Different regions of the latent space generate distinct kinds of tissue.}
    		 \label{fig:latent_space_crc}
    	\end{figure}  
        
    	\begin{figure}[!t]
            \centering
            \includegraphics[scale=0.17]{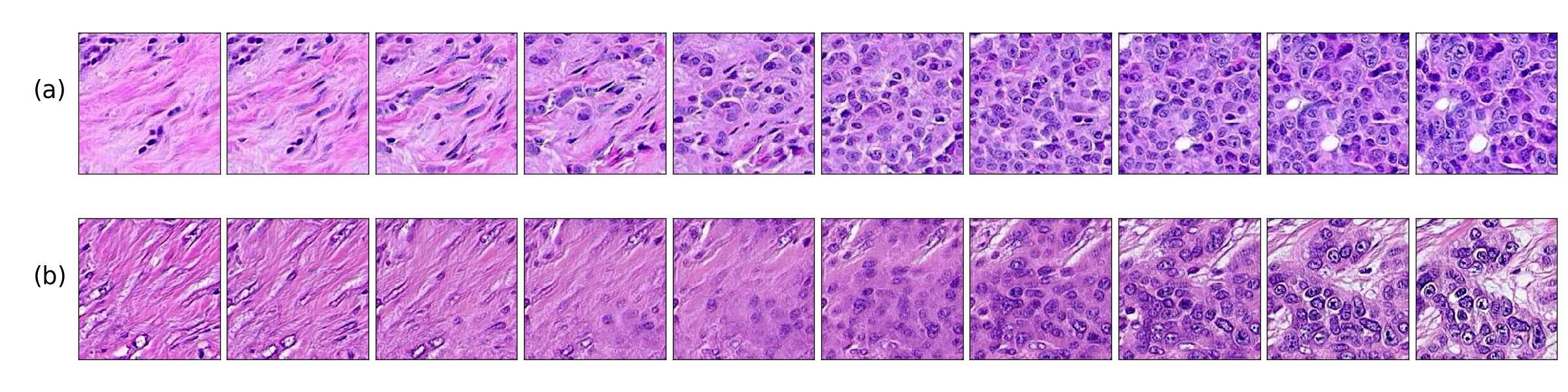}
            \includegraphics[scale=0.17]{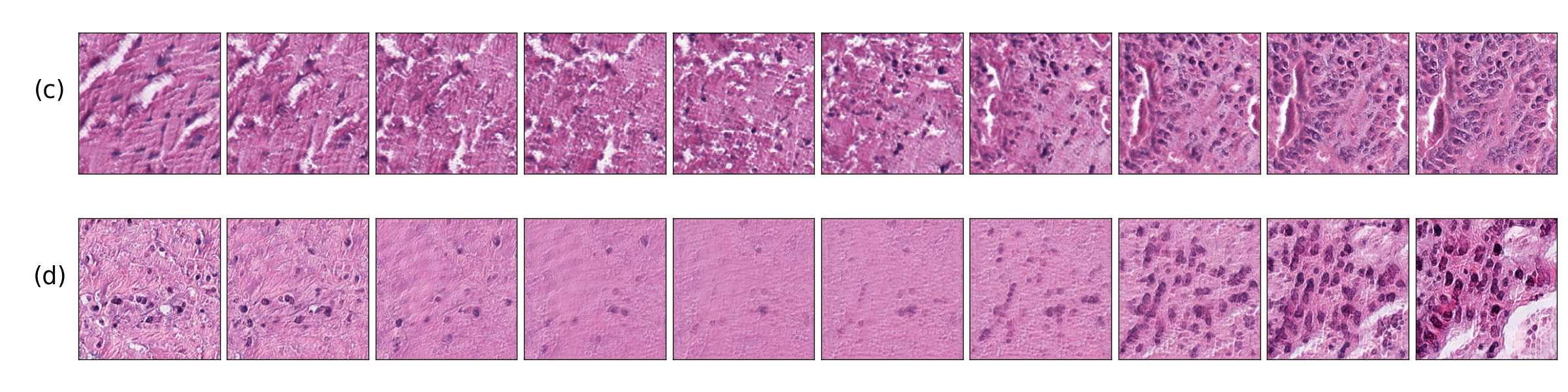}
            \caption{Linear interpolation in the latent space $w$ from a benign (less cancer cells, left end) to a malignant tissue (more cancer cells, right end) on breast cancer (a, b) and colorectal cancer (c, d). (a, c) PathologyGAN model interpolations with a mapping network and style mixing regularization. (b, d) PathologyGAN model interpolations without a mapping network and style mixing regularization. (a, c) includes an increasing population of cancer cells rather than a fading effect from model (b, d), this shows that model (a, c) better translates high level features of images from latent space vectors.}
            \label{fig:linear_interpolation}
        \end{figure}

        \begin{figure}[!t]
            \centering
            \includegraphics[scale=0.15, trim=0 0 0 100, clip]{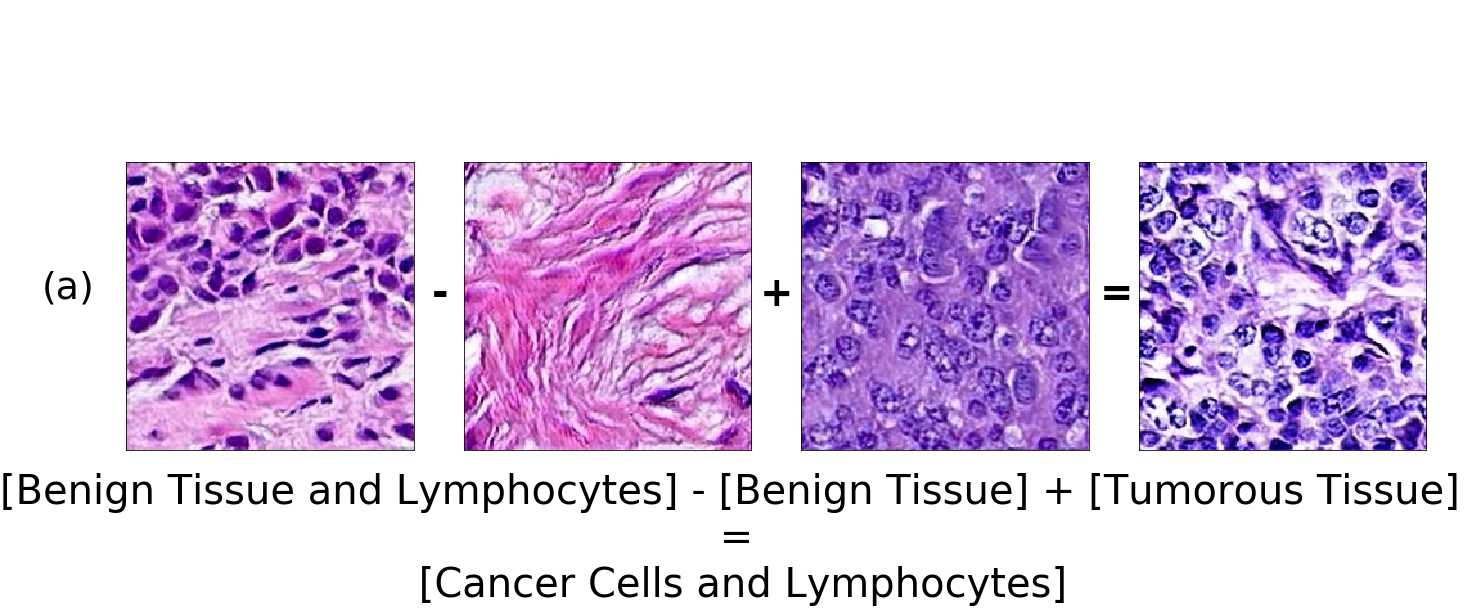}
            \includegraphics[scale=0.15, trim=0 0 0 100, clip]{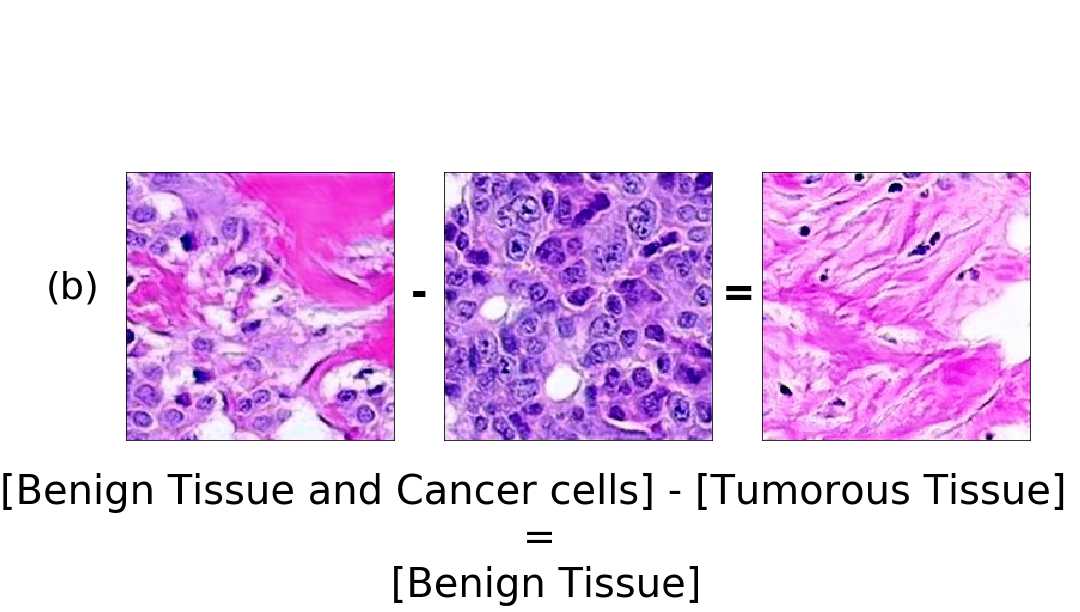}
            \includegraphics[scale=0.15, trim=0 0 0 100, clip]{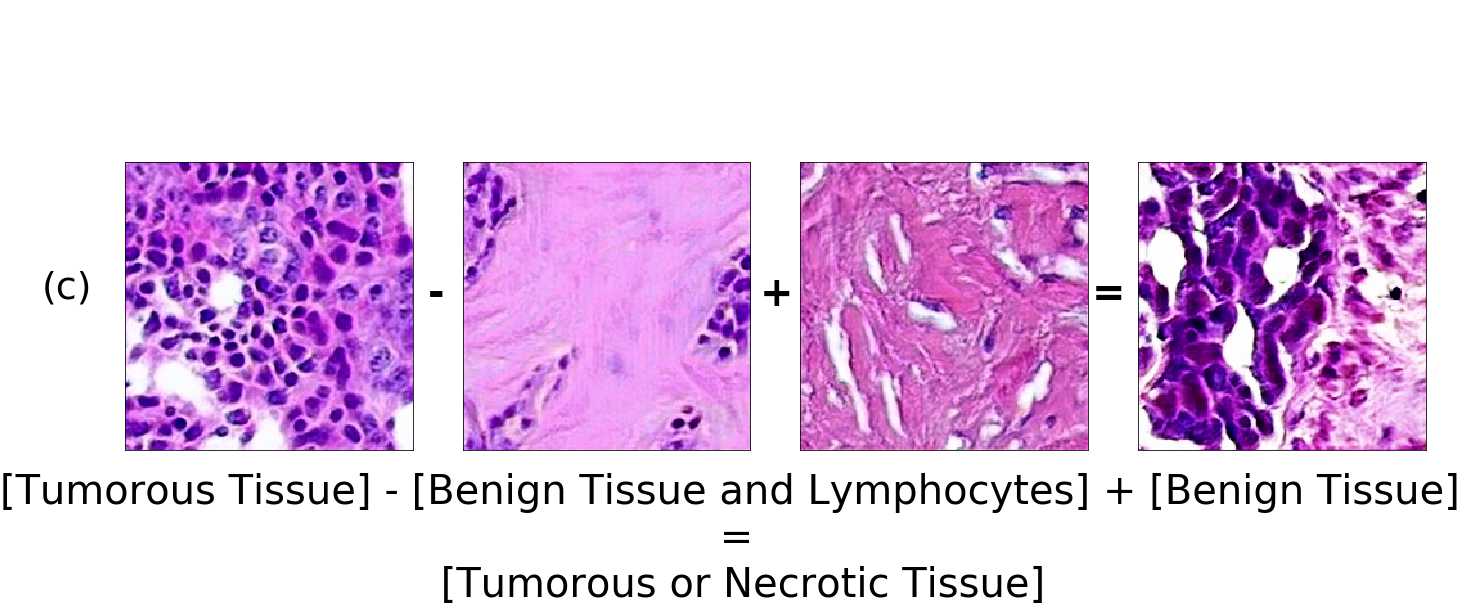}        
            \includegraphics[scale=0.15, trim=0 0 0 100, clip]{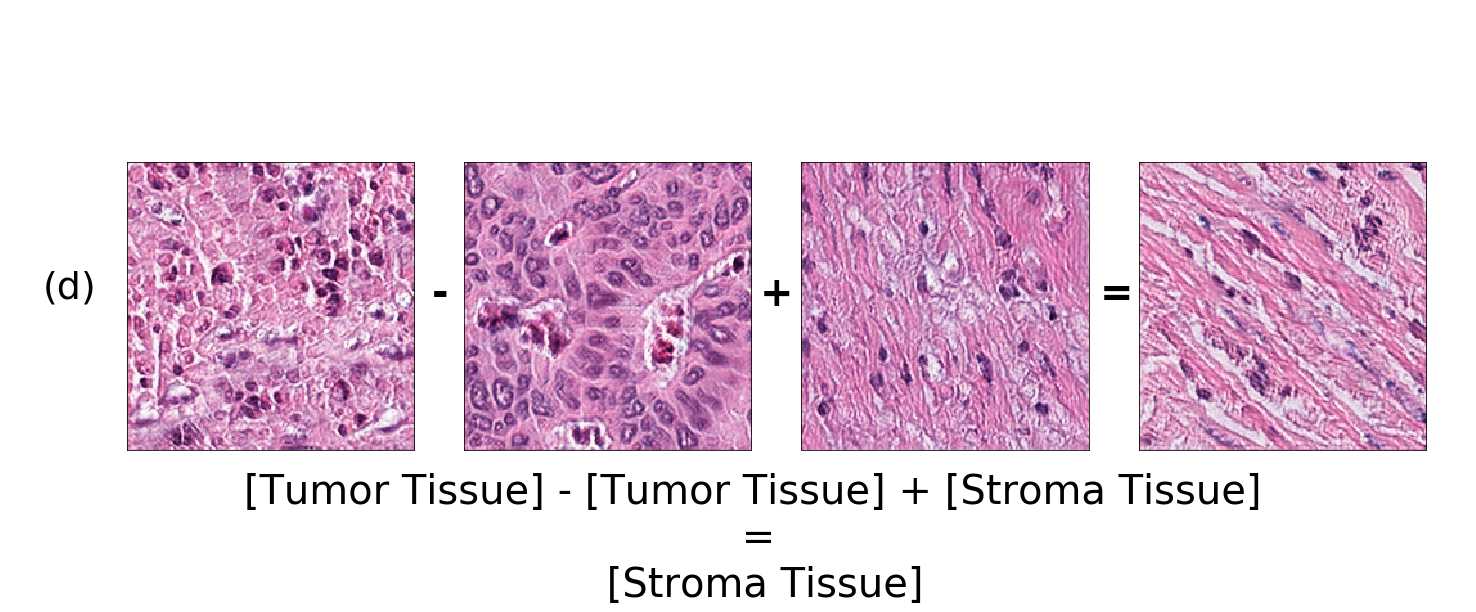}
            \includegraphics[scale=0.15, trim=0 0 0 100, clip]{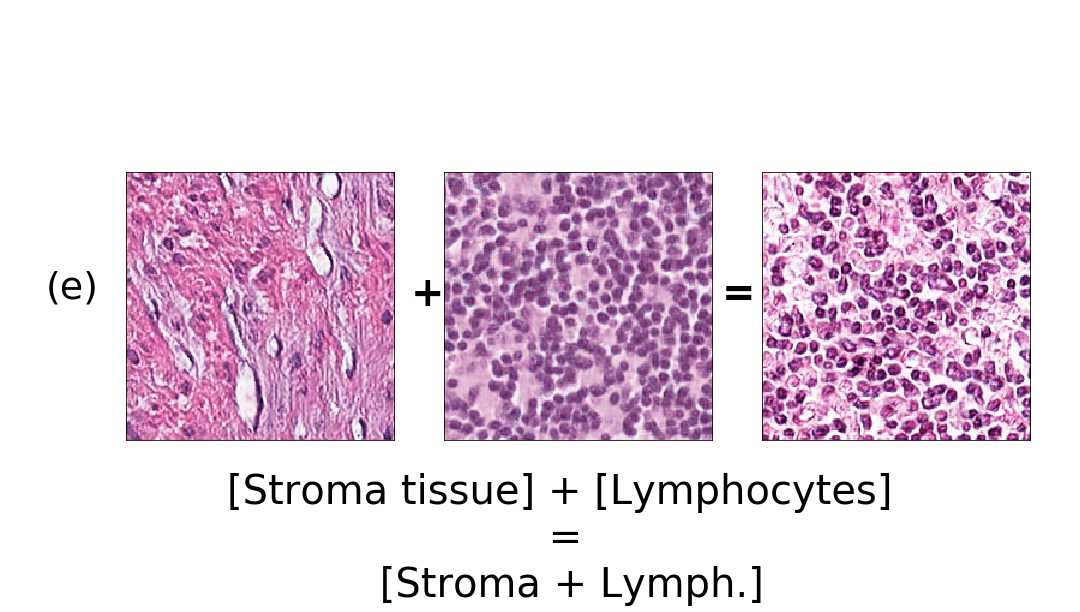}
            \includegraphics[scale=0.15, trim=0 0 0 100, clip]{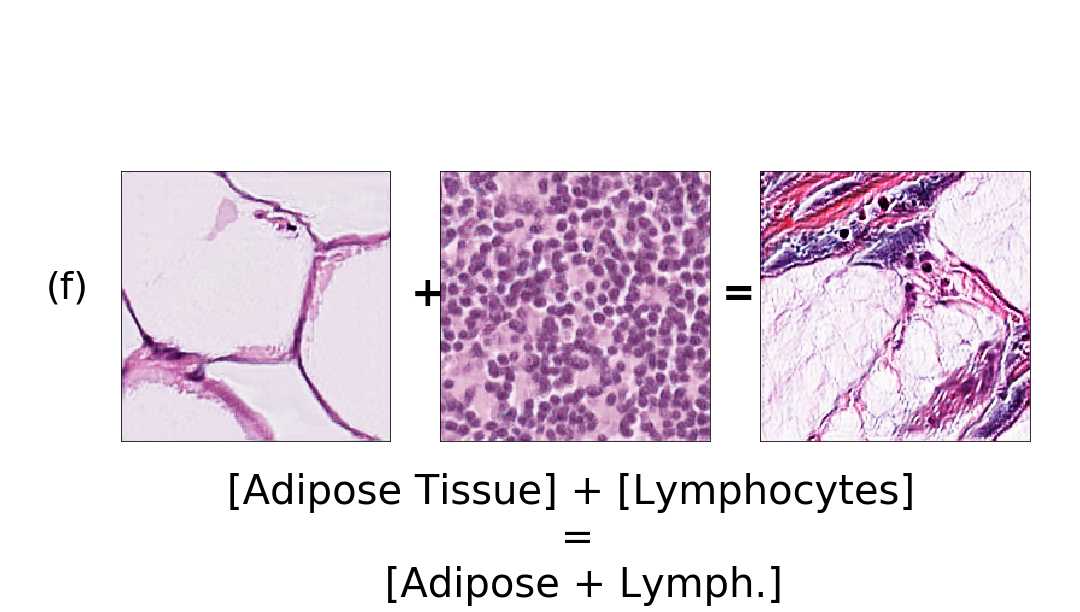}
            \includegraphics[scale=0.15, trim=0 0 0 100, clip]{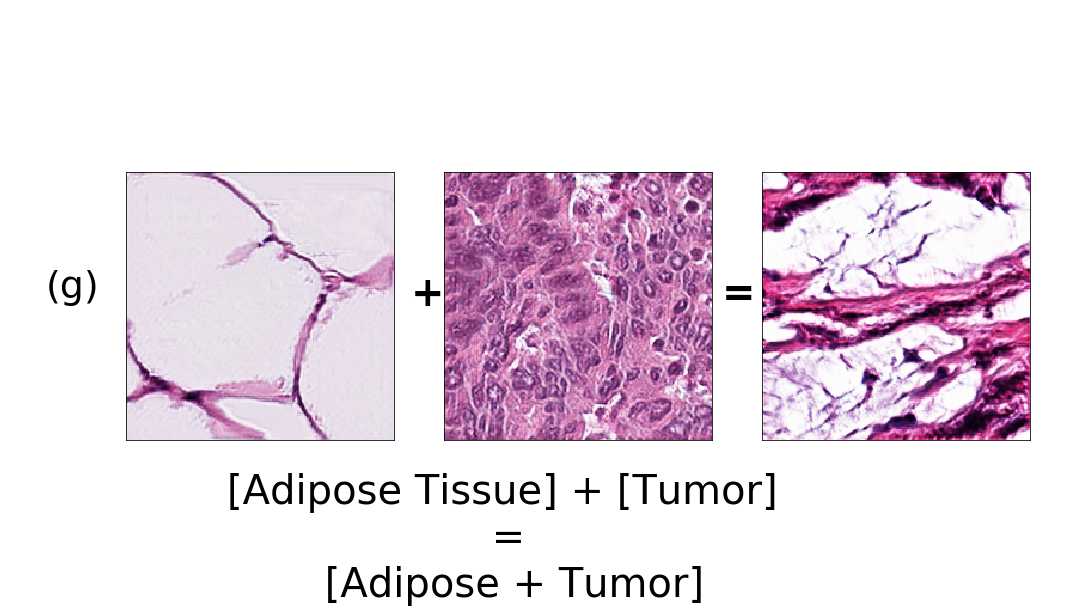}
            \vspace{-0.5cm}
            \caption{Linear vector operations on the latent space $w$ translate into image feature transformations. We gather latent vectors that generate images with different high level features and perform linear operations on the vectors before we feed the generator, resulting into semantic translations of the characteristics of the images. We perform the arithmetic operations (a, b, and c) on breast cancer tissue and (d, e, f, and g) on colorectal cancer tissue.}
            \label{fig:vector_operations}
        \end{figure}
    
        
        In this section we focus on the PathologyGAN's latent space, exploring the impact of introducing a mapping network in the generator and using style mixing regularization. Here we will provide examples of its impact on linear interpolations and vector operations on the latent space $w$, as well as visualizations on the latent space $w$. We conclude that PathologyGAN holds representation learning properties over cancer tissue morphologies. 
    
        Figures \ref{fig:latent_space_vgh_nki} and \ref{fig:latent_space_crc} capture how the latent space $w$ has a structure that shows direct relationship with tissue properties. To create this figures, we generated $10K$ images and labeled them accordingly to Sections \ref{quantification_images} and \ref{tissue_images}, along with each tissue image we also have the corresponding latent vector $w$ and we used UMAP \citep{McInnes2018} to project them to a two dimensional space. 
        
        Figure \ref{fig:latent_space_vgh_nki} reveals the relationship between number of cancer cells in the breast cancer tissue and regions of the latent space, low counts of cancer cells (class 0) are concentrated at quadrant $II$ while they increase as we move to quadrant $IV$ (class 7). Figure \ref{fig:latent_space_crc} displays how the distinct regions of the latent space generate different kinds of tissue. These examples provide evidence of a structured latent space according to tissue cellular characteristics and tissue type. We include a further detailed exploration with density and scatter plots in the Appendix \ref{appendix:map_style}.
        
        We also found that linear interpolations between two latent vectors $w$ have better feature transformations when the mapping network and style mixing regularization are introduced. Figure \ref{fig:linear_interpolation} shows linear interpolations in latent space $w$ between images with malignant tissue and benign tissue. (a, c) correspond to a model with a mapping network and style mixing regularization and (b, d) to a model without those features, we can see that transitions on (a, c) include an increasing population of cancer cells rather than the fading effect observed in images of (b, d). This result indicates that (a, c) better translates interpolations in the latent space, as real cells do not fade away. 
        
        In addition, we performed linear vector operations in $w$, that translated into semantic image features transformations. In Figure \ref{fig:vector_operations} we provide examples of three vector operations that result into feature alterations in the images. This evidence shows further support on the relation between a structured latent space and tissue characteristics. 
        
\begin{figure}[!t]
    		\centering
    		\minipage{0.5\textwidth}
    		  \includegraphics[scale=0.2]{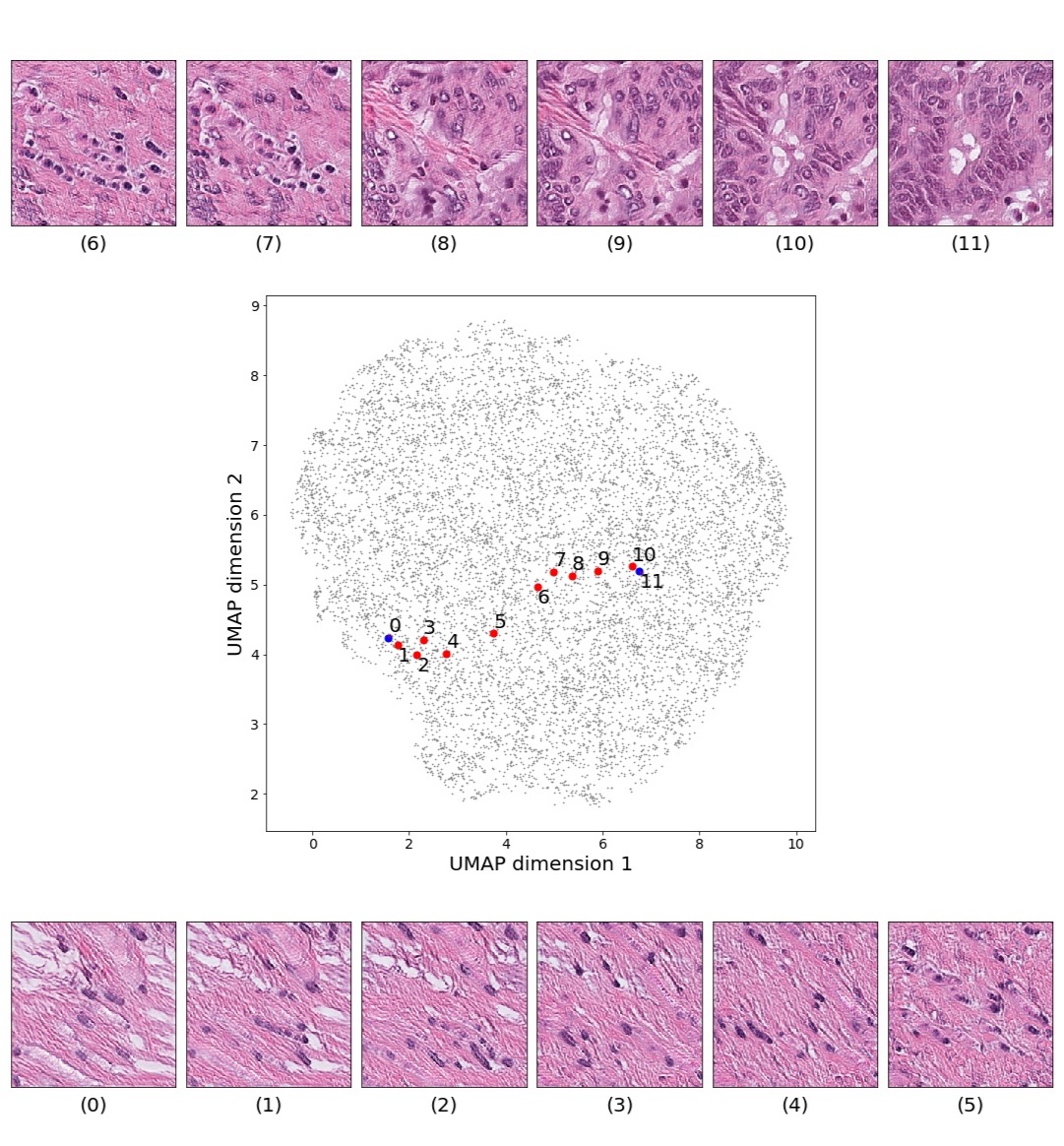}
    		\endminipage\hfill
    		\minipage{0.5\textwidth}
    		  \includegraphics[scale=0.2]{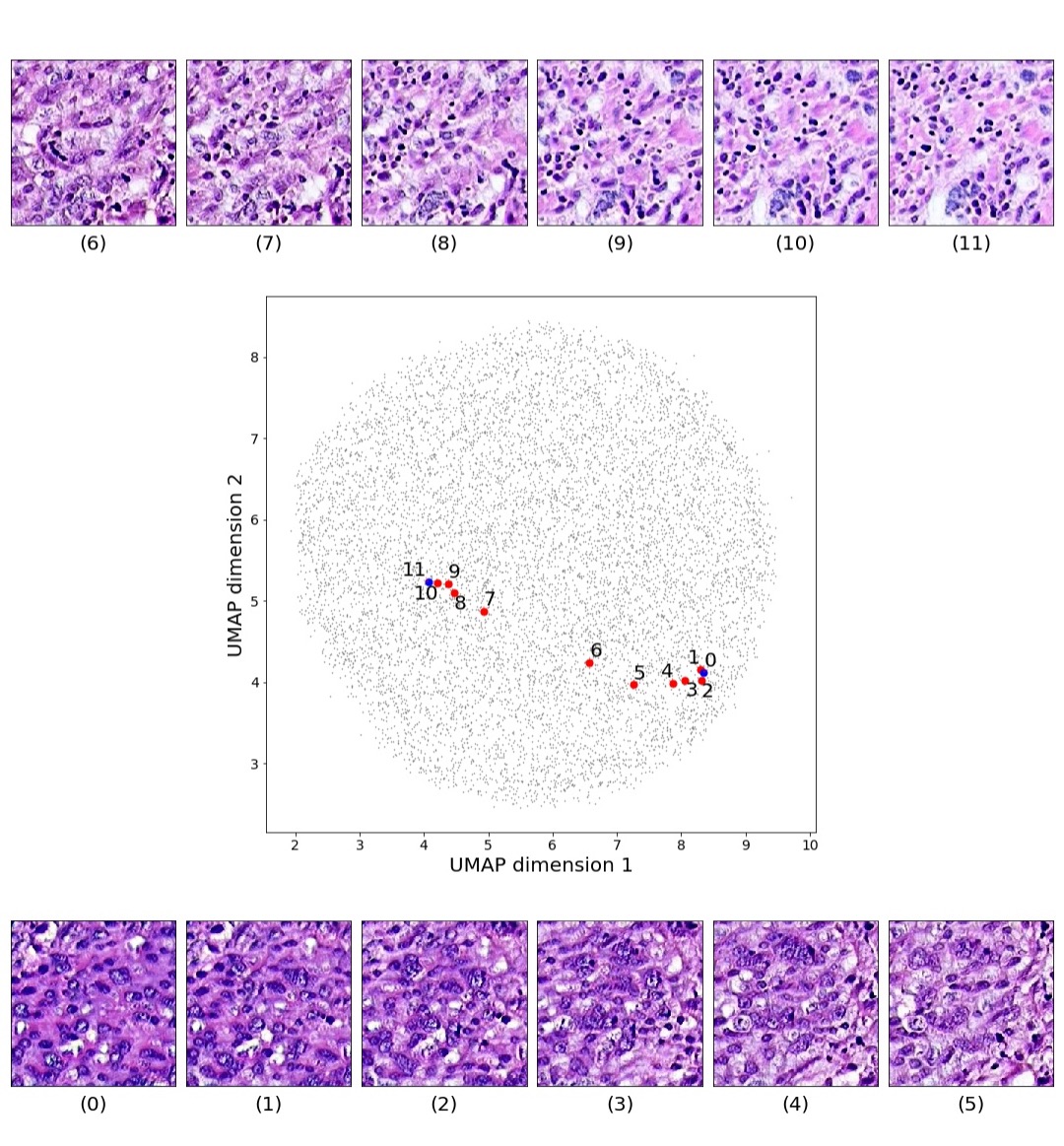}
    		\endminipage
            \vspace{-0.25cm}
    		 \caption{Uniform Manifold Approximation and Projection (UMAP) representations of generated tissue samples where linear interpolations in the latent space are highlighted. Colorectal cancer (left) shows a transition between stroma and tumor, while Brest cancer (right) shows a transition between tumor and lymphocytes. Starting vectors are colored in blue while intermediate points of the interpolations are colored in red. Through the intermediate vectors we show that gradual transitions in the latent space translate into smooth feature transformations, increase/decrease of tumorous cells or increase of lymphocyte counts.}
    		 \label{fig:linear_interpolation_latent_space}
    	\end{figure}

\begin{figure}[!t]
    		\centering
    		\minipage{0.5\textwidth}
    		  \includegraphics[scale=0.25]{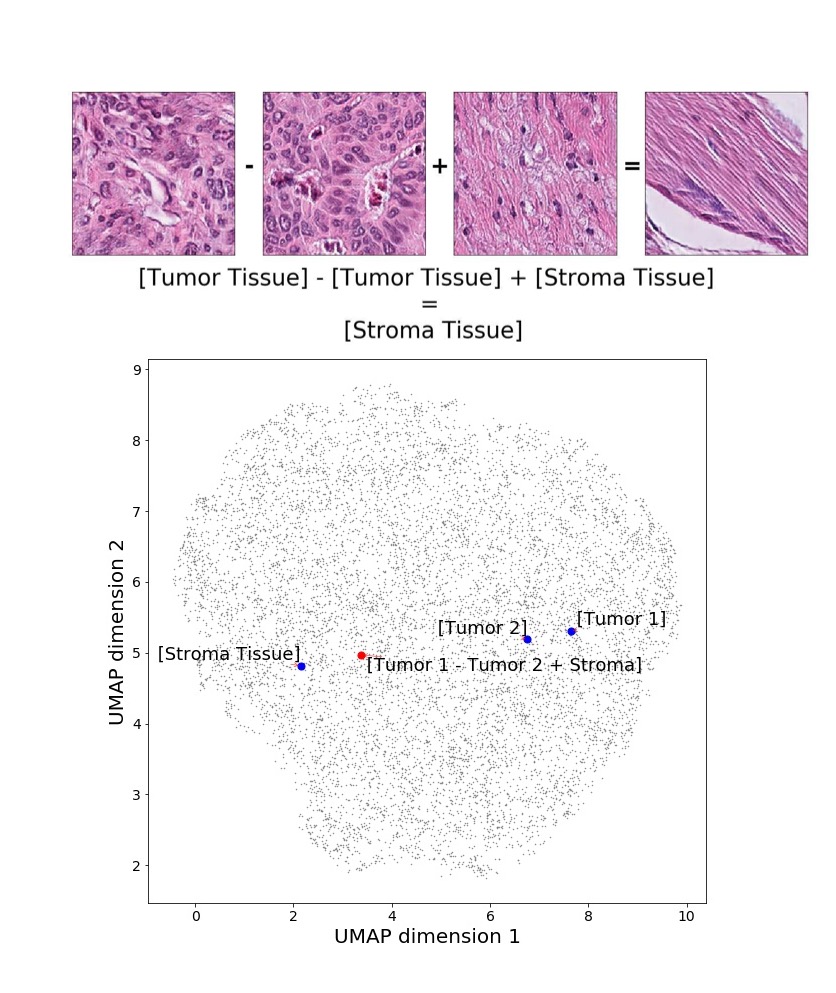}
    		\endminipage\hfill
    		\minipage{0.5\textwidth}
    		  \includegraphics[scale=0.25]{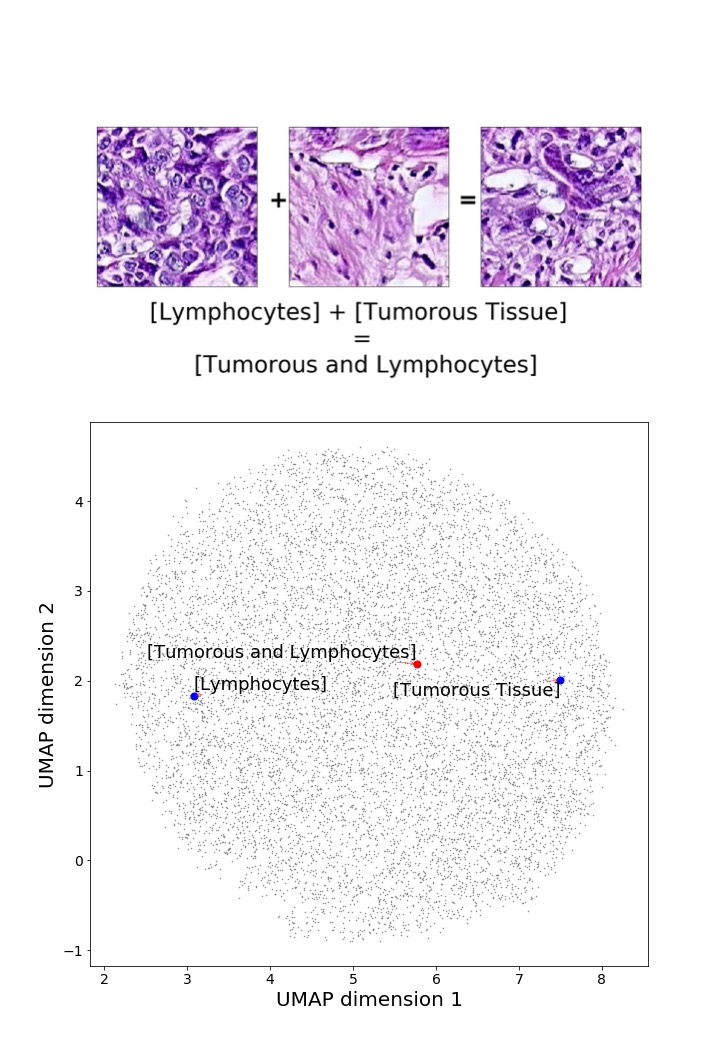}
    		\endminipage
            \vspace{-0.5cm}
    		 \caption{Uniform Manifold Approximation and Projection (UMAP) representations of generated tissue samples where vectors involved in the linear vector operations are highlighted. Original vectors are colored in blue while the results are colored in red. We show  colorectal cancer examples on the left and breast cancer examples on the right. After vector operations the results fall into regions of the latent space that correspond to the tissue type (left) or a combination of features, tumor and lymphocytes (right).}
    		 \label{fig:vector_op_latent_space}
    	\end{figure}

        Finally, we explored how linear interpolation and vector operations translate into the individual points in the latent space. In Figure \ref{fig:linear_interpolation_latent_space} we provide examples of interpolations from stroma to tumor in colorectal cancer, and tumor to lymphocytes in breast cancer. Through the intermediate vectors we show that gradual transitions in the latent space translate into smooth feature transformations, in these cases increase/decrease of tumorous cells or increase of lymphocyte counts. Alternatively, Figure \ref{fig:vector_op_latent_space} shows how results from vector operations fall into regions of the latent space that correspond to the expected tissue type or a combination of features such as tumor and lymphocytes. With these figures we visualize the meaningful representations of cells and tissue types, and also the interpretability of the latent space. Appendix \ref{appendix:latent_li_vp} contains additional examples of these visualizations.
        
    \subsection{Pathologists' results}
    
        \begin{figure}[!t]
            \centering
            \includegraphics[scale=0.17]{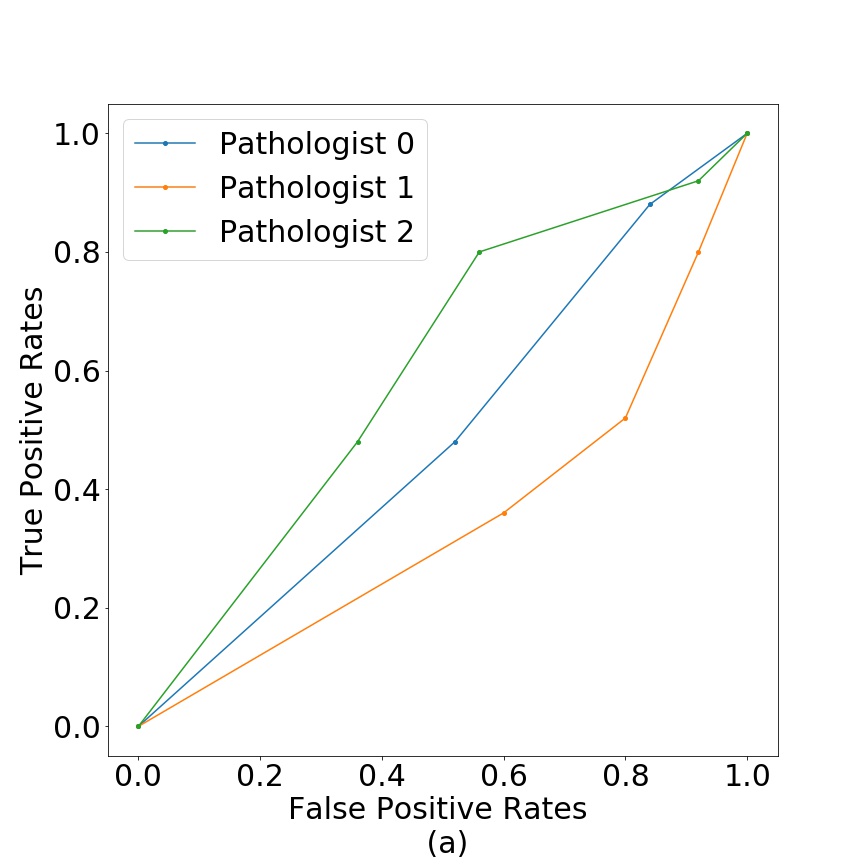}
            \includegraphics[scale=0.17]{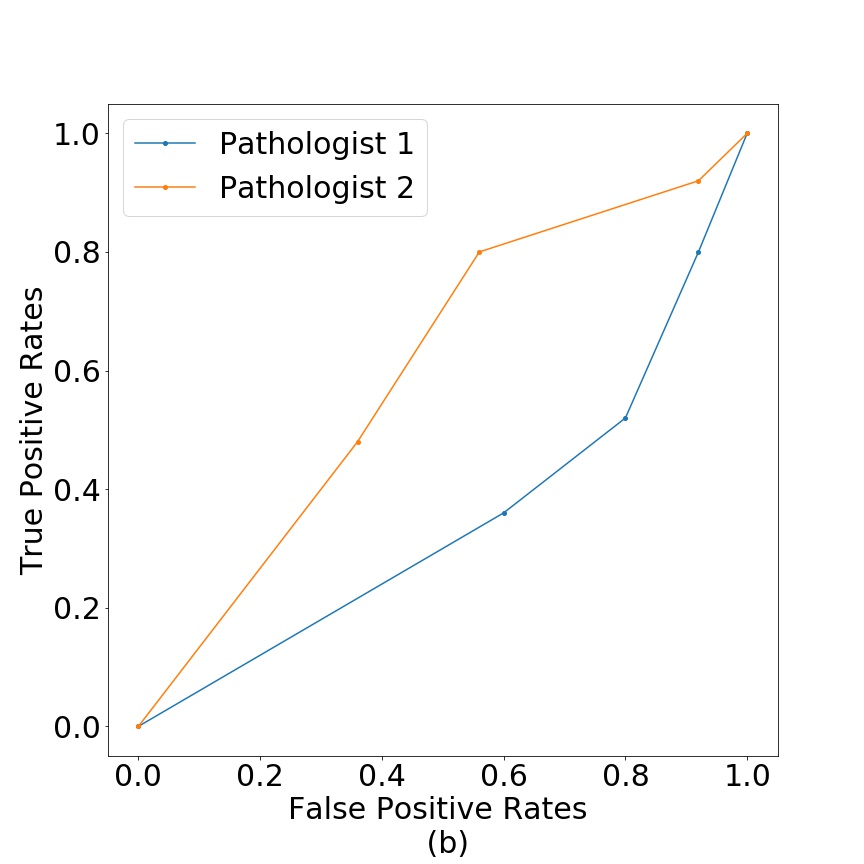}
            \vspace{-0.25cm}
            \caption{ROC curve of Pathologists' real/fake classification for breast (a) and colorectal cancer tissue (b).  The near random classification performance from both expert pathologists suggests that generated tissue images do not present artifacts that give away the tissue as generated.}
            \label{fig:roc_curve}
        \end{figure}   
        
    
    
    
    To demonstrate that the generated images can sustain the scrutiny of clinical examination, we asked expert pathologists to take a test, setup as follows:
    \begin{itemize}
        \item 50 Individual images - Pathologists were asked to rate all individual images from 1 to 5, where 5 meant the image appeared the most real.
    \end{itemize}
    
    We chose fake images in two ways, with half of them hand-selected and the other half with fake images that had the smallest Euclidean distance to real images in the convolutional feature space (Inception-V1). All the real images are randomly selected between the three closest neighbors of the fake images. 
    
    Figure~\ref{fig:roc_curve} shows the test results in terms of false positive vs true positive for breast (a) and colorectal cancer tissue (b). We can see that pathologist classification is close to random. The pathologists mentioned that the usual procedure is to work with larger images with bigger resolution, but that the generated fake images were of a quality, that at the $224\times224$ size used in this work, they were not able to differentiate between real and fake tissue.

\section{Discussion}
    
    Our goal is to develop a generative model that is able to capture and create representations based tissue architectures and cellular characteristics that define phenotype, for this reason our results are focused on testing two features of our model: Image quality of generated tissue and  interpretability of representations/structure of the latent space.
    
    Through image quality we tested the model's ability to learn and reproduce the distribution of tissue and cellular information, we argue that by doing so it would have capture phenotypes. The FID results on Table \ref{GAN_results-table} show that PathologyGAN does reproduce these distributions not only when it is judged by convolutional features ($32.05/16.65$) but also when we use an external tool to directly quantify cellular information in the tissue ($9.86$). We would like to highlight that the FID score measures the distribution difference between real and generated samples, and given the low values we argue that the model is also capturing the abundance or scarcity of different tissue patterns.
    
    Figure \ref{fig:nearest_neighbors} shows generated tissue samples and their closest real neighbors in the Inception-V1 convolutional space, allowing us to visually inspect the similarity between them. In the colorectal cancer samples we can see that the different types of tissue (from tumor (i) to background (p)) clearly resemble the real tissue. Breast cancer tissue samples (a-h) also show the same result where the real and generated tissue present the same patterns and shapes.
    
    As an additional image quality verification, we tested the generated images of our model against pathologists' interpretation. We aimed to test that the generated tissues do not contain any artifacts that give them away as fake through the eyes of professionals. Given the near random classification between real/fake in Figure \ref{fig:roc_curve}, we conclude that generated samples are realistic enough to pass as real through the examination of pathologists. We argue this is relevant because the model is able to reproduce tissue patterns to which pathologists are accustom to. 
    
    With the previous results we conclude that the model is able to reproduce the detail and distribution of tissue, not only from the convolutional features or cellular characteristics perspective, but also by the interpretation of pathologists. 
    
    In relation to PathologyGAN's latent space structure and interpretability of its representations, Figures \ref{fig:latent_space_vgh_nki} and \ref{fig:latent_space_crc} show how distinct regions of the latent space hold tissue and cellular information about the generated images. Figure \ref{fig:latent_space_vgh_nki} shows a clear relationship between regions of the latent space and the cancer cell density in the tissue, while Figure \ref{fig:latent_space_crc} directly links the region of the space with the different tissue types (e.g. tumor, stroma, muscle, lymphocytes, debris, mucus, adipose, and background). In Figures \ref{fig:linear_interpolation} and \ref{fig:linear_interpolation_latent_space} we relate linear interpolations between latent vectors and generated tissue attributes, showing that gradual transitions in the latent space transfer into smooth feature transformations, in these cases increase/decrease of tumorous cells or increase of lymphocyte counts. Finally, Figures \ref{fig:vector_operations} and \ref{fig:vector_op_latent_space} provide examples of linear vector operations and their translation into tissue characteristic changes, with different tissue changes. These results provide support on the representation learning properties of the model, not only holding meaningful information of cell and tissue types but also an interpretable insight to the representations themselves. 
    
    As future research, we consider that our model could be used in different settings. PathologyGAN could be extended to achieve higher resolutions such as $1024\times1024$, a level of resolution which could include complete TMAs which hold value for diagnosis and prognosis. In these cases, pathologists' insight will hold a higher value, since they are used to working at WSI or TMA level. In addition, the study with pathologists could be extended to include a larger number of experts and to use a random sample of generated images which would give a less biased result. 
    
    The representation learning properties of the model can also contribute as a educational tool, providing tissue samples with certain cellular characteristics that are rare, elucidating possible transitions of tissue (e.g. from tumor to high lymphocyte infiltration), or enabling the study of borderline cases (e.g. atypia) between generated images and pathologists interpretations.
    
    Furthermore, the  model could be used to generate synthetic samples to improve classifiers performance, helping to avoid overfitting and providing samples with rare tissue pathologies. 
    
    Finally, we consider the model can contribute to characterizing phenotype patterns, creating representations by cellular and tissue morphologies, this is where we think the tissue representation learning properties are key. Linking these phenotypes to patient related information such as genomic, transcriptomic information, or survival expectancy. Ultimately this could give insight to the tumor microenvironment recorded in the WSIs and a better understanding of the disease. In order to achieve this goal, it will require exploring the addition of an encoder to map real images into the GAN's latent space \citep{quiros2020learning} and verify its converge when large amounts of samples are used, since WSIs in datasets like TCGA could amount to millions of tissue samples. 
    
\section{Conclusion}
    We presented a new approach to the use of machine learning in digital pathology, using GANs to learn cancer tissue representations. We assessed the quality of the generated images through the FID metric, using the convolutional features of a Inception-V1 network and quantitative cellular information of the tissue, both showed consistent state-of-the-art values for different kinds of tissue, breast and colorectal cancer. We showed that our model allows high level interpretation of its latent space, even performing linear operations that translate into feature tissue transformations. Finally, we demonstrate that the quality of the generated images do not allow pathologists to reliably find differences between real and generated images.
    
    With PathologyGAN we proposed a generative model that captures representations of entire tissue architectures and defines an interpretable latent space (e.g. colour, texture, spatial features of cancer and normal cells, and their interaction), contributing to generative models ability to capture phenotype representations.

\acks{We would like to thank Joanne Edwards, Christopher Bigley, and Elizabeth Mallon for helpful insights and discussions on this work.

We will also like to acknowledge funding support from University of Glasgow on A.C.Q scholarship, K.Y from EPSRC grant EP/R018634/1., and R.M-S. from EPSRC grants EP/T00097X/1 and EP/R018634/1.}

\ethics{The work follows appropriate ethical standards in conducting research and writing the manuscript. Our models were trained with publicly available data, for which no ethical approval was required.}

\coi{We declare we don't have conflicts of interest.}

\bibliography{sample}

\begin{thebibliography}{60}
\providecommand{\natexlab}[1]{#1}
\providecommand{\url}[1]{\texttt{#1}}
\expandafter\ifx\csname urlstyle\endcsname\relax
  \providecommand{\doi}[1]{doi: #1}\else
  \providecommand{\doi}{doi: \begingroup \urlstyle{rm}\Url}\fi

\bibitem[Barratt and Sharma(2018)]{Barratt2018}
Shane Barratt and Rishi Sharma.
\newblock A note on the {Inception} score.
\newblock \emph{CoRR}, abs/1801.01973, 2018.
\newblock URL \url{http://arxiv.org/abs/1801.01973}.

\bibitem[Beck et~al.(2011)Beck, Sangoi, Leung, Marinelli, Nielsen, van~de
  Vijver, West, van~de Rijn, and Koller]{Beck2011}
Andrew~H Beck, Ankur~R Sangoi, Samuel Leung, Robert~J Marinelli, Torsten~O
  Nielsen, Marc~J van~de Vijver, Robert~B West, Matt van~de Rijn, and Daphne
  Koller.
\newblock Systematic analysis of breast cancer morphology uncovers stromal
  features associated with survival.
\newblock \emph{Sci Transl Med}, 3\penalty0 (108):\penalty0 108ra113, Nov 2011.
\newblock ISSN 1946-6242 (Electronic); 1946-6234 (Linking).
\newblock \doi{10.1126/scitranslmed.3002564}.

\bibitem[Bergenstr{\aa}hle et~al.(2020{\natexlab{a}})Bergenstr{\aa}hle,
  Larsson, and Lundeberg]{Bergenstrahle2020}
Joseph Bergenstr{\aa}hle, Ludvig Larsson, and Joakim Lundeberg.
\newblock Seamless integration of image and molecular analysis for spatial
  transcriptomics workflows.
\newblock \emph{BMC Genomics}, 21\penalty0 (1):\penalty0 482,
  2020{\natexlab{a}}.
\newblock \doi{10.1186/s12864-020-06832-3}.
\newblock URL \url{https://doi.org/10.1186/s12864-020-06832-3}.

\bibitem[Bergenstr{\aa}hle et~al.(2020{\natexlab{b}})Bergenstr{\aa}hle, He,
  Bergenstr{\aa}hle, Andersson, Lundeberg, Zou, and
  Maaskola]{Bergenstrahle2020_2}
Ludvig Bergenstr{\aa}hle, Bryan He, Joseph Bergenstr{\aa}hle, Alma Andersson,
  Joakim Lundeberg, James Zou, and Jonas Maaskola.
\newblock Super-resolved spatial transcriptomics by deep data fusion.
\newblock \emph{bioRxiv}, page 2020.02.28.963413, 01 2020{\natexlab{b}}.

\bibitem[Bińkowski et~al.(2018)Bińkowski, Sutherland, Arbel, and
  Gretton]{Binkowski2018}
Mikołaj Bińkowski, Dougal~J. Sutherland, Michael Arbel, and Arthur Gretton.
\newblock Demystifying {MMD} {GAN}s.
\newblock In \emph{International Conference on Learning Representations}, 2018.
\newblock URL \url{https://openreview.net/forum?id=r1lUOzWCW}.

\bibitem[Brock et~al.(2019)Brock, Donahue, and Simonyan]{Brock2018}
Andrew Brock, Jeff Donahue, and Karen Simonyan.
\newblock Large scale {GAN} training for high fidelity natural image synthesis.
\newblock In \emph{International Conference on Learning Representations}, 2019.
\newblock URL \url{https://openreview.net/forum?id=B1xsqj09Fm}.

\bibitem[Campbell et~al.(2020)Campbell, Getz, Korbel, and et~al.]{Campbell2020}
Peter~J. Campbell, Gad Getz, Jan~O. Korbel, and et~al.
\newblock Pan-cancer analysis of whole genomes.
\newblock \emph{Nature}, 578\penalty0 (7793):\penalty0 82--93, 2020.

\bibitem[Coudray and Tsirigos(2020)]{Coudray2020}
Nicolas Coudray and Aristotelis Tsirigos.
\newblock Deep learning links histology, molecular signatures and prognosis in
  cancer.
\newblock \emph{Nature Cancer}, 1\penalty0 (8):\penalty0 755--757, 2020.
\newblock \doi{10.1038/s43018-020-0099-2}.
\newblock URL \url{https://doi.org/10.1038/s43018-020-0099-2}.

\bibitem[Coudray et~al.(2018)Coudray, Ocampo, Sakellaropoulos, Narula, Snuderl,
  Feny{\"o}, Moreira, Razavian, and Tsirigos]{Coudray2018}
Nicolas Coudray, Paolo~Santiago Ocampo, Theodore Sakellaropoulos, Navneet
  Narula, Matija Snuderl, David Feny{\"o}, Andre~L. Moreira, Narges Razavian,
  and Aristotelis Tsirigos.
\newblock Classification and mutation prediction from non--small cell lung
  cancer histopathology images using deep learning.
\newblock \emph{Nature Medicine}, 24\penalty0 (10):\penalty0 1559--1567, 2018.
\newblock \doi{10.1038/s41591-018-0177-5}.
\newblock URL \url{https://doi.org/10.1038/s41591-018-0177-5}.

\bibitem[de~Bel et~al.(2018)de~Bel, Hermsen, Smeets, Hilbrands, van~der Laak,
  and Litjens]{deBel2018}
Thomas de~Bel, Meyke Hermsen, Bart Smeets, Luuk Hilbrands, Jeroen van~der Laak,
  and Geert Litjens.
\newblock Automatic segmentation of histopathological slides of renal tissue
  using deep learning.
\newblock In \emph{Medical Imaging 2018: Digital Pathology}, volume 10581, page
  1058112. International Society for Optics and Photonics, 2018.

\bibitem[Deng et~al.(2009)Deng, Dong, Socher, Li, Li, and
  Fei-Fei]{imagenet_cvpr09}
J.~Deng, W.~Dong, R.~Socher, L.-J. Li, K.~Li, and L.~Fei-Fei.
\newblock {ImageNet: A Large-Scale Hierarchical Image Database}.
\newblock In \emph{CVPR09}, 2009.

\bibitem[Dentro et~al.(2020)Dentro, Leshchiner, and et~al.]{Dentro2020}
Stefan~C. Dentro, Ignaty Leshchiner, and et~al.
\newblock Characterizing genetic intra-tumor heterogeneity across 2,658 human
  cancer genomes.
\newblock \emph{bioRxiv}, 2020.
\newblock \doi{10.1101/312041}.
\newblock URL \url{https://www.biorxiv.org/content/early/2020/04/22/312041}.

\bibitem[Fr{\'e}chet(1957)]{frechet1957}
Maurice Fr{\'e}chet.
\newblock Sur la distance de deux lois de probabilit{\'e}.
\newblock \emph{COMPTES RENDUS HEBDOMADAIRES DES SEANCES DE L ACADEMIE DES
  SCIENCES}, 244\penalty0 (6):\penalty0 689--692, 1957.

\bibitem[Fu et~al.(2020)Fu, Jung, Torne, Gonzalez, V{\"o}hringer, Shmatko,
  Yates, Jimenez-Linan, Moore, and Gerstung]{Fu2020}
Yu~Fu, Alexander~W. Jung, Ramon~Vi{\~n}as Torne, Santiago Gonzalez, Harald
  V{\"o}hringer, Artem Shmatko, Lucy~R. Yates, Mercedes Jimenez-Linan, Luiza
  Moore, and Moritz Gerstung.
\newblock Pan-cancer computational histopathology reveals mutations, tumor
  composition and prognosis.
\newblock \emph{Nature Cancer}, 1\penalty0 (8):\penalty0 800--810, 2020.
\newblock \doi{10.1038/s43018-020-0085-8}.
\newblock URL \url{https://doi.org/10.1038/s43018-020-0085-8}.

\bibitem[Gadermayr et~al.(2018)Gadermayr, Gupta, Klinkhammer, Boor, and
  Merhof]{Gadermayr2018}
Michael Gadermayr, Laxmi Gupta, Barbara~M Klinkhammer, Peter Boor, and Dorit
  Merhof.
\newblock Unsupervisedly training gans for segmenting digital pathology with
  automatically generated annotations.
\newblock \emph{arXiv preprint arXiv:1805.10059}, 2018.

\bibitem[Gadermayr et~al.(2019)Gadermayr, Gupta, Appel, Boor, Klinkhammer, and
  Merhof]{Gadermayr2019}
Michael Gadermayr, Laxmi Gupta, Vitus Appel, Peter Boor, Barbara~M Klinkhammer,
  and Dorit Merhof.
\newblock Generative adversarial networks for facilitating stain-independent
  supervised and unsupervised segmentation: a study on kidney histology.
\newblock \emph{IEEE transactions on medical imaging}, 38\penalty0
  (10):\penalty0 2293--2302, 2019.

\bibitem[Gerstung et~al.(2020)Gerstung, Jolly, and et~al.]{Gerstung2020}
Moritz Gerstung, Clemency Jolly, and et~al.
\newblock The evolutionary history of 2,658 cancers.
\newblock \emph{Nature}, 578\penalty0 (7793):\penalty0 122--128, 2020.

\bibitem[Goodfellow et~al.(2014)Goodfellow, Pouget-Abadie, Mirza, Xu,
  Warde-Farley, Ozair, Courville, and Bengio]{Goodfellow2014}
Ian Goodfellow, Jean Pouget-Abadie, Mehdi Mirza, Bing Xu, David Warde-Farley,
  Sherjil Ozair, Aaron Courville, and Yoshua Bengio.
\newblock Generative adversarial nets.
\newblock In Z.~Ghahramani, M.~Welling, C.~Cortes, N.~D. Lawrence, and K.~Q.
  Weinberger, editors, \emph{Advances in Neural Information Processing Systems
  27}, pages 2672--2680. Curran Associates, Inc., 2014.
\newblock URL
  \url{http://papers.nips.cc/paper/5423-generative-adversarial-nets.pdf}.

\bibitem[Gretton et~al.(2012)Gretton, Borgwardt, Rasch, Sch{{\"o}}lkopf, and
  Smola]{Gretton2012}
Arthur Gretton, Karsten~M. Borgwardt, Malte~J. Rasch, Bernhard Sch{{\"o}}lkopf,
  and Alexander Smola.
\newblock A kernel two-sample test.
\newblock \emph{Journal of Machine Learning Research}, 13\penalty0
  (25):\penalty0 723--773, 2012.
\newblock URL \url{http://jmlr.org/papers/v13/gretton12a.html}.

\bibitem[He et~al.(2020)He, Bergenstr{\aa}hle, Stenbeck, Abid, Andersson, Borg,
  Maaskola, Lundeberg, and Zou]{He2020}
Bryan He, Ludvig Bergenstr{\aa}hle, Linnea Stenbeck, Abubakar Abid, Alma
  Andersson, {\AA}ke Borg, Jonas Maaskola, Joakim Lundeberg, and James Zou.
\newblock Integrating spatial gene expression and breast tumour morphology via
  deep learning.
\newblock \emph{Nature Biomedical Engineering}, 4\penalty0 (8):\penalty0
  827--834, 2020.
\newblock \doi{10.1038/s41551-020-0578-x}.
\newblock URL \url{https://doi.org/10.1038/s41551-020-0578-x}.

\bibitem[He et~al.(2016)He, Zhang, Ren, and Sun]{He2016}
Kaiming He, Xiangyu Zhang, Shaoqing Ren, and Jian Sun.
\newblock Deep residual learning for image recognition.
\newblock \emph{2016 IEEE Conference on Computer Vision and Pattern Recognition
  (CVPR)}, Jun 2016.
\newblock \doi{10.1109/cvpr.2016.90}.
\newblock URL \url{http://dx.doi.org/10.1109/cvpr.2016.90}.

\bibitem[Heusel et~al.(2017)Heusel, Ramsauer, Unterthiner, Nessler, and
  Hochreiter]{Heusel2017}
Martin Heusel, Hubert Ramsauer, Thomas Unterthiner, Bernhard Nessler, and Sepp
  Hochreiter.
\newblock Gans trained by a two time-scale update rule converge to a local nash
  equilibrium.
\newblock In I.~Guyon, U.~V. Luxburg, S.~Bengio, H.~Wallach, R.~Fergus,
  S.~Vishwanathan, and R.~Garnett, editors, \emph{Advances in Neural
  Information Processing Systems 30}, pages 6626--6637. Curran Associates,
  Inc., 2017.

\bibitem[Hou et~al.(2019)Hou, Nguyen, Kanevsky, Samaras, Kurc, Zhao, Gupta,
  Gao, Chen, Foran, and Saltz]{Hou2019}
Le~Hou, Vu~Nguyen, Ariel~B Kanevsky, Dimitris Samaras, Tahsin~M Kurc, Tianhao
  Zhao, Rajarsi~R Gupta, Yi~Gao, Wenjin Chen, David Foran, and Joel~H Saltz.
\newblock Sparse autoencoder for unsupervised nucleus detection and
  representation in histopathology images.
\newblock \emph{Pattern recognition}, 86:\penalty0 188--200, 02 2019.
\newblock \doi{10.1016/j.patcog.2018.09.007}.
\newblock URL \url{https://pubmed.ncbi.nlm.nih.gov/30631215}.

\bibitem[Hu et~al.(2019)Hu, Tang, Chang, Fan, Lai, and Xu]{Hu2019}
Bo~Hu, Ye~Tang, Eric I-Chao Chang, Yubo Fan, Maode Lai, and Yan Xu.
\newblock Unsupervised learning for cell-level visual representation in
  histopathology images with generative adversarial networks.
\newblock \emph{IEEE Journal of Biomedical and Health Informatics}, 23\penalty0
  (3):\penalty0 1316–1328, May 2019.
\newblock ISSN 2168-2208.
\newblock \doi{10.1109/jbhi.2018.2852639}.
\newblock URL \url{http://dx.doi.org/10.1109/JBHI.2018.2852639}.

\bibitem[Huang et~al.(2018)Huang, Yuan, Xu, Guo, Sun, Wu, and
  Weinberger]{Huang2018}
Gao Huang, Yang Yuan, Qiantong Xu, Chuan Guo, Yu~Sun, Felix Wu, and Kilian
  Weinberger.
\newblock An empirical study on evaluation metrics of generative adversarial
  networks, 2018.
\newblock URL \url{https://openreview.net/forum?id=Sy1f0e-R-}.

\bibitem[Jolicoeur-Martineau(2019)]{Jolicoeur-Martineau2018}
Alexia Jolicoeur-Martineau.
\newblock The relativistic discriminator: a key element missing from standard
  {GAN}.
\newblock In \emph{International Conference on Learning Representations}, 2019.
\newblock URL \url{https://openreview.net/forum?id=S1erHoR5t7}.

\bibitem[Karras et~al.(2019)Karras, Laine, and Aila]{Karras2019}
Tero Karras, Samuli Laine, and Timo Aila.
\newblock A style-based generator architecture for generative adversarial
  networks.
\newblock \emph{2019 IEEE/CVF Conference on Computer Vision and Pattern
  Recognition (CVPR)}, Jun 2019.
\newblock \doi{10.1109/cvpr.2019.00453}.
\newblock URL \url{http://dx.doi.org/10.1109/CVPR.2019.00453}.

\bibitem[Kather et~al.(2018)Kather, Halama, and Marx]{kather_2018}
Jakob~Nikolas Kather, Niels Halama, and Alexander Marx.
\newblock {100,000 histological images of human colorectal cancer and healthy
  tissue}, April 2018.
\newblock URL \url{https://doi.org/10.5281/zenodo.1214456}.

\bibitem[Kather et~al.(2020)Kather, Heij, Grabsch, and et~al.]{Kather2020}
Jakob~Nikolas Kather, Lara~R. Heij, Heike~I. Grabsch, and et~al.
\newblock Pan-cancer image-based detection of clinically actionable genetic
  alterations.
\newblock \emph{Nature Cancer}, 1\penalty0 (8):\penalty0 789--799, 2020.
\newblock \doi{10.1038/s43018-020-0087-6}.
\newblock URL \url{https://doi.org/10.1038/s43018-020-0087-6}.

\bibitem[Katzman et~al.(2018)Katzman, Shaham, Cloninger, Bates, Jiang, and
  Kluger]{Katzman2018}
Jared~L. Katzman, Uri Shaham, Alexander Cloninger, Jonathan Bates, Tingting
  Jiang, and Yuval Kluger.
\newblock Deepsurv: personalized treatment recommender system using a cox
  proportional hazards deep neural network.
\newblock \emph{BMC Medical Research Methodology}, 18\penalty0 (1), Feb 2018.
\newblock ISSN 1471-2288.
\newblock \doi{10.1186/s12874-018-0482-1}.
\newblock URL \url{http://dx.doi.org/10.1186/s12874-018-0482-1}.

\bibitem[Kingma and Ba(2014)]{Kingma2014}
Diederik~P Kingma and Jimmy Ba.
\newblock Adam: A method for stochastic optimization.
\newblock \emph{arXiv preprint arXiv:1412.6980}, 2014.

\bibitem[Lee et~al.(2018)Lee, Zame, Yoon, and Schaar]{Lee2018}
C.~Lee, W.~Zame, Jinsung Yoon, and M.~V.~D. Schaar.
\newblock Deephit: A deep learning approach to survival analysis with competing
  risks.
\newblock In \emph{AAAI}, 2018.

\bibitem[Levine et~al.(2020)Levine, Peng, Farnell, Nursey, and
  et~al.]{Levine2020}
Adrian~B Levine, Jason Peng, David Farnell, Mitchell Nursey, and et~al.
\newblock Synthesis of diagnostic quality cancer pathology images by generative
  adversarial networks.
\newblock \emph{The Journal of Pathology}, 252\penalty0 (2):\penalty0 178--188,
  2020.
\newblock \doi{https://doi.org/10.1002/path.5509}.
\newblock URL \url{https://onlinelibrary.wiley.com/doi/abs/10.1002/path.5509}.

\bibitem[Lim and Ye(2017)]{Lim2017}
Jae~Hyun Lim and Jong~Chul Ye.
\newblock Geometric {GAN}, 2017.

\bibitem[Lopez-Paz and Oquab(2016)]{Lopezpaz2016}
David Lopez-Paz and Maxime Oquab.
\newblock Revisiting classifier two-sample tests, 2016.

\bibitem[Mahmood et~al.(2018)Mahmood, Borders, Chen, McKay, J~Salimian, Baras,
  and Durr]{Mahmood2018}
Faisal Mahmood, Daniel Borders, Richard Chen, Gregory McKay, Kevan J~Salimian,
  Alexander Baras, and Nicholas Durr.
\newblock Deep adversarial training for multi-organ nuclei segmentation in
  histopathology images, 09 2018.

\bibitem[Marinelli et~al.(2008)Marinelli, Montgomery, Long~Liu, Shah, Prapong,
  Nitzberg, K~Zachariah, Sherlock, Natkunam, B~West, van~de Rijn, O~Brown, and
  A~Ball]{stanford_tma}
Robert~J. Marinelli, Kelli Montgomery, Chih Long~Liu, Nigam Shah, Wijan
  Prapong, Michael Nitzberg, Zachariah K~Zachariah, Gavin Sherlock, Yasodha
  Natkunam, Robert B~West, Matt van~de Rijn, Patrick O~Brown, and Catherine
  A~Ball.
\newblock The stanford tissue microarray database.
\newblock \emph{Nucleic acids research}, 36:\penalty0 D871--7, 02 2008.
\newblock \doi{10.1093/nar/gkm861}.

\bibitem[McInnes et~al.(2018)McInnes, Healy, Saul, and
  Gro{\ss}berger]{McInnes2018}
Leland McInnes, John Healy, Nathaniel Saul, and Lukas Gro{\ss}berger.
\newblock {UMAP: Uniform Manifold Approximation and Projection}.
\newblock \emph{Journal of Open Source Software}, 3\penalty0 (29), 2018.

\bibitem[Miyato et~al.(2018)Miyato, Kataoka, Koyama, and Yoshida]{Miyato2018}
Takeru Miyato, Toshiki Kataoka, Masanori Koyama, and Yuichi Yoshida.
\newblock Spectral normalization for generative adversarial networks.
\newblock In \emph{International Conference on Learning Representations}, 2018.
\newblock URL \url{https://openreview.net/forum?id=B1QRgziT-}.

\bibitem[Qaiser et~al.(2019)Qaiser, Tsang, Taniyama, Sakamoto, Nakane, Epstein,
  and Rajpoot]{Qaiser2019}
Talha Qaiser, Yee-Wah Tsang, Daiki Taniyama, Naoya Sakamoto, Kazuaki Nakane,
  David Epstein, and Nasir Rajpoot.
\newblock Fast and accurate tumor segmentation of histology images using
  persistent homology and deep convolutional features.
\newblock \emph{Medical image analysis}, 55:\penalty0 1--14, 2019.

\bibitem[Qu et~al.(2019)Qu, Riedlinger, Wu, Huang, Yi, De, and Metaxas]{Qu2019}
Hui Qu, Gregory Riedlinger, Pengxiang Wu, Qiaoying Huang, Jingru Yi, Subhajyoti
  De, and Dimitris Metaxas.
\newblock Joint segmentation and fine-grained classification of nuclei in
  histopathology images.
\newblock In \emph{2019 IEEE 16th International Symposium on Biomedical Imaging
  (ISBI 2019)}, pages 900--904. IEEE, 2019.

\bibitem[Quiros et~al.(2020)Quiros, Murray-Smith, and
  YuCoudan]{quiros2020learning}
Adalberto~Claudio Quiros, Roderick Murray-Smith, and Ke~YuCoudan.
\newblock Learning a low dimensional manifold of real cancer tissue with
  pathologygan, 2020.

\bibitem[Radford et~al.(2015)Radford, Metz, and Chintala]{Radford2015}
Alec Radford, Luke Metz, and Soumith Chintala.
\newblock Unsupervised representation learning with deep convolutional
  generative adversarial networks, 2015.

\bibitem[Rana et~al.(2018)Rana, Yauney, Lowe, and Shah]{Rana2018}
Aman Rana, Gregory Yauney, Alarice Lowe, and Pratik Shah.
\newblock Computational histological staining and destaining of prostate core
  biopsy {RGB} images with {Generative Adversarial Neural Networks}.
\newblock \emph{2018 17th IEEE International Conference on Machine Learning and
  Applications (ICMLA)}, Dec 2018.
\newblock \doi{10.1109/icmla.2018.00133}.
\newblock URL \url{http://dx.doi.org/10.1109/ICMLA.2018.00133}.

\bibitem[Salimans et~al.(2016)Salimans, Goodfellow, Zaremba, Cheung, Radford,
  Chen, and Chen]{Salimans2016}
Tim Salimans, Ian Goodfellow, Wojciech Zaremba, Vicki Cheung, Alec Radford,
  Xi~Chen, and Xi~Chen.
\newblock Improved techniques for training gans.
\newblock In D.~D. Lee, M.~Sugiyama, U.~V. Luxburg, I.~Guyon, and R.~Garnett,
  editors, \emph{Advances in Neural Information Processing Systems 29}, pages
  2234--2242. Curran Associates, Inc., 2016.
\newblock URL
  \url{http://papers.nips.cc/paper/6125-improved-techniques-for-training-gans.pdf}.

\bibitem[Schmauch et~al.(2020)Schmauch, Romagnoni, Pronier, Saillard,
  Maill{\'e}, Calderaro, Kamoun, Sefta, Toldo, Zaslavskiy, Clozel, Moarii,
  Courtiol, and Wainrib]{Schmauch2020}
Beno{\^\i}t Schmauch, Alberto Romagnoni, Elodie Pronier, Charlie Saillard,
  Pascale Maill{\'e}, Julien Calderaro, Aur{\'e}lie Kamoun, Meriem Sefta,
  Sylvain Toldo, Mikhail Zaslavskiy, Thomas Clozel, Matahi Moarii, Pierre
  Courtiol, and Gilles Wainrib.
\newblock A deep learning model to predict rna-seq expression of tumours from
  whole slide images.
\newblock \emph{Nature Communications}, 11\penalty0 (1):\penalty0 3877, 2020.
\newblock \doi{10.1038/s41467-020-17678-4}.
\newblock URL \url{https://doi.org/10.1038/s41467-020-17678-4}.

\bibitem[{Szegedy} et~al.(2016){Szegedy}, {Vanhoucke}, {Ioffe}, {Shlens}, and
  {Wojna}]{Szegedy2016}
C.~{Szegedy}, V.~{Vanhoucke}, S.~{Ioffe}, J.~{Shlens}, and Z.~{Wojna}.
\newblock Rethinking the inception architecture for computer vision.
\newblock In \emph{2016 IEEE Conference on Computer Vision and Pattern
  Recognition (CVPR)}, pages 2818--2826, 2016.
\newblock \doi{10.1109/CVPR.2016.308}.

\bibitem[Tellez et~al.(2018)Tellez, Balkenhol, Otte-H{\"o}ller, van~de Loo,
  Vogels, Bult, Wauters, Vreuls, Mol, Karssemeijer, et~al.]{Tellez2018}
David Tellez, Maschenka Balkenhol, Irene Otte-H{\"o}ller, Rob van~de Loo, Rob
  Vogels, Peter Bult, Carla Wauters, Willem Vreuls, Suzanne Mol, Nico
  Karssemeijer, et~al.
\newblock Whole-slide mitosis detection in h\&e breast histology using phh3 as
  a reference to train distilled stain-invariant convolutional networks.
\newblock \emph{IEEE transactions on medical imaging}, 37\penalty0
  (9):\penalty0 2126--2136, 2018.

\bibitem[Vickovic et~al.(2019)Vickovic, Eraslan, Salm{\'e}n, and
  et~al.]{Vickovic2019}
Sanja Vickovic, G{\"o}kcen Eraslan, Fredrik Salm{\'e}n, and et~al.
\newblock High-definition spatial transcriptomics for in situ tissue profiling.
\newblock \emph{Nature Methods}, 16\penalty0 (10):\penalty0 987--990, 2019.

\bibitem[Wang et~al.(2020)Wang, Ma, and Ruzzo]{Wang2020}
Yuliang Wang, Shuyi Ma, and Walter~L. Ruzzo.
\newblock Spatial modeling of prostate cancer metabolic gene expression reveals
  extensive heterogeneity and selective vulnerabilities.
\newblock \emph{Scientific Reports}, 10\penalty0 (1):\penalty0 3490, 2020.
\newblock \doi{10.1038/s41598-020-60384-w}.
\newblock URL \url{https://doi.org/10.1038/s41598-020-60384-w}.

\bibitem[Wei et~al.(2019)Wei, Tafe, Linnik, Vaickus, Tomita, and
  Hassanpour]{Wei2019}
Jason~W. Wei, Laura~J. Tafe, Yevgeniy~A. Linnik, Louis~J. Vaickus, Naofumi
  Tomita, and Saeed Hassanpour.
\newblock Pathologist-level classification of histologic patterns on resected
  lung adenocarcinoma slides with deep neural networks.
\newblock \emph{Scientific Reports}, 9\penalty0 (1):\penalty0 3358, 2019.
\newblock \doi{10.1038/s41598-019-40041-7}.
\newblock URL \url{https://doi.org/10.1038/s41598-019-40041-7}.

\bibitem[Woerl et~al.(2020)Woerl, Eckstein, Geiger, Wagner, Daher, Stenzel,
  Fernandez, Hartmann, Wand, Roth, and Foersch]{Woerl2020}
Ann-Christin Woerl, Markus Eckstein, Josephine Geiger, Daniel~C. Wagner, Tamas
  Daher, Philipp Stenzel, Aurélie Fernandez, Arndt Hartmann, Michael Wand,
  Wilfried Roth, and Sebastian Foersch.
\newblock Deep learning predicts molecular subtype of muscle-invasive bladder
  cancer from conventional histopathological slides.
\newblock \emph{European Urology}, 78\penalty0 (2):\penalty0 256 -- 264, 2020.
\newblock ISSN 0302-2838.
\newblock \doi{https://doi.org/10.1016/j.eururo.2020.04.023}.
\newblock URL
  \url{http://www.sciencedirect.com/science/article/pii/S0302283820302554}.

\bibitem[Xu et~al.(2019{\natexlab{a}})Xu, Liu, Hou, Liu, Garibaldi, Ellis,
  Green, Shen, and Qiu]{Xu2019b}
Bolei Xu, Jingxin Liu, Xianxu Hou, Bozhi Liu, Jon Garibaldi, Ian~O Ellis, Andy
  Green, Linlin Shen, and Guoping Qiu.
\newblock Look, investigate, and classify: A deep hybrid attention method for
  breast cancer classification.
\newblock In \emph{2019 IEEE 16th international symposium on biomedical imaging
  (ISBI 2019)}, pages 914--918. IEEE, 2019{\natexlab{a}}.

\bibitem[Xu et~al.(2016)Xu, Xiang, Liu, Gilmore, Wu, Tang, and
  Madabhushi]{Xu2015}
Jun Xu, Lei Xiang, Qingshan Liu, Hannah Gilmore, Jianzhong Wu, Jinghai Tang,
  and Anant Madabhushi.
\newblock Stacked sparse autoencoder (ssae) for nuclei detection on breast
  cancer histopathology images.
\newblock \emph{IEEE transactions on medical imaging}, 35\penalty0
  (1):\penalty0 119--130, 01 2016.
\newblock \doi{10.1109/TMI.2015.2458702}.
\newblock URL \url{https://pubmed.ncbi.nlm.nih.gov/26208307}.

\bibitem[Xu et~al.(2018)Xu, Huang, Yuan, Guo, Sun, Wu, and Weinberger]{Xu2018}
Qiantong Xu, Gao Huang, Yang Yuan, Chuan Guo, Yu~Sun, Felix Wu, and Kilian~Q.
  Weinberger.
\newblock An empirical study on evaluation metrics of generative adversarial
  networks.
\newblock \emph{CoRR}, abs/1806.07755, 2018.
\newblock URL \url{http://arxiv.org/abs/1806.07755}.

\bibitem[Xu et~al.(2019{\natexlab{b}})Xu, Moro, Bozóky, and Zhang]{Xu2019}
Zhaoyang Xu, Carlos~Fernández Moro, Béla Bozóky, and Qianni Zhang.
\newblock {GAN-based Virtual Re-Staining}: A promising solution for whole slide
  image analysis, 2019{\natexlab{b}}.

\bibitem[Yuan et~al.(2012)Yuan, Failmezger, Rueda, Ali, Gr{\"a}f, Chin,
  Schwarz, Curtis, Dunning, Bardwell, Johnson, Doyle, Turashvili, Provenzano,
  Aparicio, Caldas, and Markowetz]{Yuan2012}
Yinyin Yuan, Henrik Failmezger, Oscar~M Rueda, H~Raza Ali, Stefan Gr{\"a}f,
  Suet-Feung Chin, Roland~F Schwarz, Christina Curtis, Mark~J Dunning, Helen
  Bardwell, Nicola Johnson, Sarah Doyle, Gulisa Turashvili, Elena Provenzano,
  Sam Aparicio, Carlos Caldas, and Florian Markowetz.
\newblock Quantitative image analysis of cellular heterogeneity in breast
  tumors complements genomic profiling.
\newblock \emph{Sci Transl Med}, 4\penalty0 (157):\penalty0 157ra143, Oct 2012.
\newblock ISSN 1946-6242 (Electronic); 1946-6234 (Linking).
\newblock \doi{10.1126/scitranslmed.3004330}.

\bibitem[{Zanjani} et~al.(2018){Zanjani}, {Zinger}, {Bejnordi}, {van der Laak},
  and {de With}]{Zanjani2018}
F.~G. {Zanjani}, S.~{Zinger}, B.~E. {Bejnordi}, J.~A. W.~M. {van der Laak}, and
  P.~H.~N. {de With}.
\newblock Stain normalization of histopathology images using generative
  adversarial networks.
\newblock In \emph{2018 IEEE 15th International Symposium on Biomedical Imaging
  (ISBI 2018)}, pages 573--577, 2018.

\bibitem[Zhang et~al.(2019{\natexlab{a}})Zhang, Goodfellow, Metaxas, and
  Odena]{Zhang2019}
Han Zhang, Ian Goodfellow, Dimitris Metaxas, and Augustus Odena.
\newblock Self-attention generative adversarial networks.
\newblock In \emph{International Conference on Machine Learning}, pages
  7354--7363. PMLR, 2019{\natexlab{a}}.

\bibitem[Zhang et~al.(2019{\natexlab{b}})Zhang, Chen, McGough, Xing, Wang, Bui,
  Xie, Sapkota, Cui, Dhillon, et~al.]{Zhang2019b}
Zizhao Zhang, Pingjun Chen, Mason McGough, Fuyong Xing, Chunbao Wang, Marilyn
  Bui, Yuanpu Xie, Manish Sapkota, Lei Cui, Jasreman Dhillon, et~al.
\newblock Pathologist-level interpretable whole-slide cancer diagnosis with
  deep learning.
\newblock \emph{Nature Machine Intelligence}, 1\penalty0 (5):\penalty0
  236--245, 2019{\natexlab{b}}.

\end{thebibliography}

\newpage
\appendix

\section{Code}
    
    We provide the code at this location: ~\url{https://github.com/AdalbertoCq/Pathology-GAN}

\section{Dataset size impact on representations.}
\label{appendix:datasetsize}
    To further understand PathologyGAN's behavior with different dataset sizes and its ability to generalize, we sub-sample the NCT and created different dataset sizes of $5K$, $10K$, and $20K$ images. 
    
    We measure the ability to generalize and hold meaningful representations in two ways: by measuring the FID against the complete dataset of $100K$ and by exploring the latent space, in order to check its structure.
    
\begin{figure}[H]
        \centering
        \minipage{0.5\textwidth}
        \includegraphics[scale=0.145]{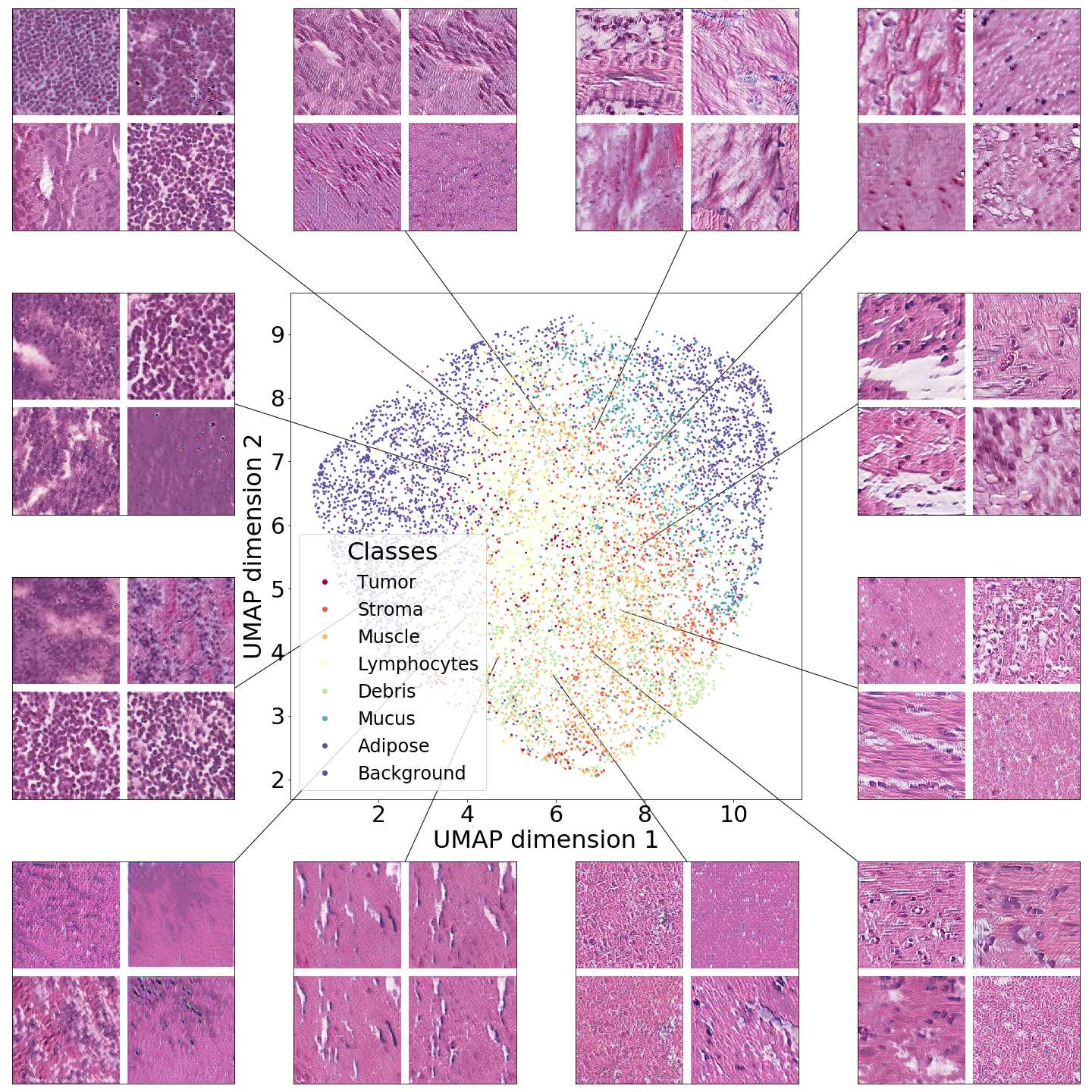}
        \endminipage\hfill
        \minipage{0.5\textwidth}
        \includegraphics[scale=0.145]{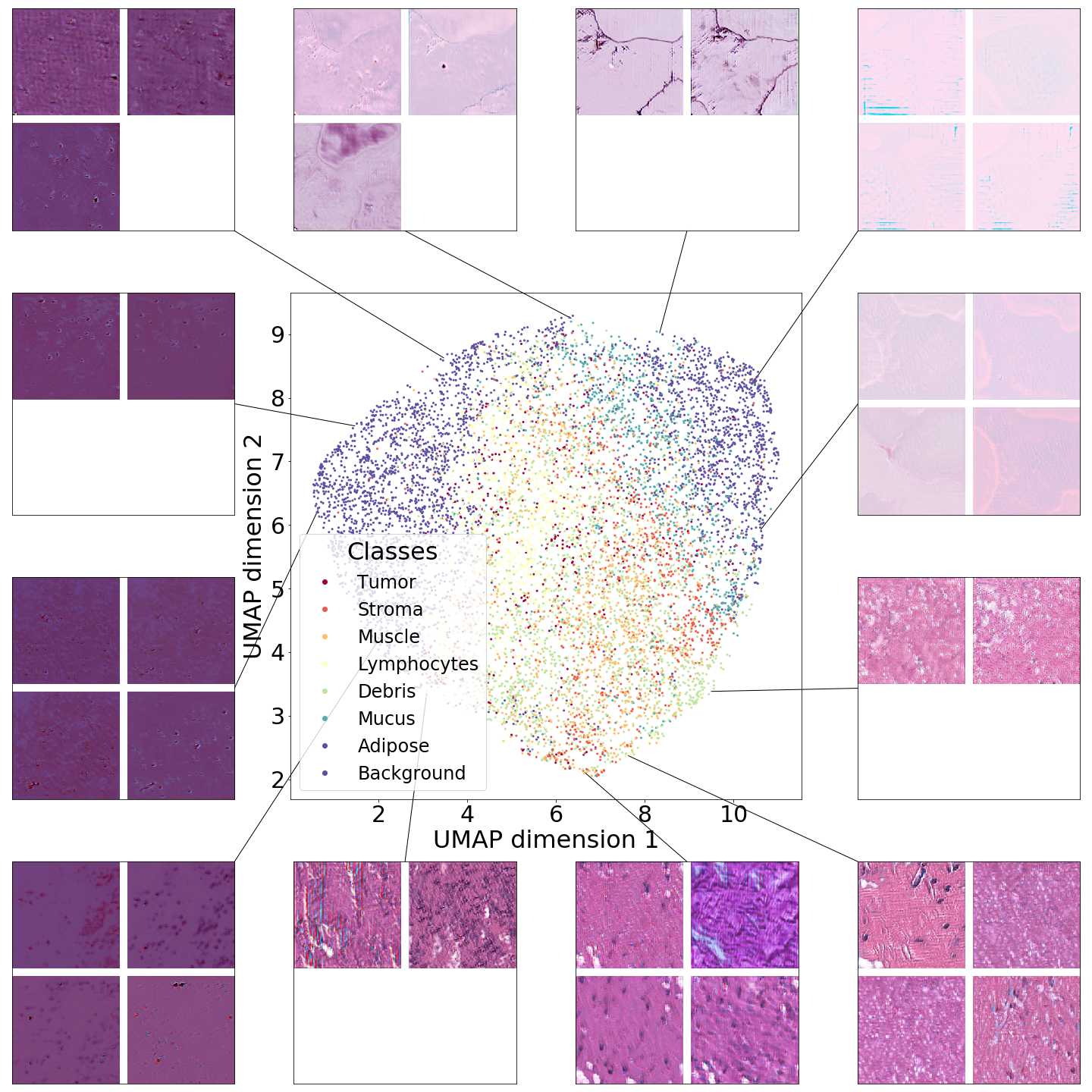}
        \endminipage
        \vspace{-0.25cm}
        \caption{Latent space of PathologyGAN trained on colorectal cancer tissue from National Center for Tumor  (NCT) dataset with $5K$ samples. Uniform Manifold Approximation and Projection (UMAP) representations of generated tissue samples, each generated image is labeled with the type of tissue. Even with $5K$ samples, PathologyGAN holds an structured latent sapce where different regions of the latent space generate distinct kinds of tissue.}
        \label{fig:latent_space_crc_5k}
    \end{figure}

    Figure \ref{fig:latent_space_crc_5k} shows that PathologyGAN's latent space shows a structure regarding tissue types even for $5K$ samples, although it is not able to reliably generalize the original dataset distribution. Table \ref{dataset_results-table} shows how $20K$ samples are enough to achieve a reasonable FID with $41.92$.
    
    \begin{table}[H]
        \centering
        \begin{tabular}{|l|l|}
        \toprule
        NCT Datase Size & FID \\
        \toprule
        \midrule
        5K & 83.52 \\
        \midrule
        10K & 56.03 \\
        \midrule
        20K & 41.92 \\
        \midrule
        Complete 100K & 32.05 \\
        \bottomrule
        \bottomrule
        \end{tabular}
        \caption{Evaluation of PathologyGANs for different dataset sizes.}
        \label{dataset_results-table}
    \end{table}

\section{Hinge vs Relativistic Average Discriminator}
\label{appendix:hinge}
	    In this section we show corresponding generated images and loss function plots for Relativistic Average Discriminator model and Hinge Loss model.
	    \begin{figure}[H]
	        \centering
	        \includegraphics[scale=0.125]{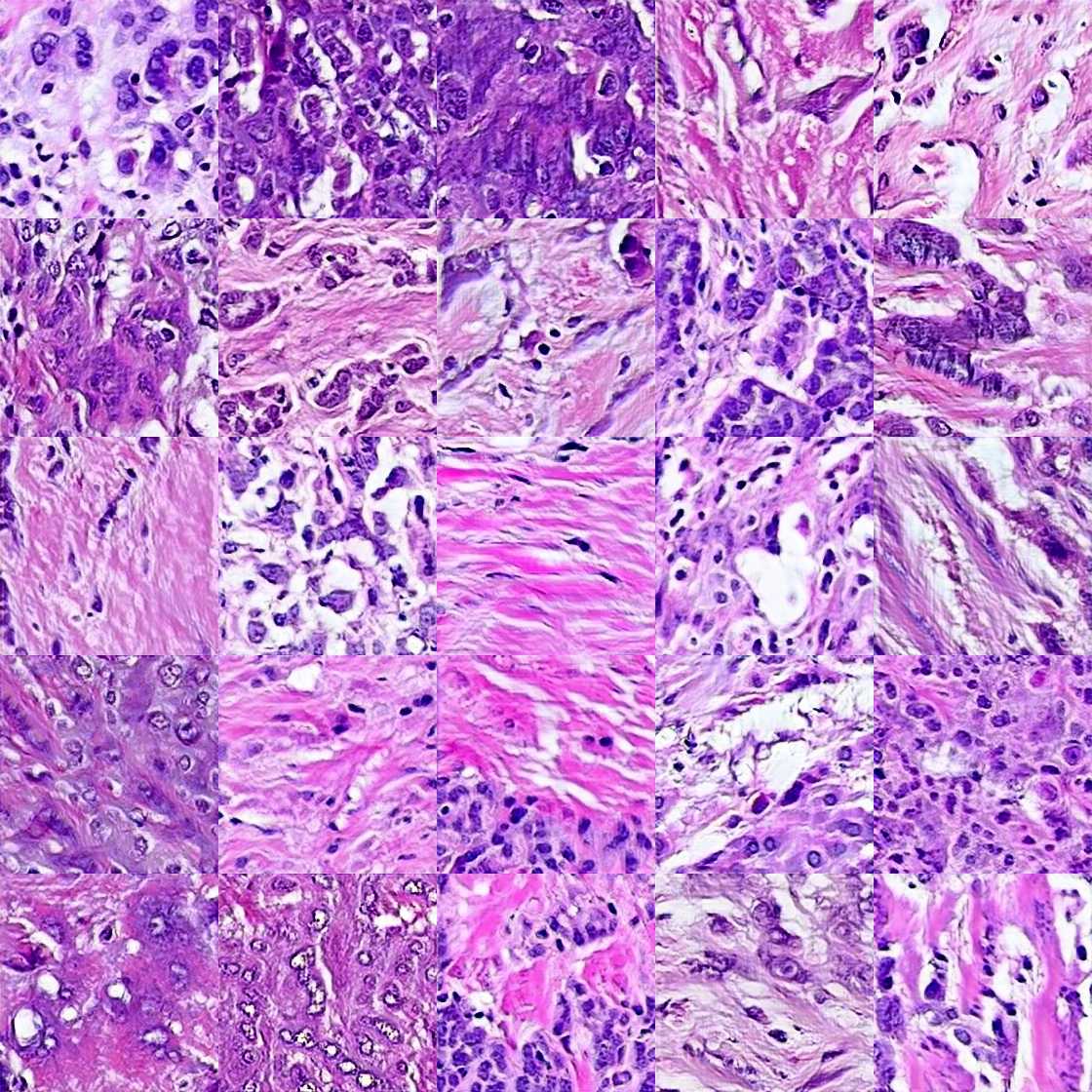}
	        \includegraphics[scale=0.125]{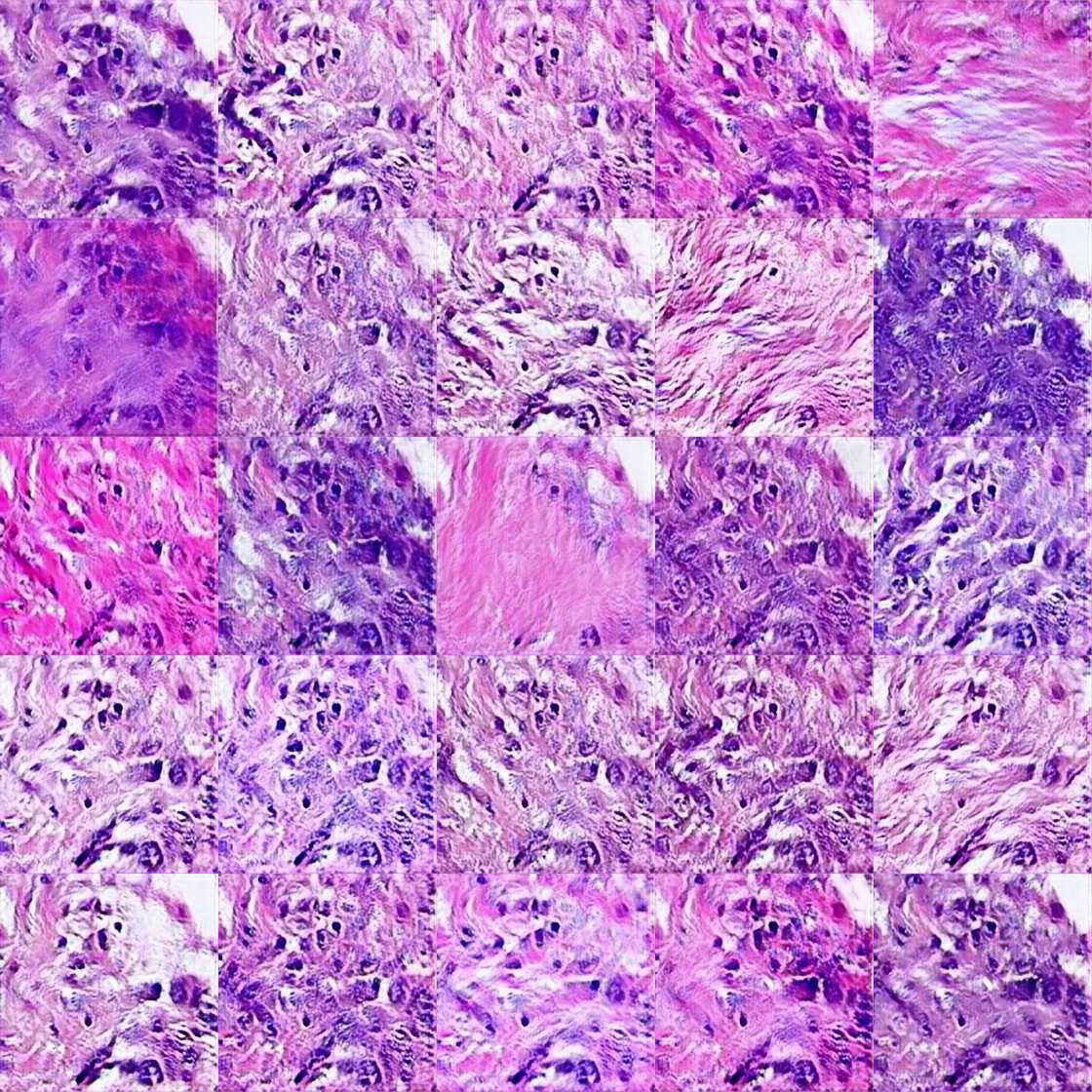}
	        
	        \caption{Left grid images correspond to Relativistic Average Discriminator model vs right grid images from the Hinge loss model. We can see that the Relativistic Average model is able to reproduce cancer tissue characteristics compared to Hinge loss, which does not.}
	        \label{fig:generated_hinge_vs_rad}
	    \end{figure}
	    \vspace{-1cm}
	    \begin{figure}[H]
	        \centering
	        \includegraphics[scale=0.25]{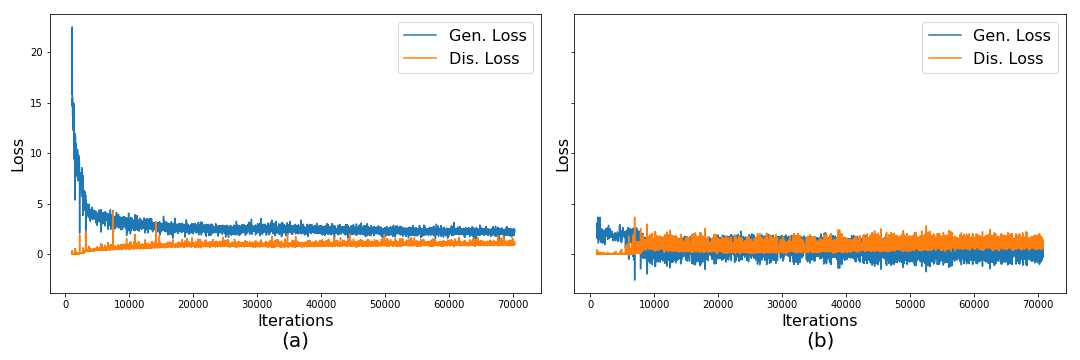}
	        \caption{(a) Generator and Discriminator loss functions of the Relativistic Average Discriminator model, (b) Generator and Discriminator loss functions from the Hinge loss model. Here we capture the corresponding loss functions to the images in Figure ~\ref{fig:generated_hinge_vs_rad}, both of them converge but only Relativistic Average Discriminator produces meaningful images.}
	        \label{fig:loss_functions}
	    \end{figure}

\section{Nearest Neighbors Additional Samples}
\label{appendix:nearest}

\begin{figure}[H]
			\centering
			\minipage{0.5\textwidth}
			  \includegraphics[scale=0.2, trim=250 0 250 140, clip]{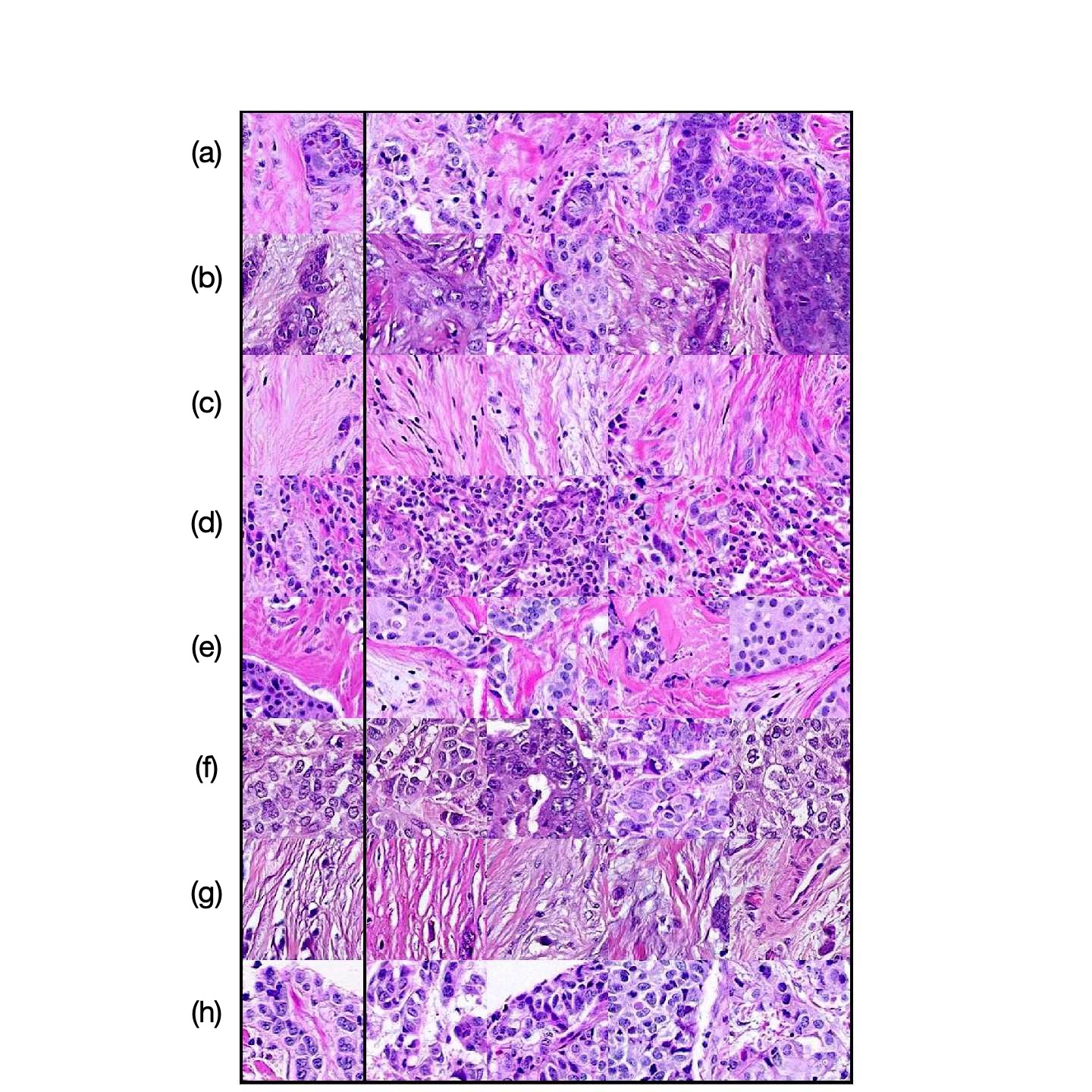}
			\endminipage\hfill
			\minipage{0.5\textwidth}
			  \includegraphics[scale=0.2, trim=250 0 250 140, clip]{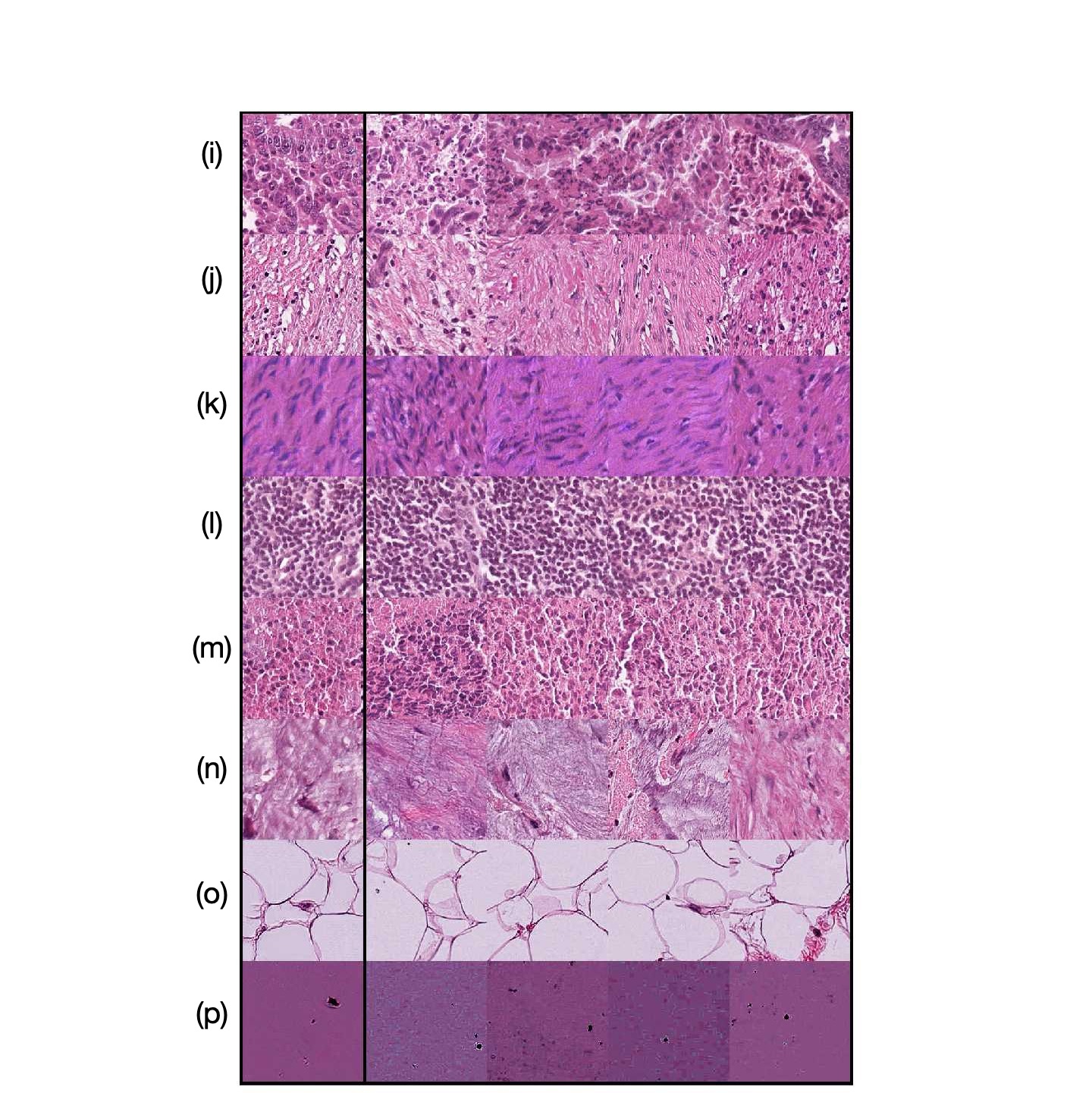}
			\endminipage
			 \caption{Nearest neighbors in Inception-V1 feature space for breast cancer (a-h) and colorectal cancer (i-p). For each row, first column images corresponds to a generated tissue samples from PathologyGAN, the remaining columns are the closest real images in feature space.}
			 \label{fig:nearest_neighbors_appendix}
		\end{figure}    

\section{Mapping Network and Style Mixing Regularization Comparison}
\label{appendix:map_style}

    To measure the impact of introducing a mapping network and using style mixing regularization during training, we provide different figures of the latent space $w$ for two PathologyGANs, one using these features and another one without them. We include both datasets, breast and colorectal cancer tissue.
    
\begin{figure}[H]
        \centering
        \includegraphics[scale=0.15]{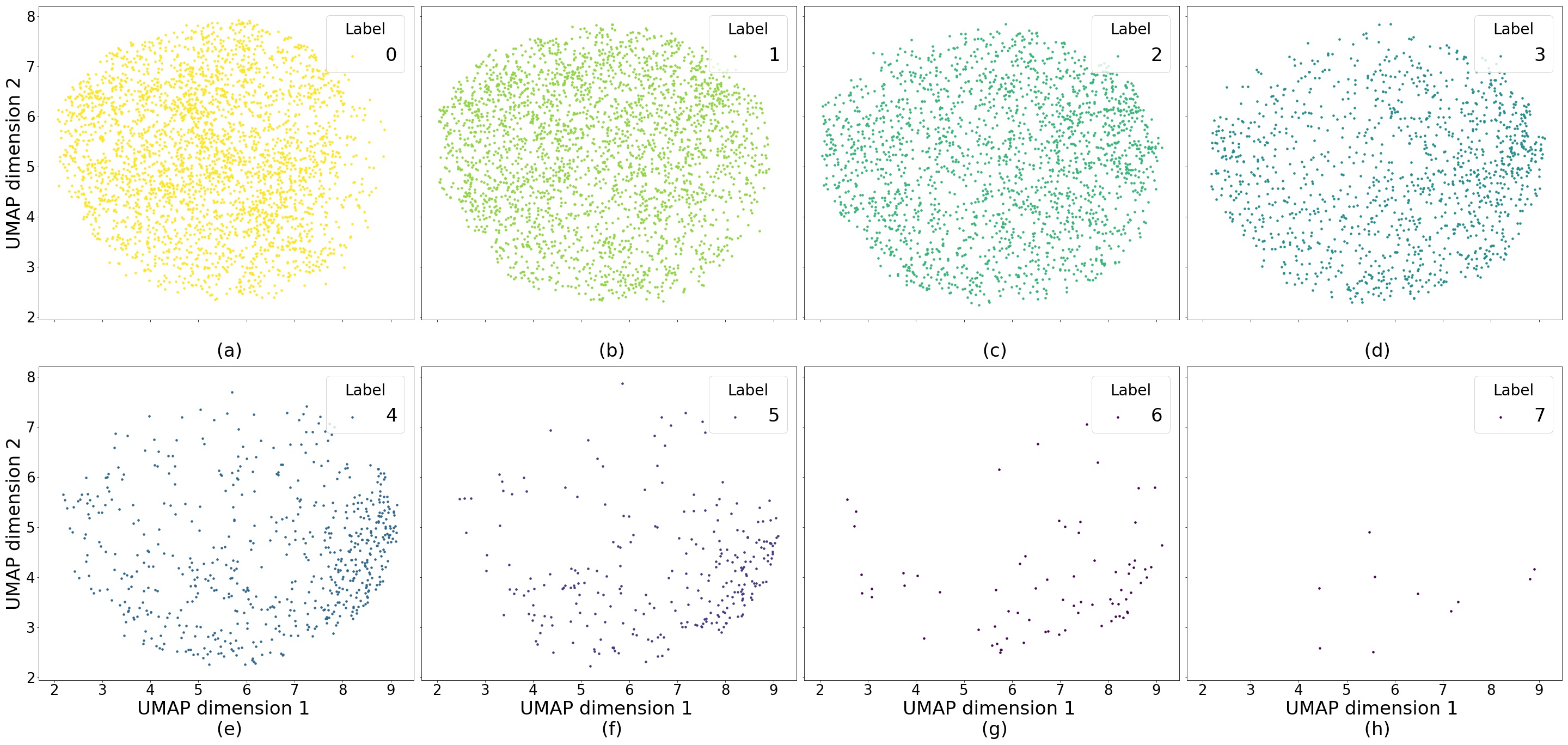}
        \includegraphics[scale=0.15]{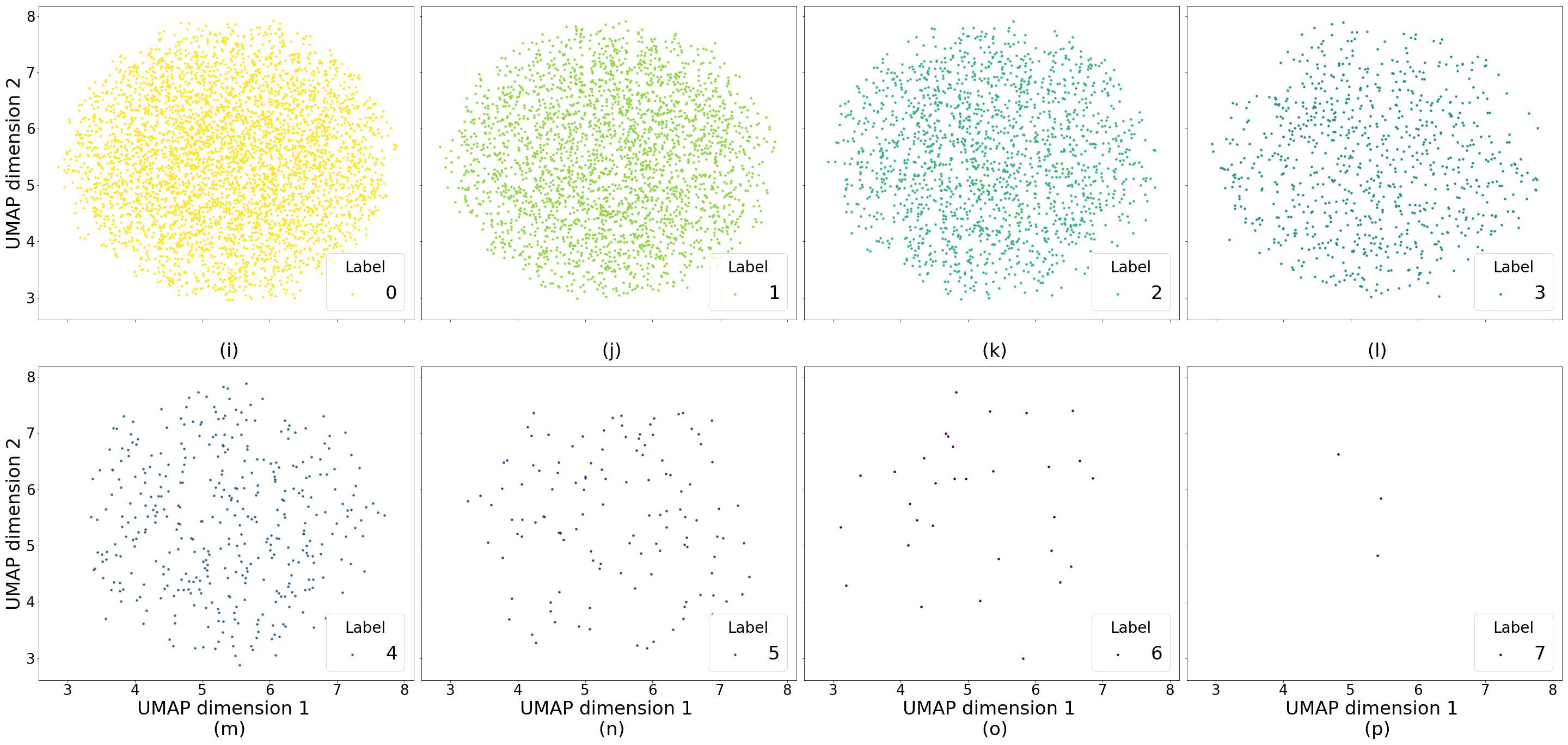}
        \caption{Breast cancer  tissue: Comparison of the latent space $w$ for two different PathologyGAN models, (a-h) include a mapping network and style mixing regularization, and (i-p) do not include them. Each sub-figure shows datapoints only related to one of the classes, and each class is subject to the count of cancer cells in the tissue image, (a) and (i) [class $0$] are associated to images with the lowest number of cancer cells, (h) and (p) [class $8$] with the largest. In the model (a-h) images with increasing number of cancer cells correspond to proportionally moving to quadrant $IV$ in the 2 dimensional space , where (i-p) are randomly placed. This figure shows how including the mapping network and style mixing regularization introduces representation learning properties.}
        \label{fig:latent_space_comp_point_label_breast}
    \end{figure}

\begin{figure}[H]
        \centering
        \includegraphics[scale=0.2]{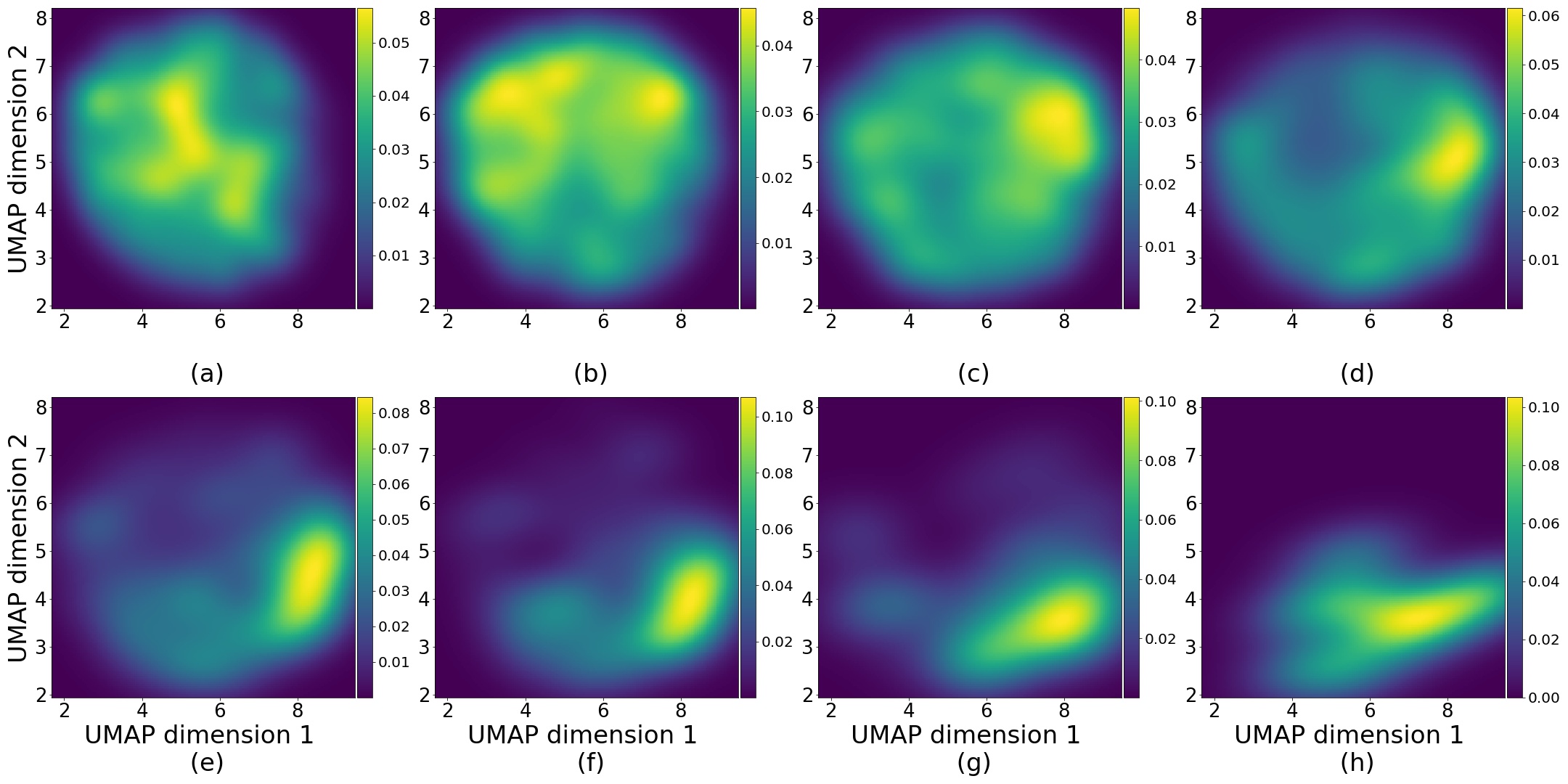}
        \includegraphics[scale=0.15]{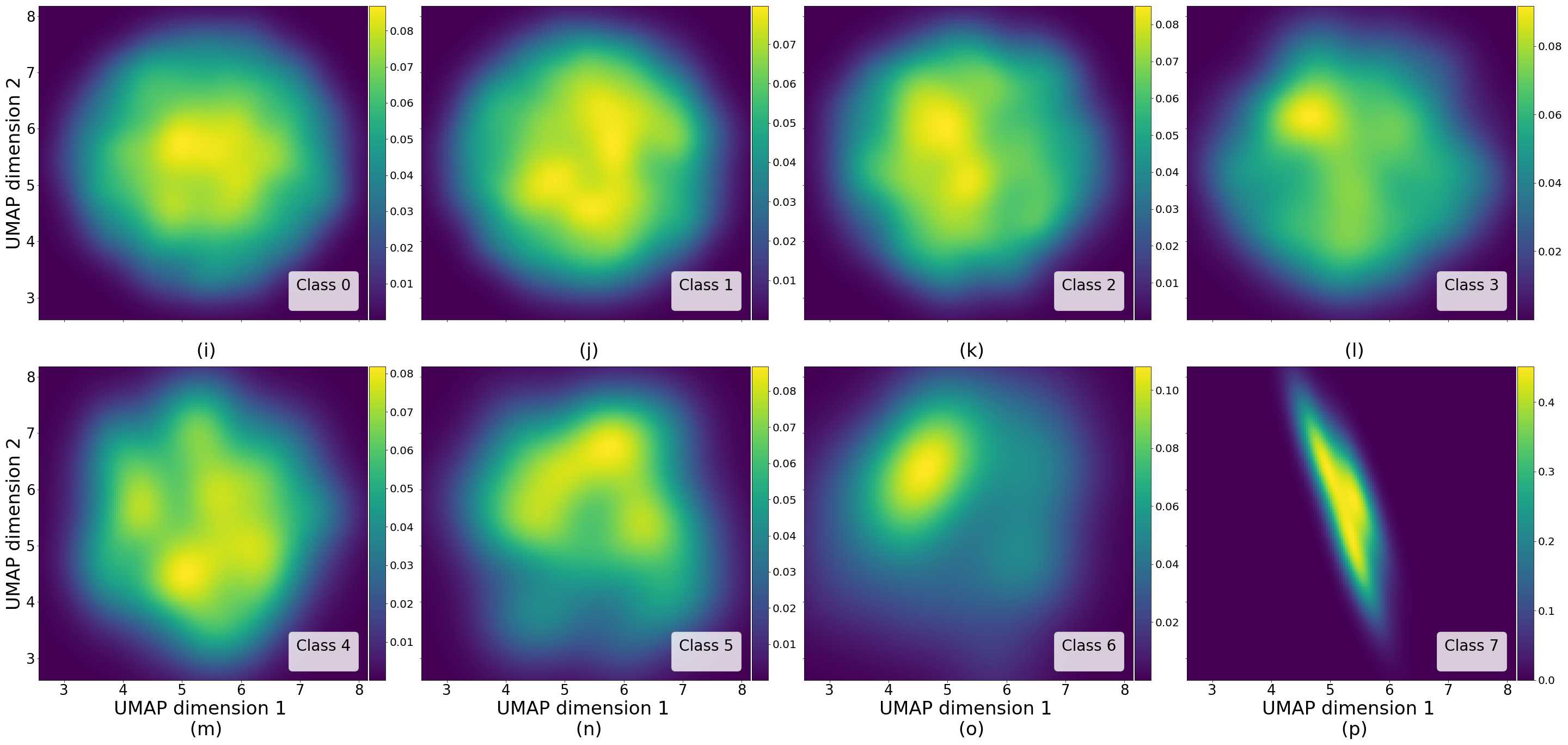}
        \caption{Breast cancer  tissue: Comparison of the latent space $w$ for two different PathologyGAN models, (a-h) include a mapping network and style mixing regularization, and (i-p) do not include them. Each sub-figure shows the density of datapoints only related to one of the classes, and each class is subject to the count of cancer cells in the tissue image, (a) and (i) [class $0$] are associated to images with the lowest number of cancer cells, (h) and (p) [class] with the largest. In the model (a-h) images with increasing number of cancer cells correspond to proportionally moving to quadrant $IV$ in the 2 dimensional space , where (i-p) are randomly placed. This figure shows how including the mapping network and style mixing regularization introduces representation learning properties.}
        \label{fig:latent_space_comp_density_label_breast}
    \end{figure}

\begin{figure}[H]
        \centering
        \includegraphics[scale=0.15]{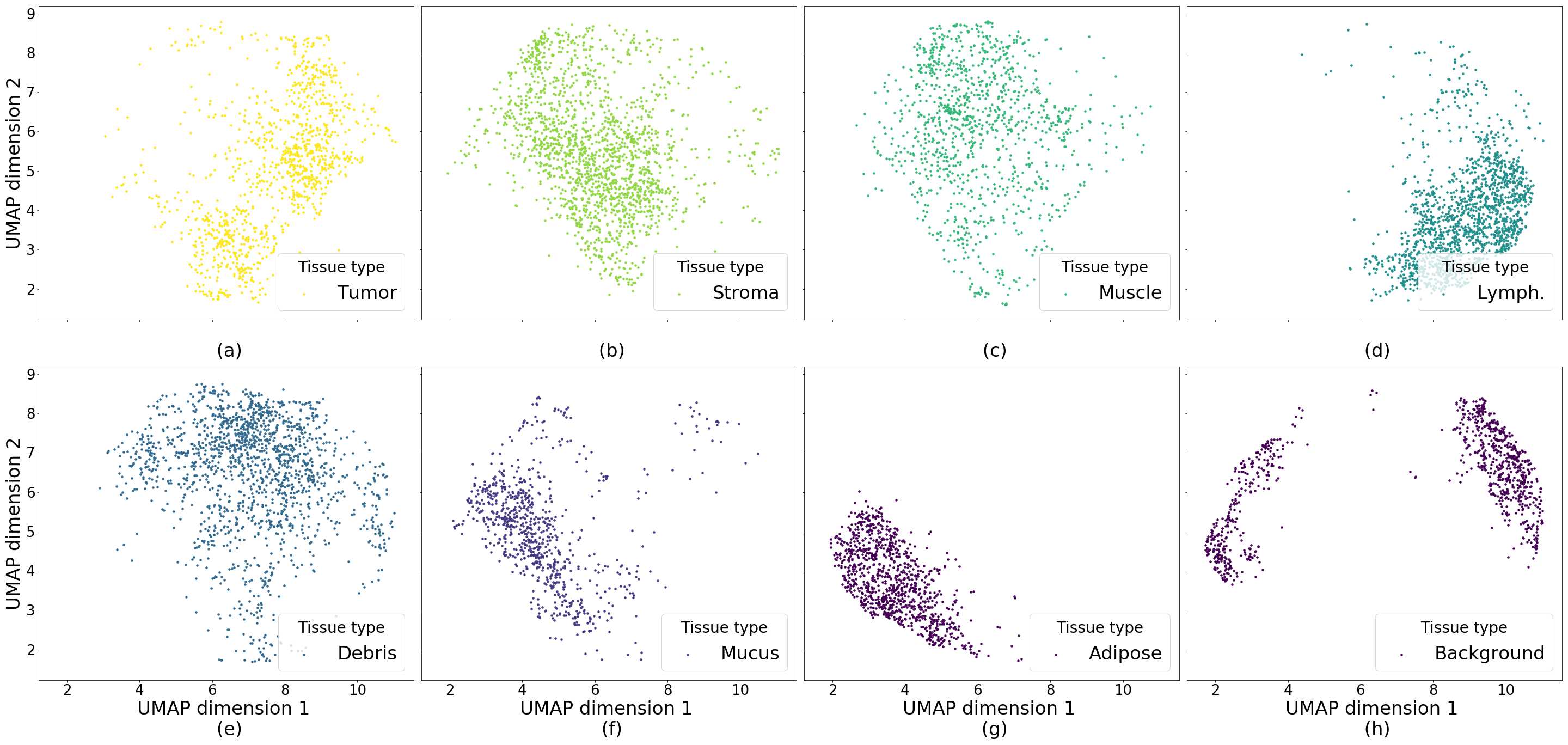}
        \includegraphics[scale=0.15]{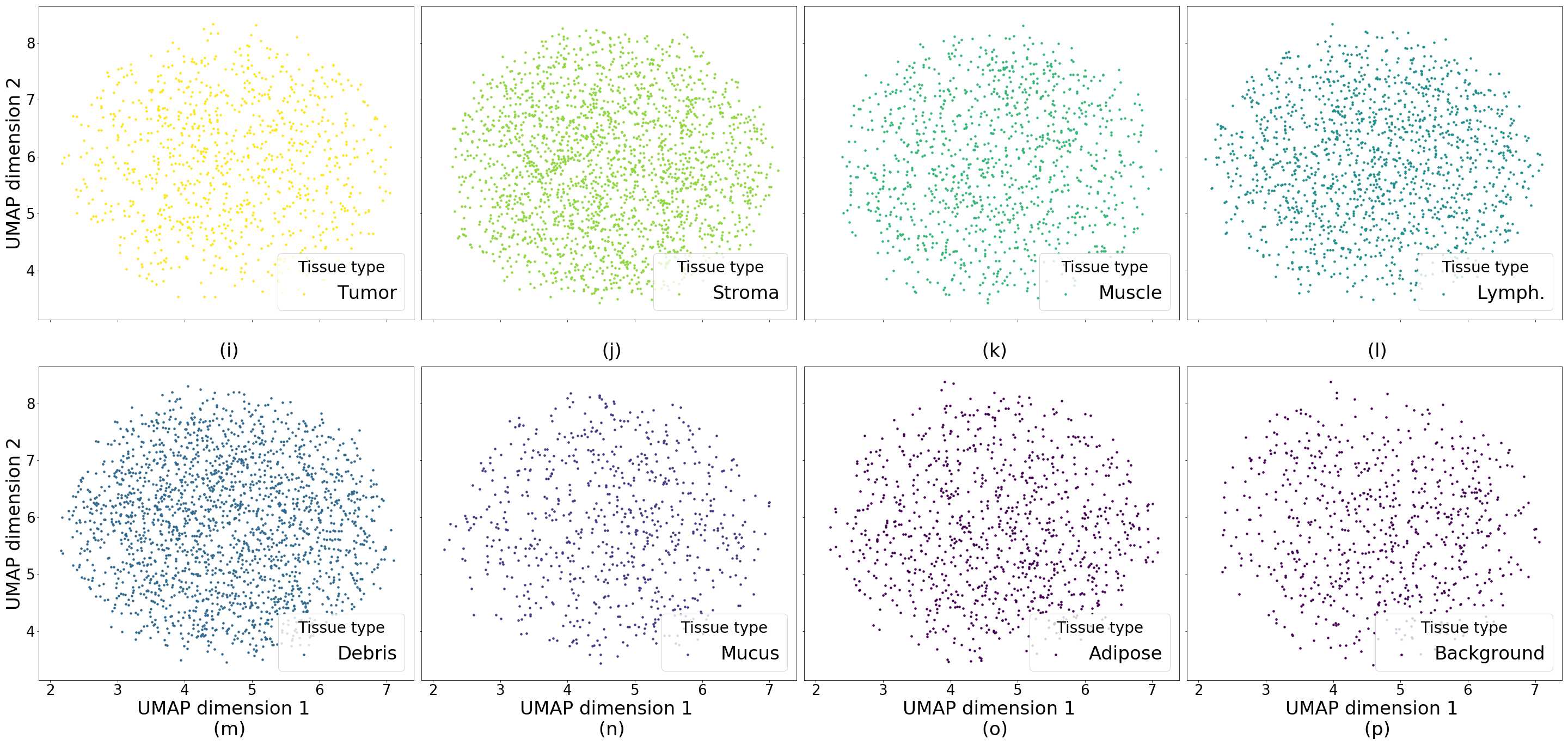}
        \caption{Colorectal cancer tissue: Comparison of the latent space $w$ for two different PathologyGAN models, (a-h) include a mapping network and style mixing regularization, and (i-p) do not include them.  Each sub-figure shows datapoints only related to to one type of tissue. In the model (a-h) distinct regions of the latent space correspond to different tissue types, while in model (i-p) they are randomly placed. This figure shows how including the mapping network and style mixing regularization introduces representation learning properties.}
        \label{fig:latent_space_comp_point_label_crc}
    \end{figure}

\begin{figure}[H]
        \centering
        \includegraphics[scale=0.15]{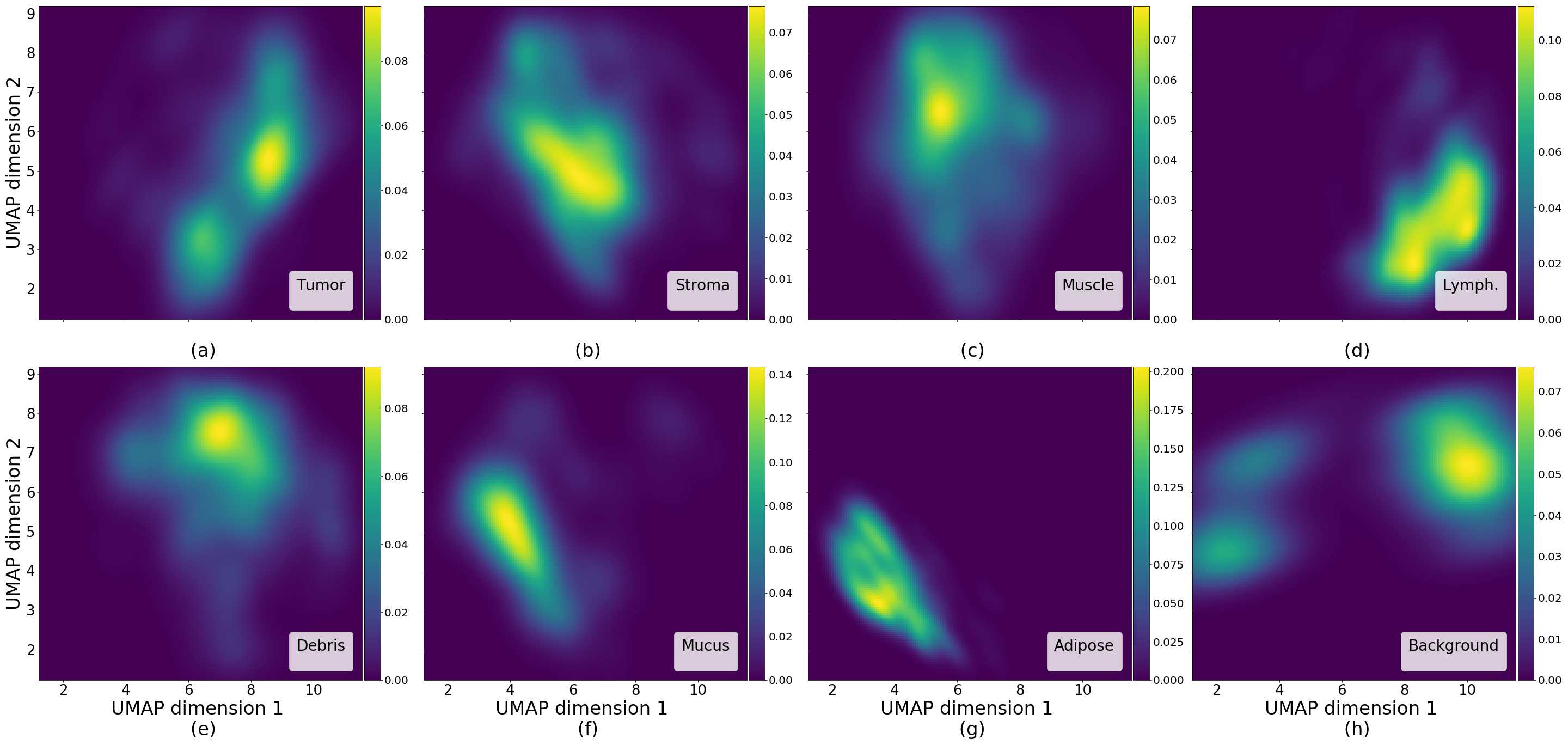}
        \includegraphics[scale=0.15]{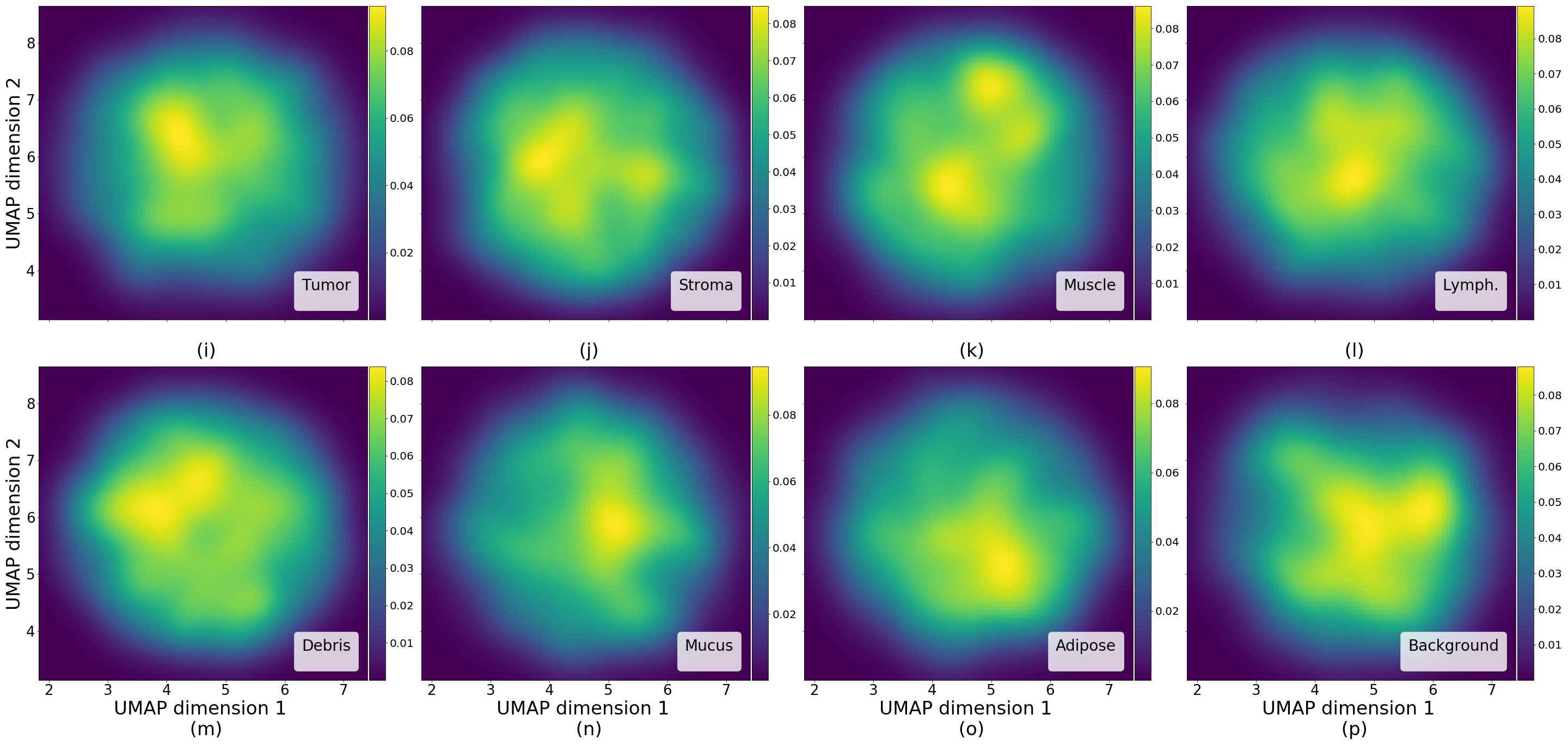}
        \caption{Colorectal cancer tissue: Comparison of the latent space $w$ for two different PathologyGAN models, (a-h) include a mapping network and style mixing regularization, and (i-p) do not include them. Each sub-figure shows the density of datapoints only related to one type of tissue. In the model (a-h) distinct regions of the latent space correspond to different tissue types, while in model (i-p) they are randomly placed. This figure shows how including the mapping network and style mixing regularization introduces representation learning properties.}
        \label{fig:latent_space_comp_density_label_crc}
    \end{figure}

    Figures \ref{fig:latent_space_comp_point_label_breast}, \ref{fig:latent_space_comp_density_label_breast}, \ref{fig:latent_space_comp_point_label_crc}, and \ref{fig:latent_space_comp_density_label_crc} capture the clear difference in the latent space ordering with respect to the counts of cancer cells and types of tissue in the image. Without a mapping network and style mixing regularization the latent space $w$ shows a random placement of the vectors subject to the tissue characteristics, when these two elements are introduced different regions of the latent space produce images with distinct characteristics.
    
\section{Vector Operation Samples}
\label{appendix:vector_op}

    \begin{figure}[H]
        \centering
        \includegraphics[scale=0.165, trim=0 0 0 100, clip]{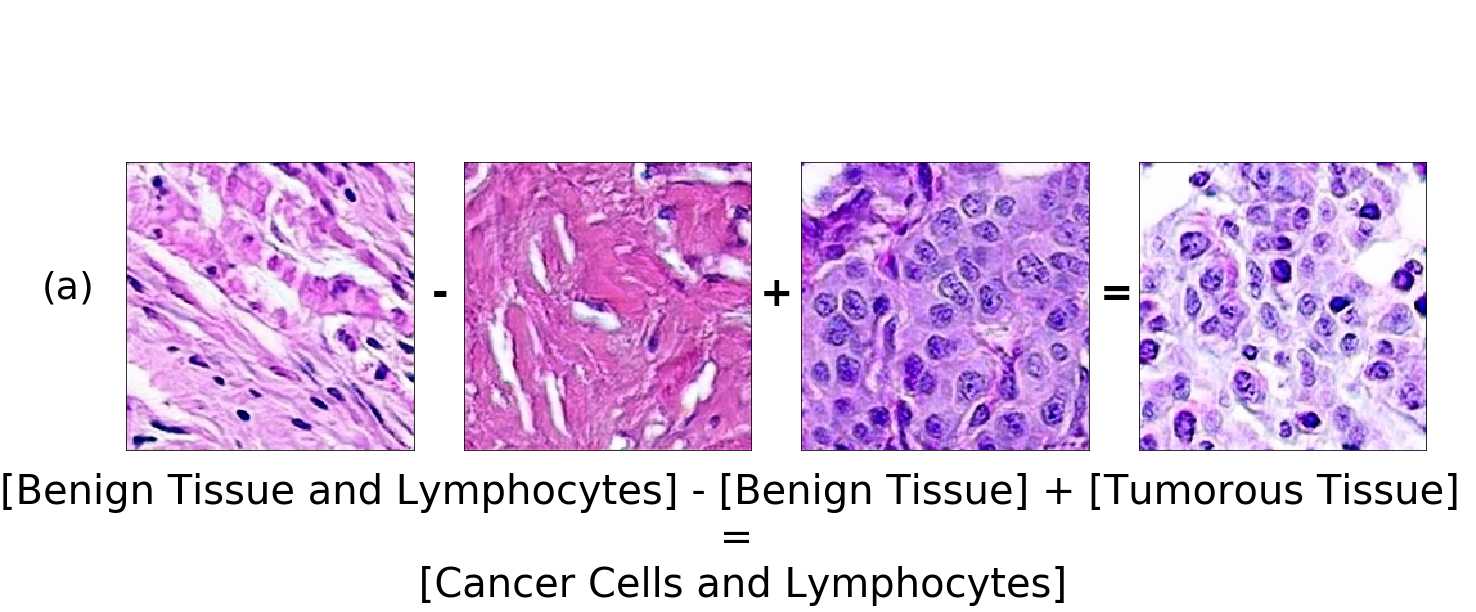}
        \includegraphics[scale=0.165, trim=0 0 0 100, clip]{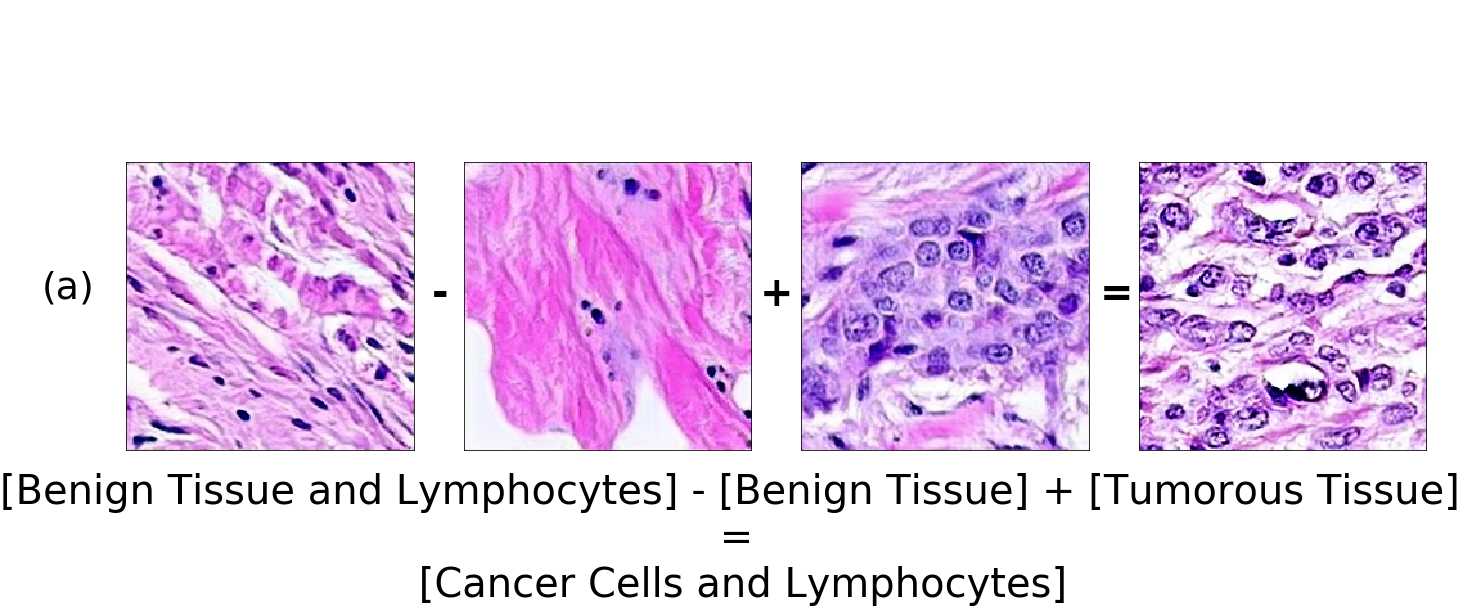}
        \includegraphics[scale=0.165, trim=0 0 0 100, clip]{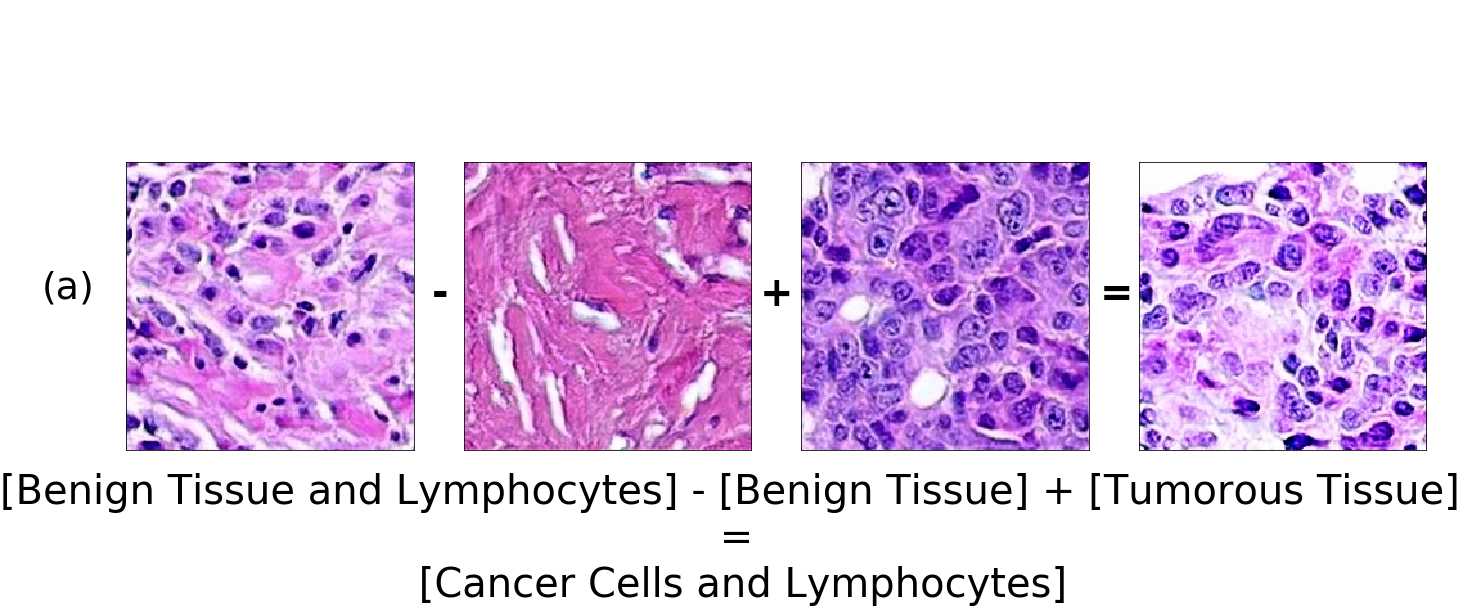}
        \caption{Breast cancer: samples of vector operations with different images, all operations correspond to: Benign  tissue  and  lymphocytes- benign tissue + tumorous tissue = cancer cells and lymphocytes.}
        \label{fig:breast_vector_op_0}
    \end{figure}
    
    \begin{figure}[H]
        \centering
        \includegraphics[scale=0.165, trim=0 0 0 100, clip]{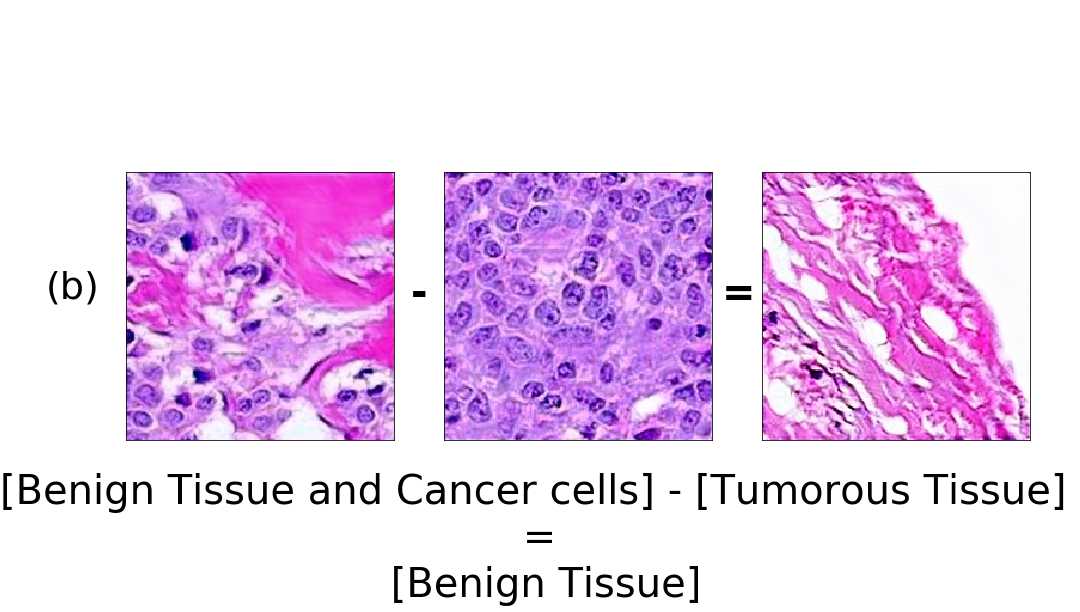}
        \includegraphics[scale=0.165, trim=0 0 0 100, clip]{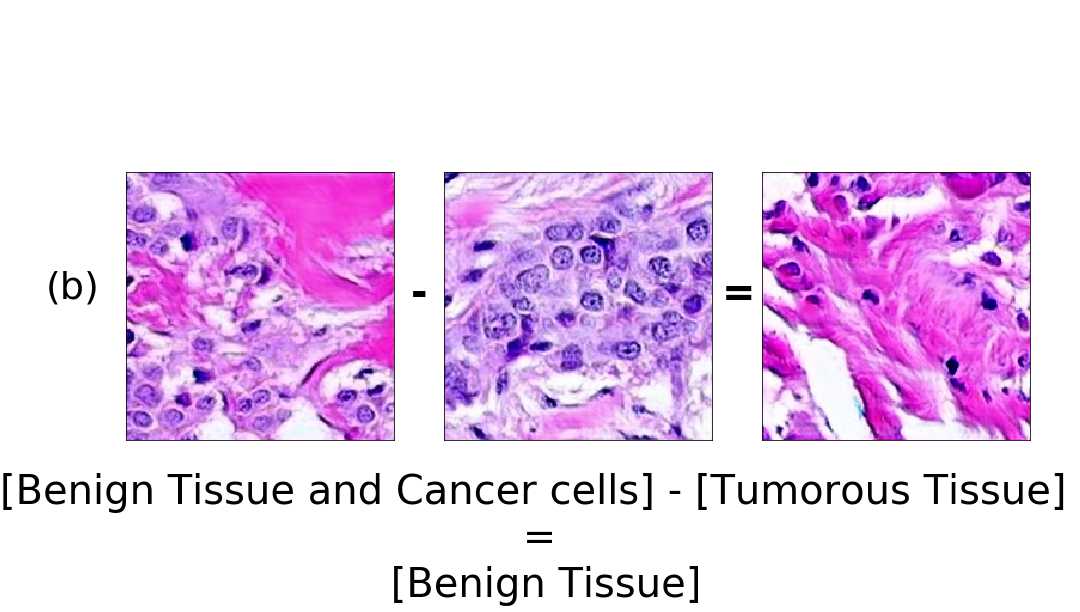}
        \includegraphics[scale=0.165, trim=0 0 0 100, clip]{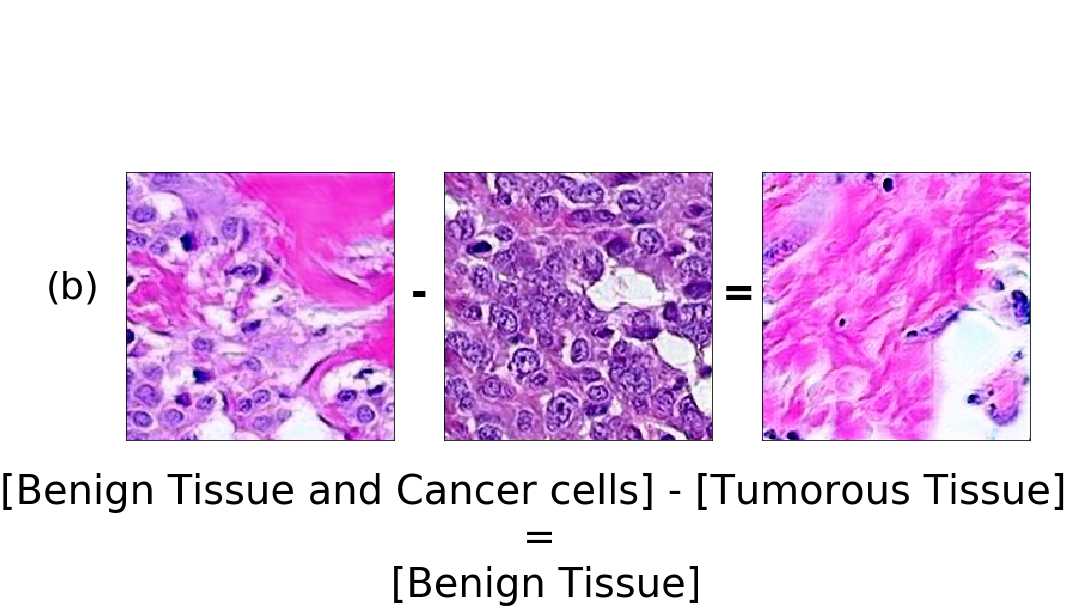}
        \includegraphics[scale=0.165, trim=0 0 0 100, clip]{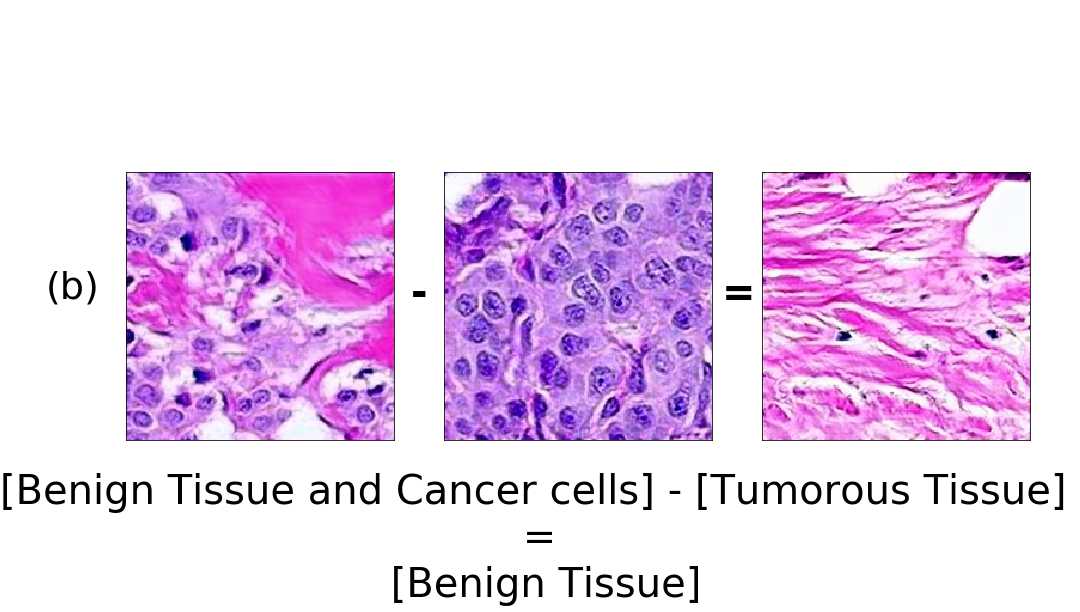}
        \caption{Breast cancer: samples of vector operations with different images, all operations correspond to: Benign tissue with patches of cancer cells - tumorous = benign tissue.}
        \label{fig:vector_op_1}
    \end{figure}
    
    \begin{figure}[H]
        \centering
        \includegraphics[scale=0.165, trim=0 0 0 100, clip]{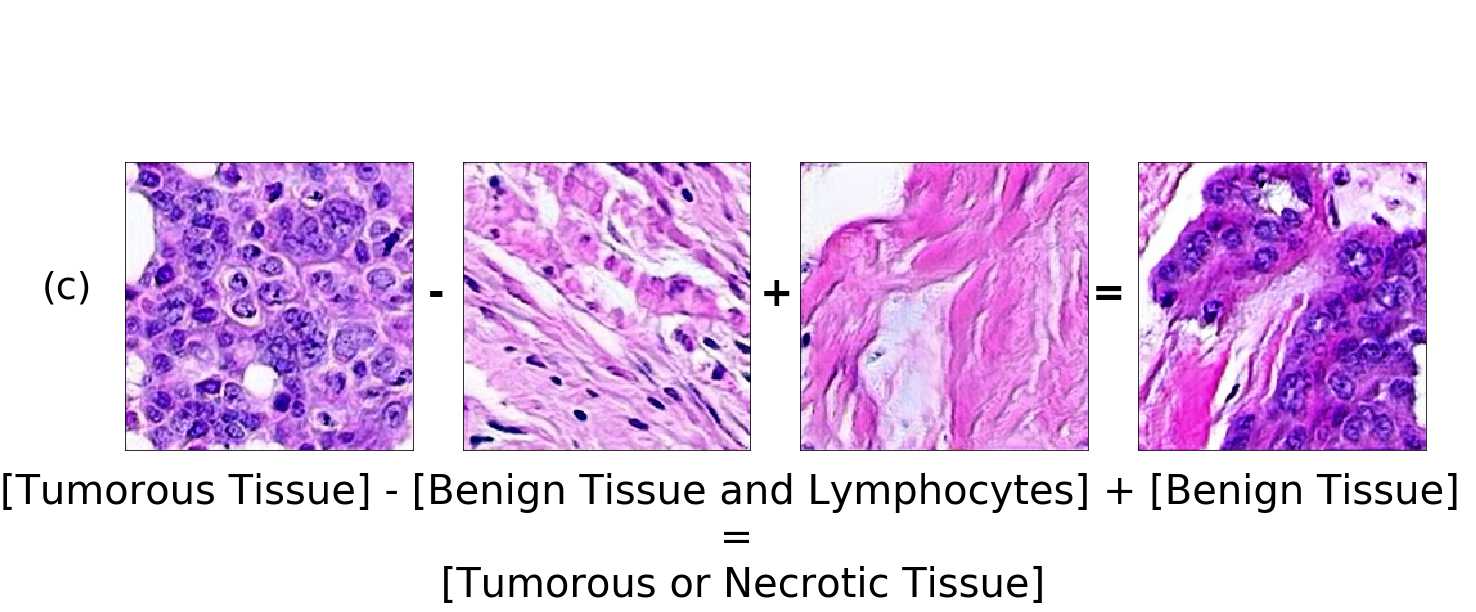}
        \includegraphics[scale=0.165, trim=0 0 0 100, clip]{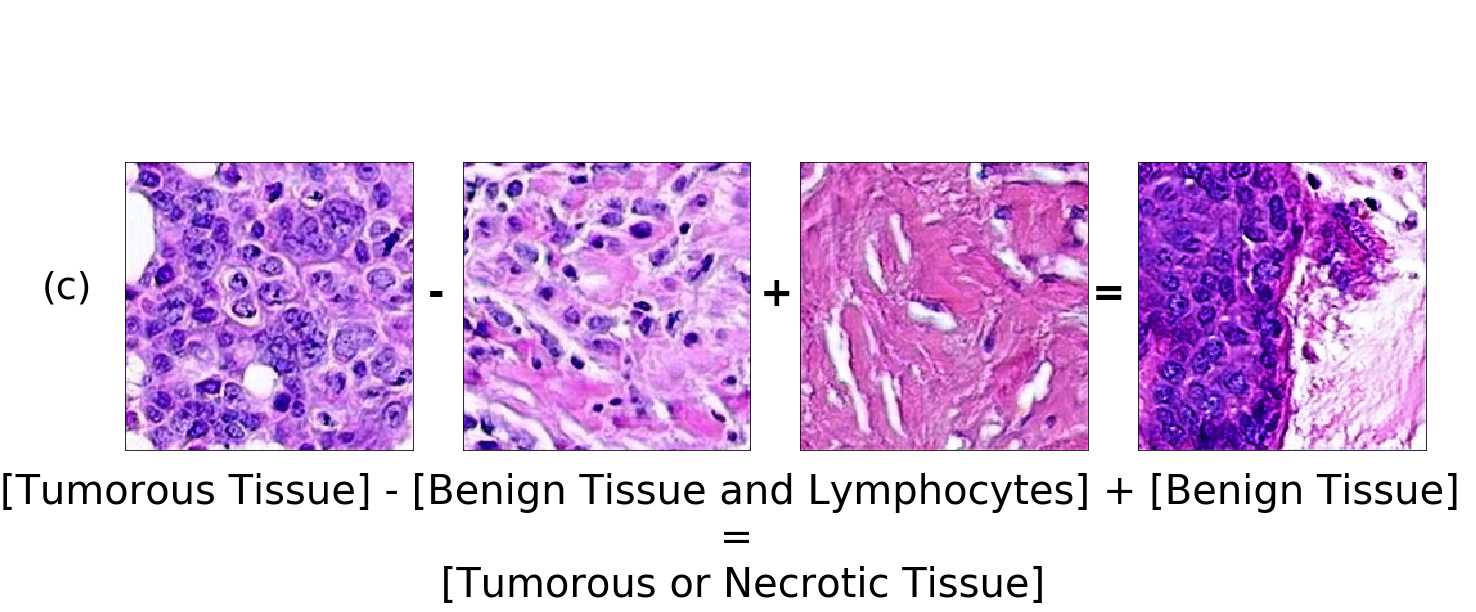}
        \includegraphics[scale=0.165, trim=0 0 0 100, clip]{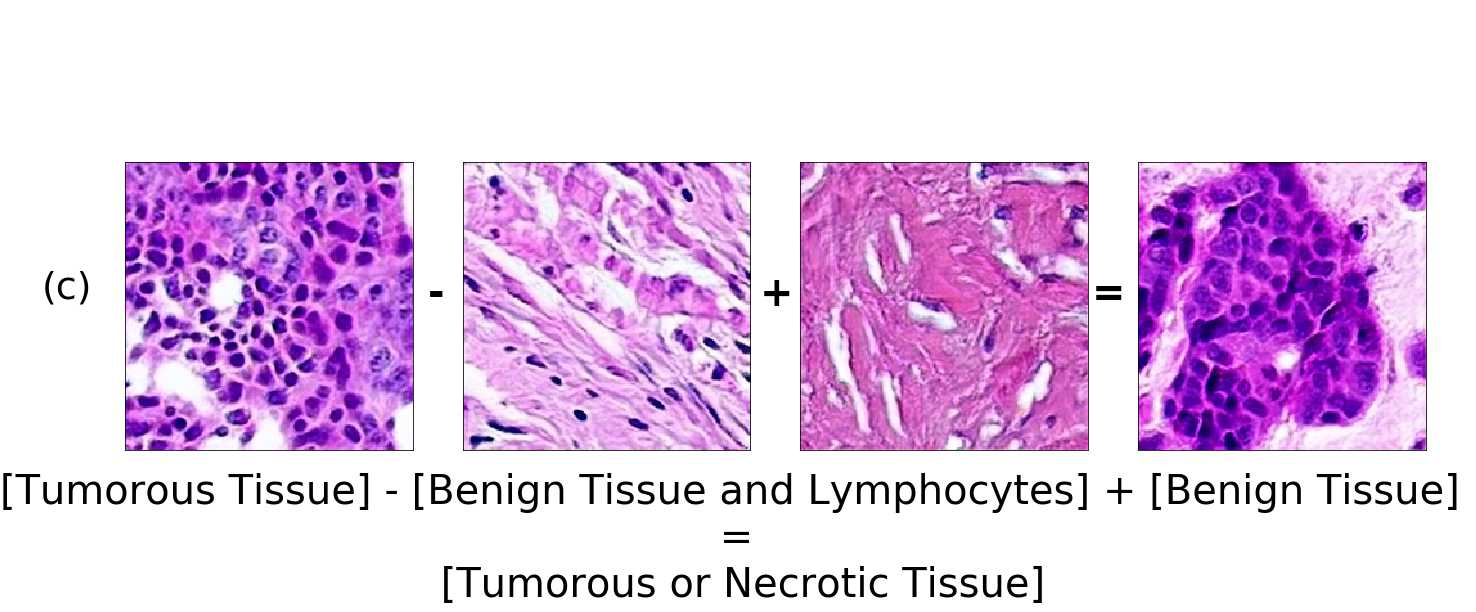}
        \includegraphics[scale=0.165, trim=0 0 0 100, clip]{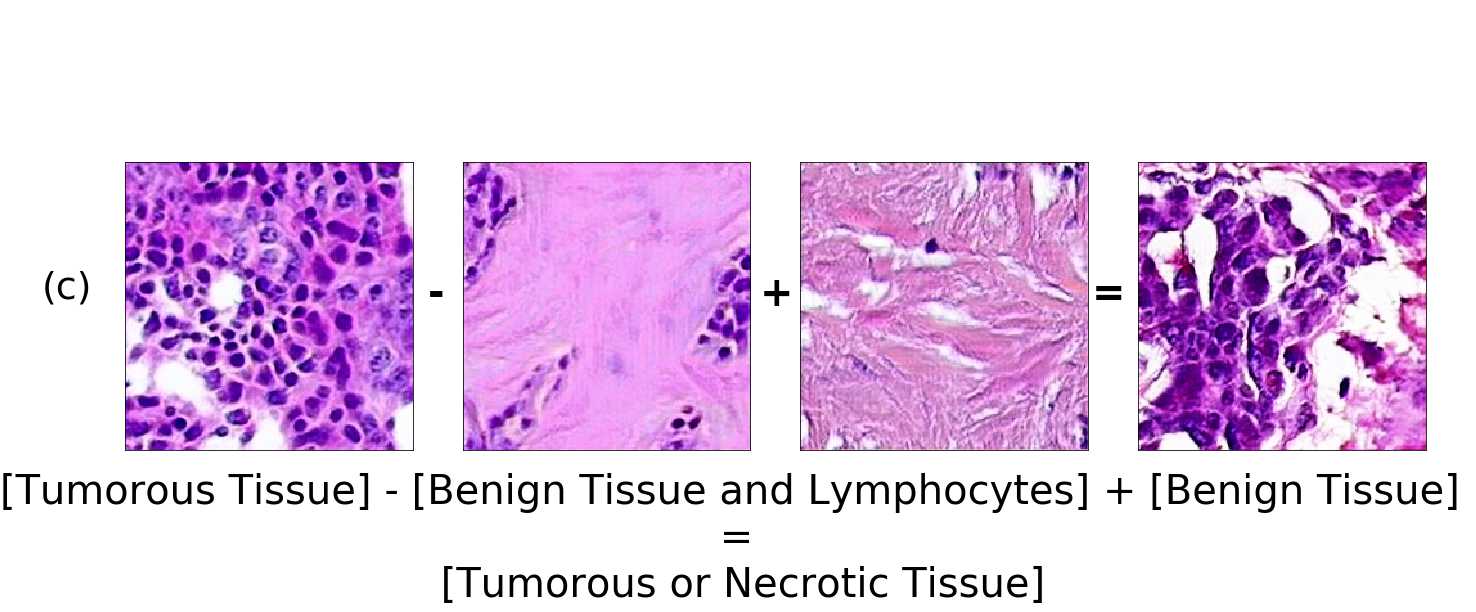}
        \caption{Breast cancer: samples of vector operations with different images, all operations correspond to: Tumorous tissue with lymphocytes - benign tissue with lymphocytes + benign tissue = tumorous or necrotic tissue.}
        \label{fig:breast_vector_op_2}
    \end{figure}
    
    \begin{figure}[H]
        \centering
        \includegraphics[scale=0.165, trim=0 0 0 100, clip]{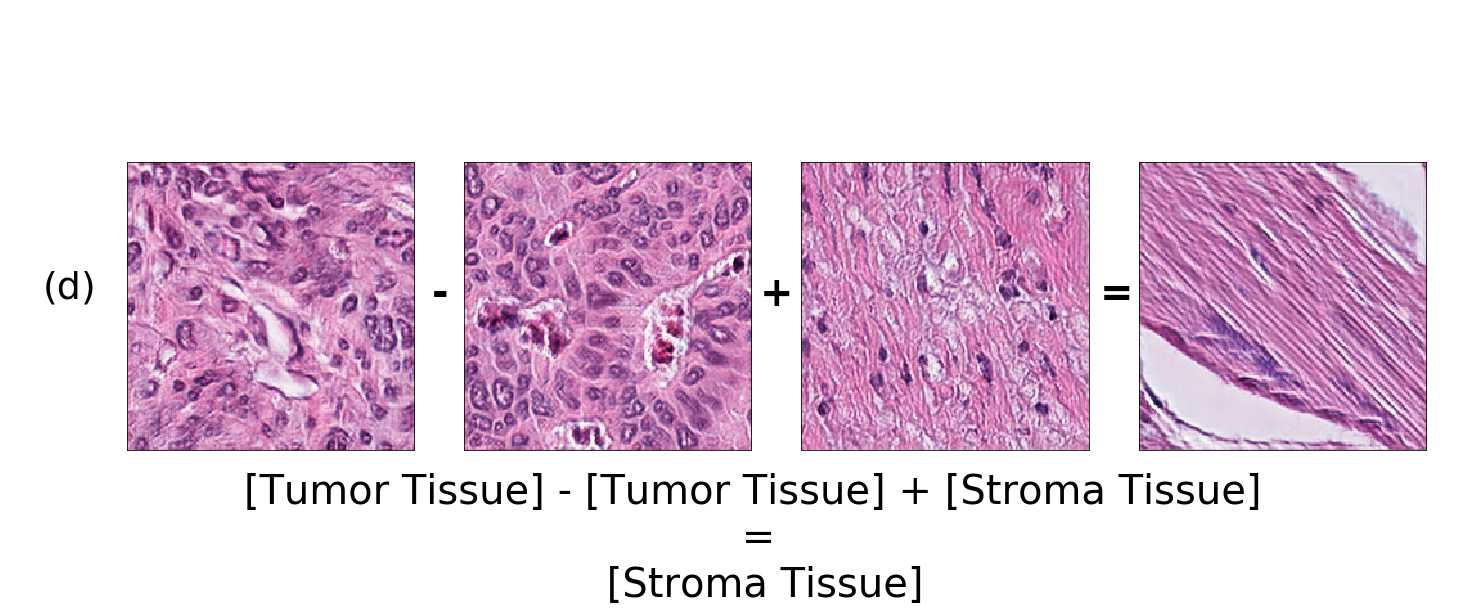}
        \includegraphics[scale=0.165, trim=0 0 0 100, clip]{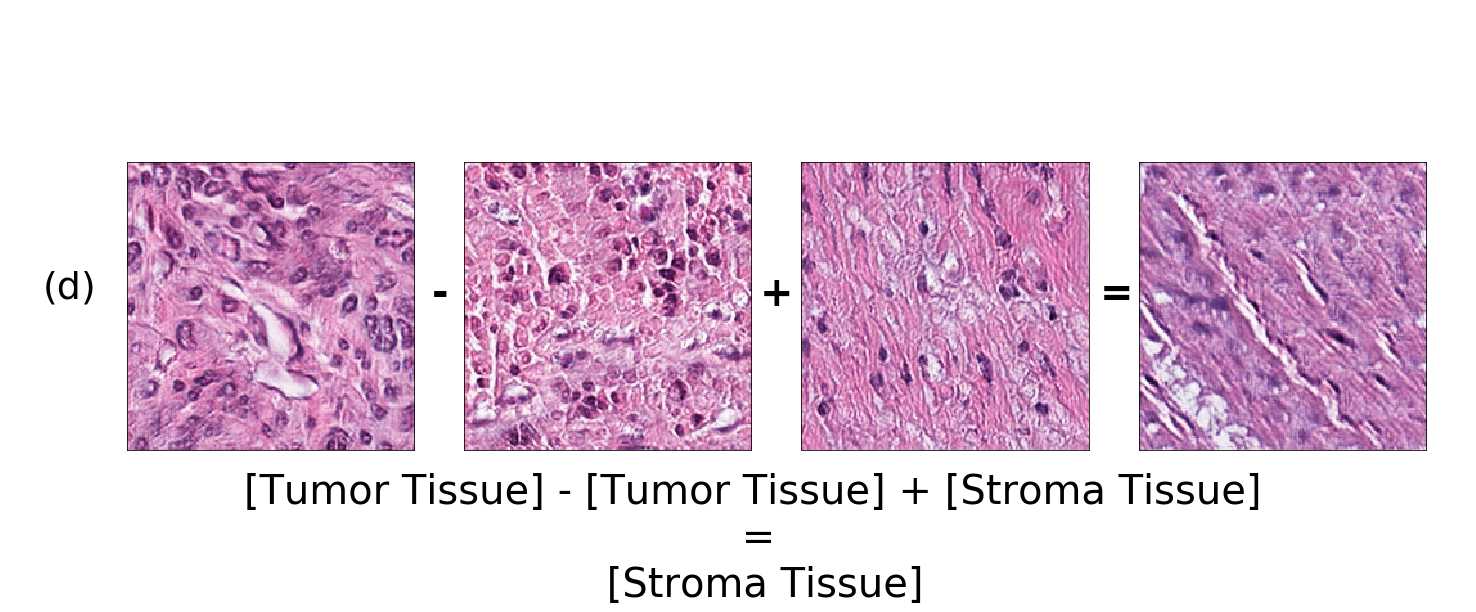}
        \includegraphics[scale=0.165, trim=0 0 0 100, clip]{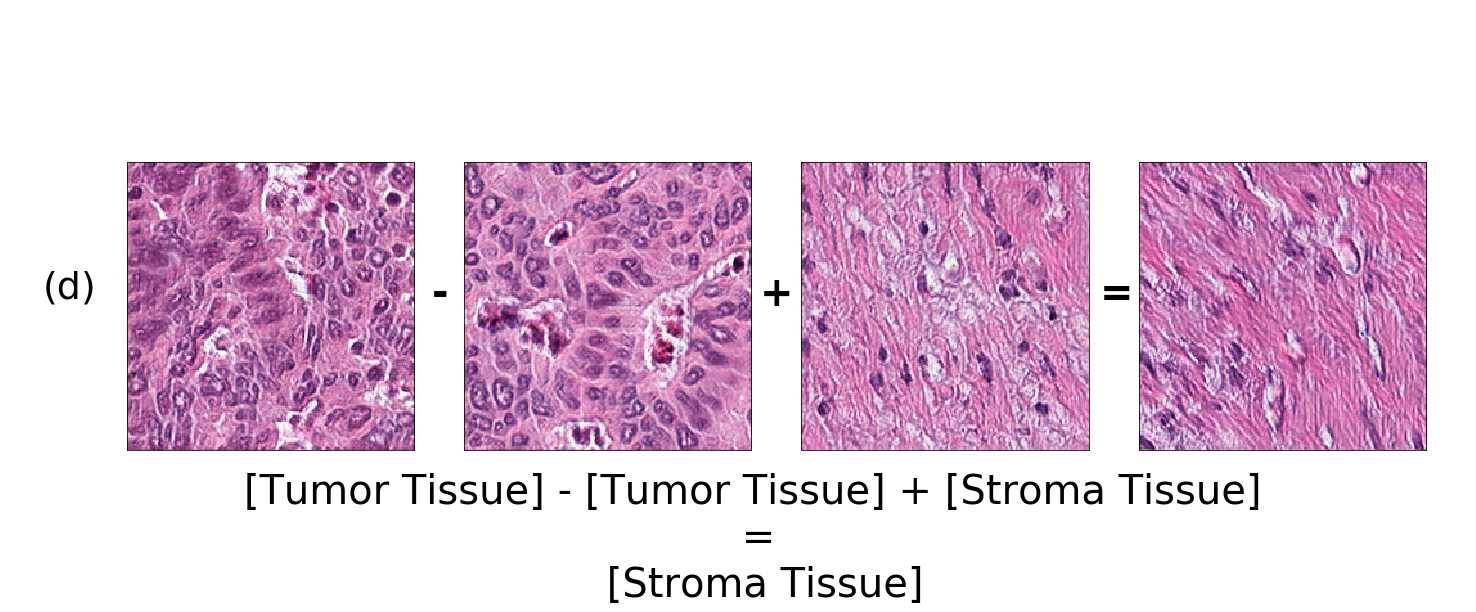}
        \includegraphics[scale=0.165, trim=0 0 0 100, clip]{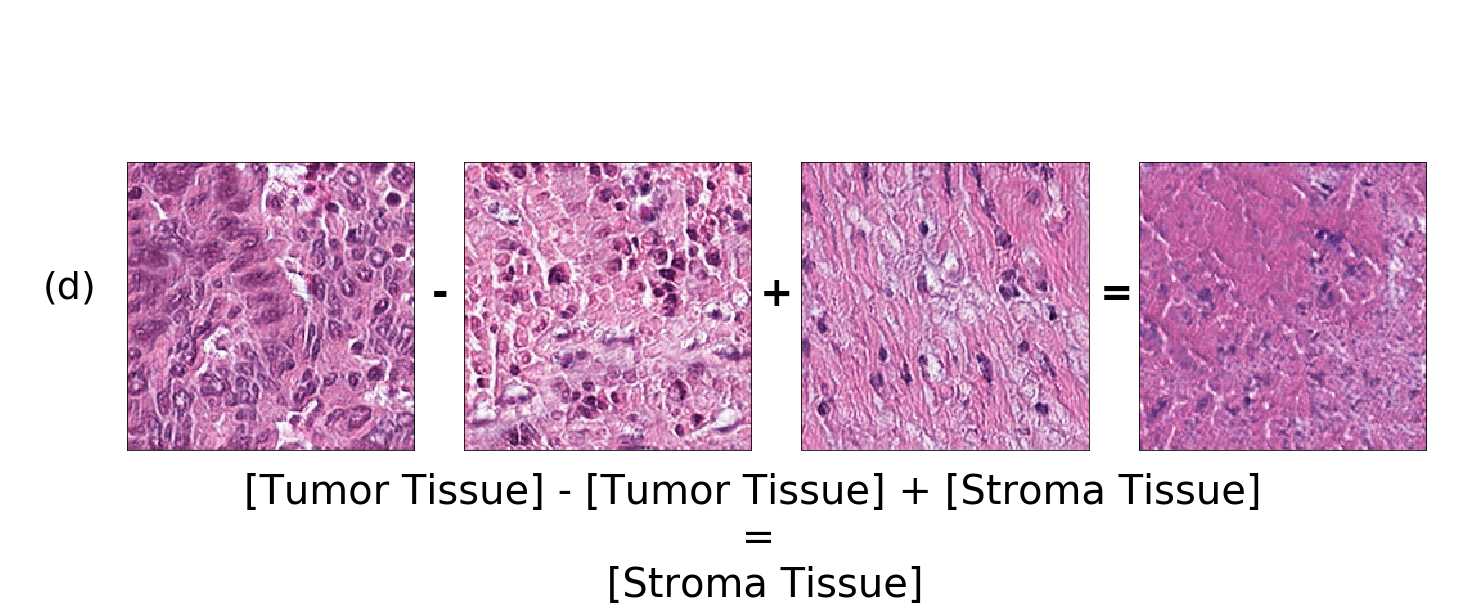}
        \caption{Colorectal cancer: samples of vector operations with different images, all operations correspond to: Tumor  tissue  - Tumor tissue + Stroma tissue = Stroma Tissue.}
        \label{fig:crc_vector_op_0}
    \end{figure}
    
    \begin{figure}[H]
        \centering
        \includegraphics[scale=0.165, trim=0 0 0 100, clip]{images/crc/vector_op/op_1_1.jpg}
        \includegraphics[scale=0.165, trim=0 0 0 100, clip]{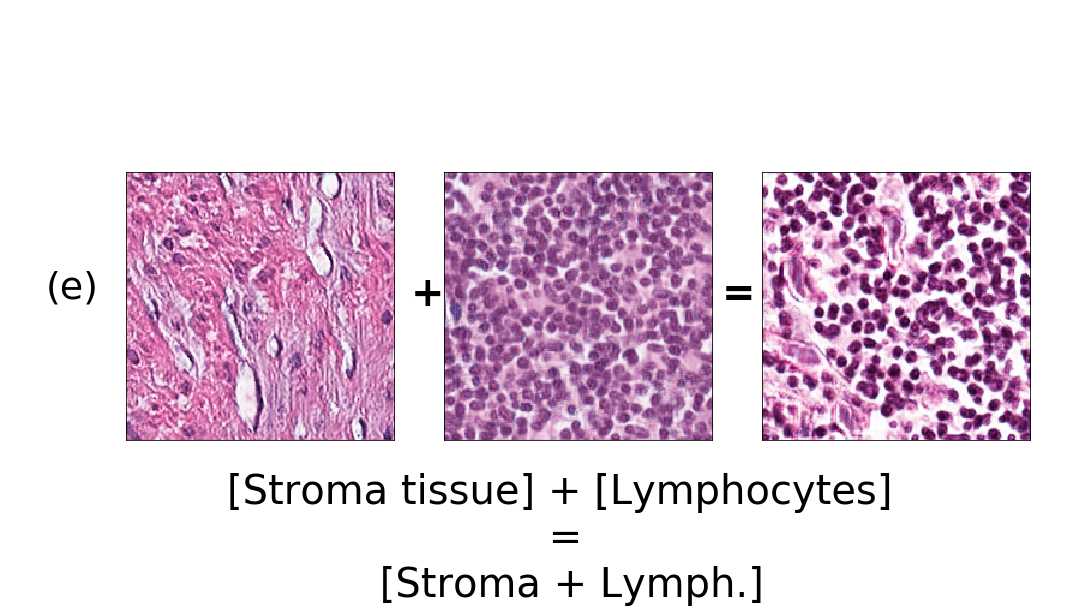}
        \includegraphics[scale=0.165, trim=0 0 0 100, clip]{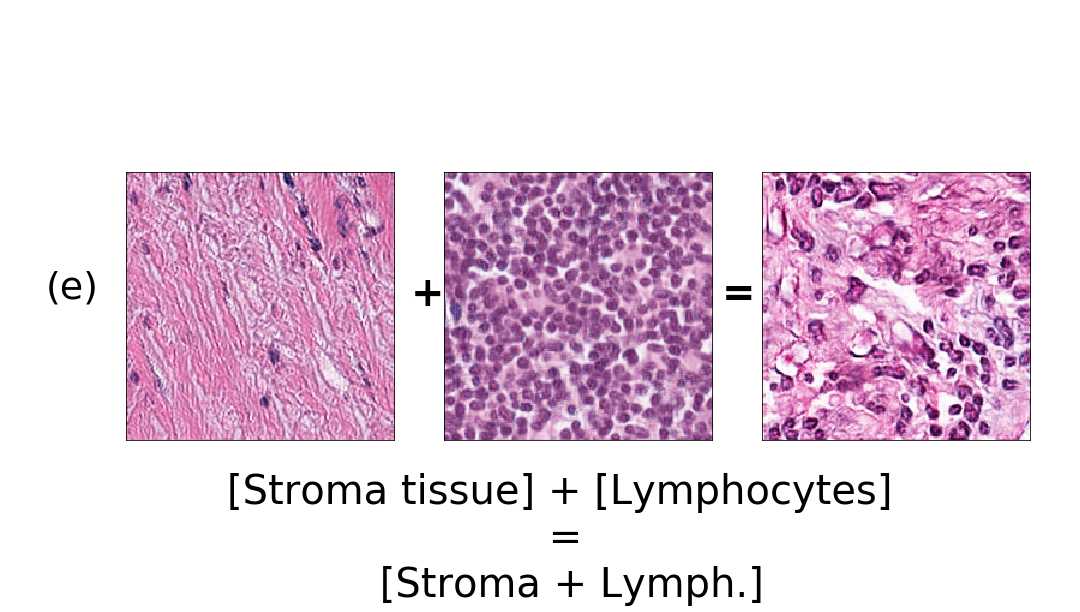}
        \includegraphics[scale=0.165, trim=0 0 0 100, clip]{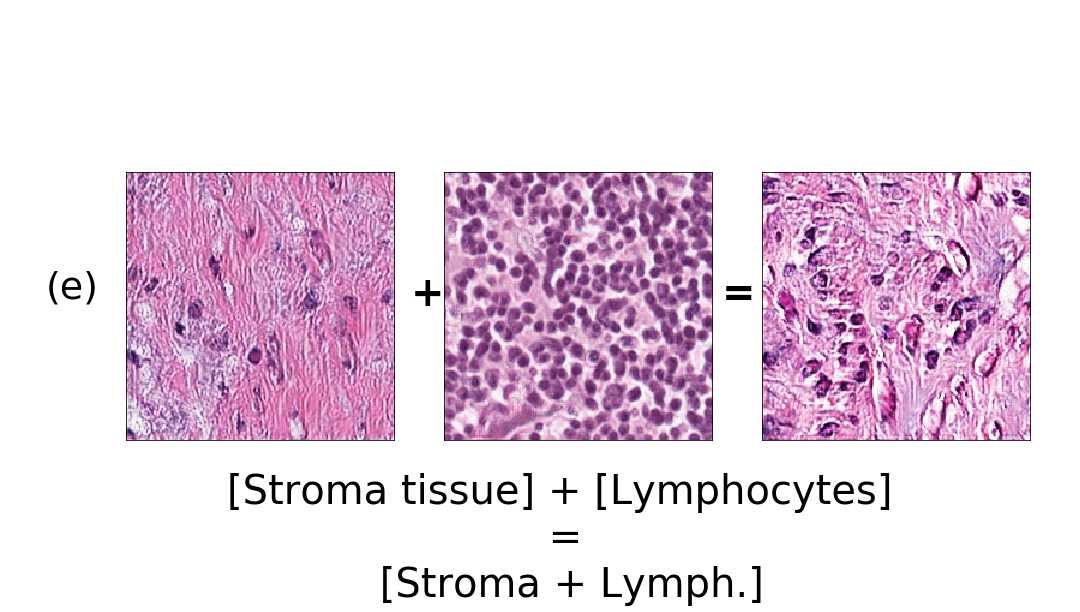}
        \caption{Colorectal cancer: samples of vector operations with different images, all operations correspond to: Stroma tissue + Lymphocytes = Stroma and Lymphocytes.}
        \label{fig:crc_vector_op_1}
    \end{figure}
    
    \begin{figure}[H]
        \centering
        \includegraphics[scale=0.165, trim=0 0 0 100, clip]{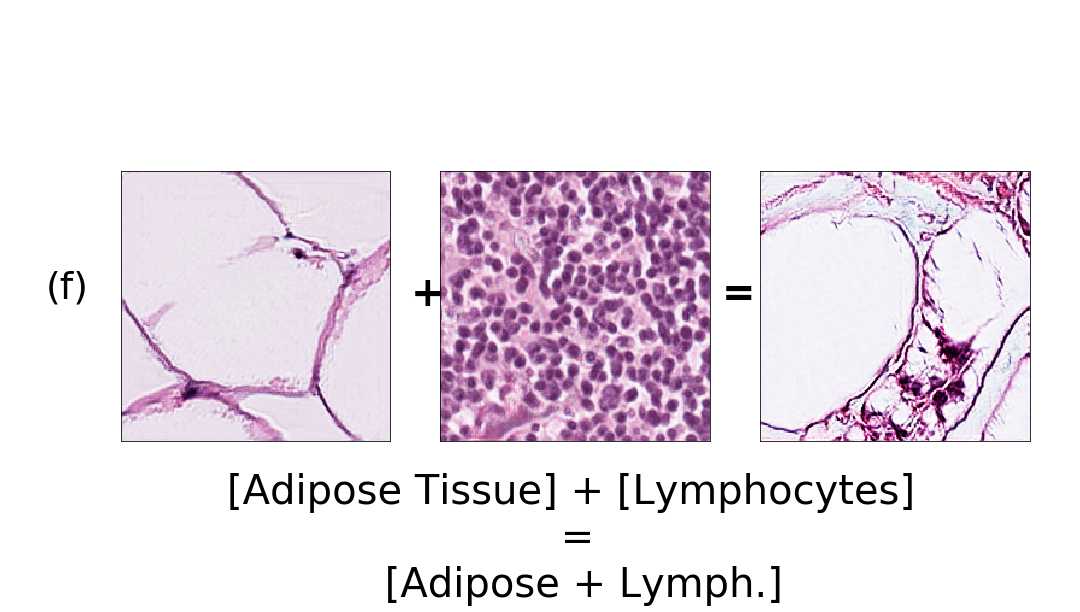}
        \includegraphics[scale=0.165, trim=0 0 0 100, clip]{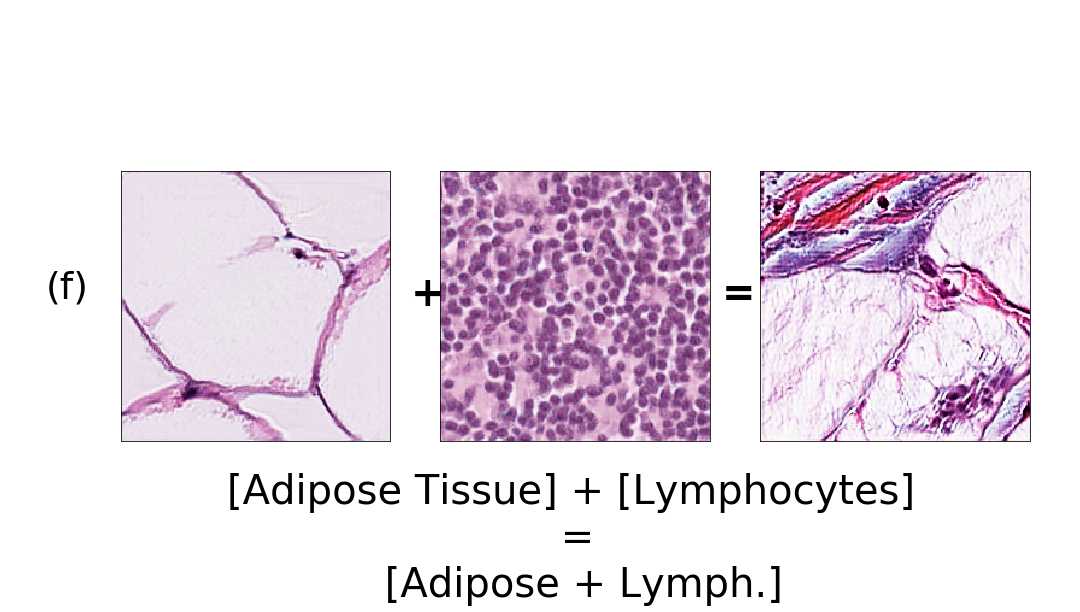}
        \includegraphics[scale=0.165, trim=0 0 0 100, clip]{images/crc/vector_op/op_2_11.jpg}
        \includegraphics[scale=0.165, trim=0 0 0 100, clip]{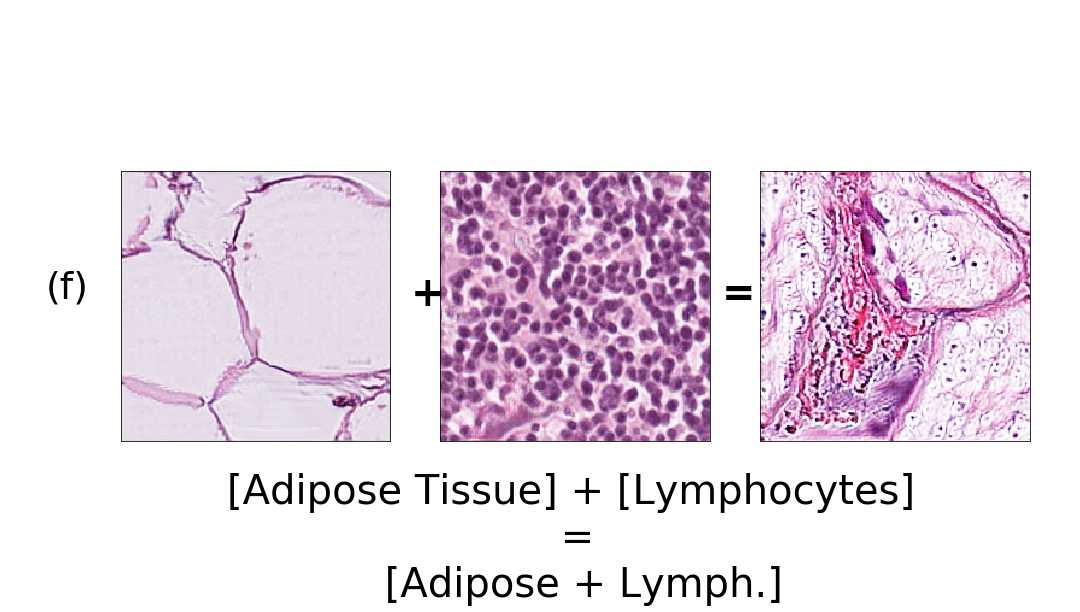}
        \caption{Colorectal cancer: samples of vector operations with different images, all operations correspond to: Adipose tissue + Lymphocytes = Adipose and Lymphocytes.}
        \label{fig:crc_vector_op_2}
    \end{figure}
    
    \begin{figure}[H]
        \centering
        \includegraphics[scale=0.165, trim=0 0 0 100, clip]{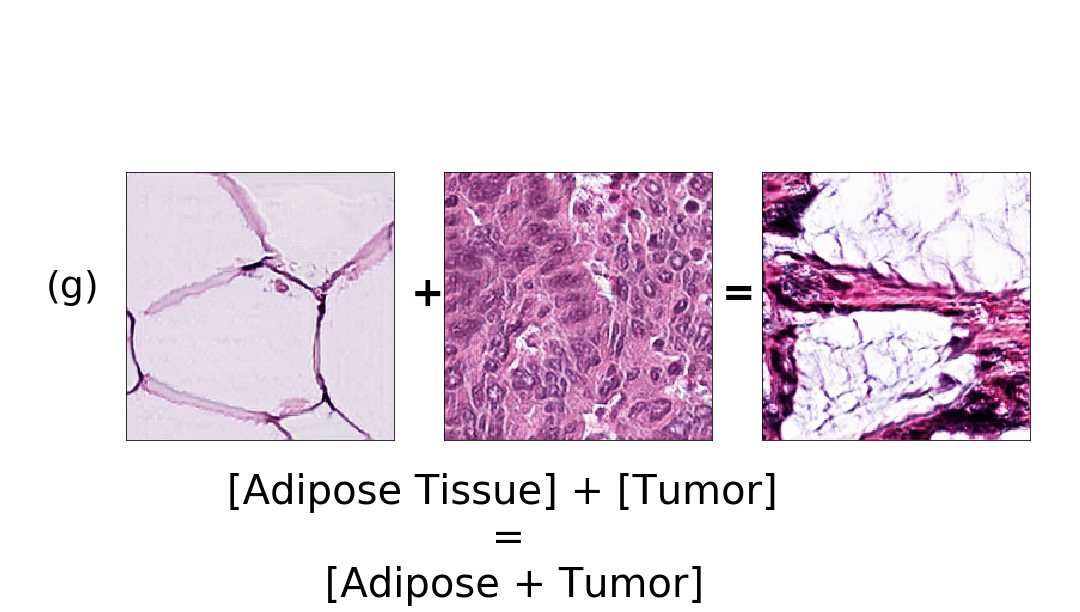}
        \includegraphics[scale=0.165, trim=0 0 0 100, clip]{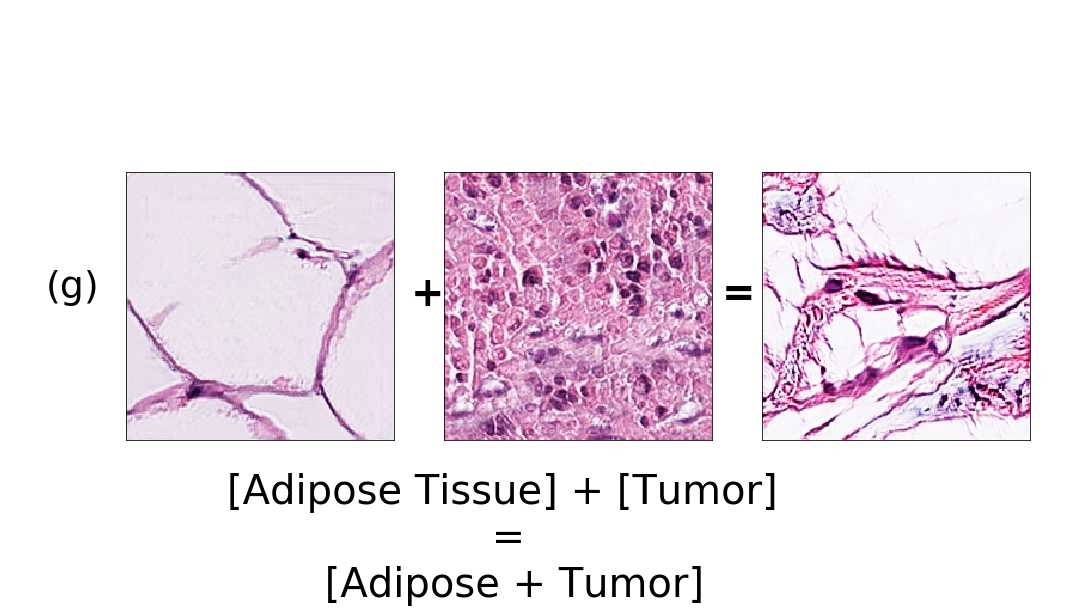}
        \includegraphics[scale=0.165, trim=0 0 0 100, clip]{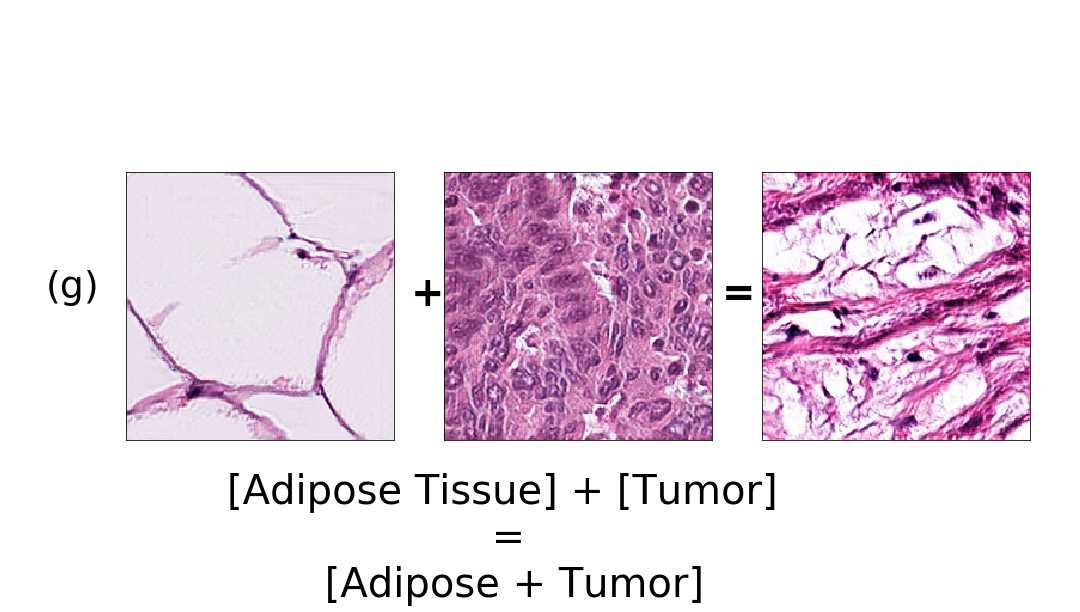}
        \includegraphics[scale=0.165, trim=0 0 0 100, clip]{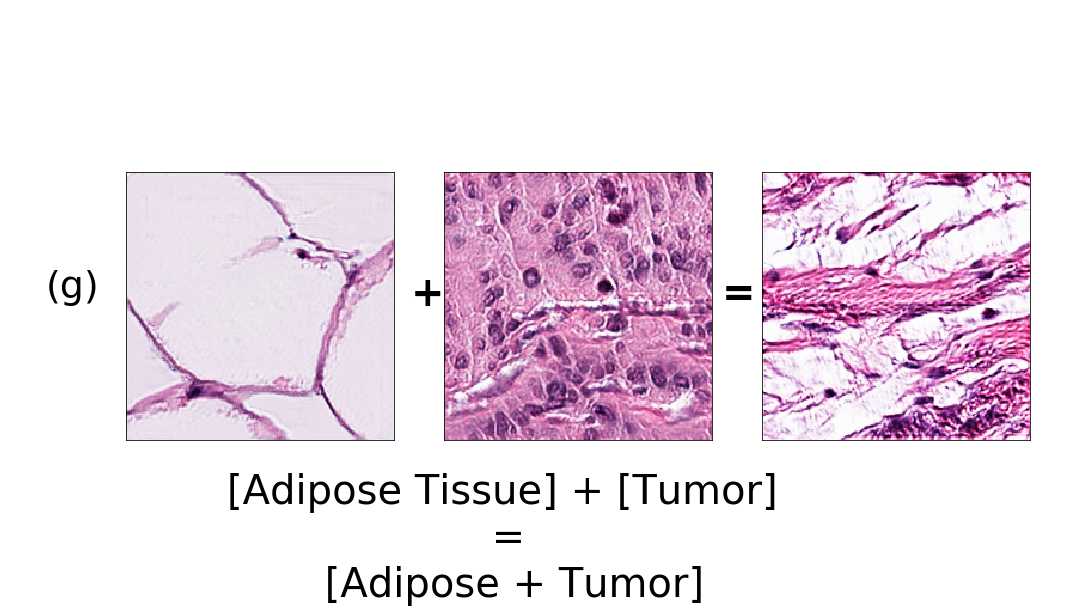}
        \caption{Colorectal cancer: samples of vector operations with different images, all operations correspond to: Adipose tissue + Tumor tissue = Adipose and Tumor.}
        \label{fig:crc_vector_op_3}
    \end{figure}
    
\section{PathologyGAN at 448x448}    
    We include in this section experimental results of a $448\times448$ image resolution model. We trained this model for 90 epochs over approximately five days, using four NVIDIA Titan RTX $24$ GB. 
    
    Over one model the results of Inception FID and CRImage FID were $29.53$ and $203$ respectively. We found that CRImage FID is highly sensitive to changes in the images since it looks for morphological shapes of cancer cells, lymphocytes, and  stroma in the tissue, at this resolution the generated tissue images don't hold the same high quality as in the $224\times224$ case. As we capture in the \textbf{Conclusion} section, this is an opportunity to improve the detail in the generated image at high resolutions.
    
\begin{figure}[H]
        \centering
        \includegraphics[scale=0.15]{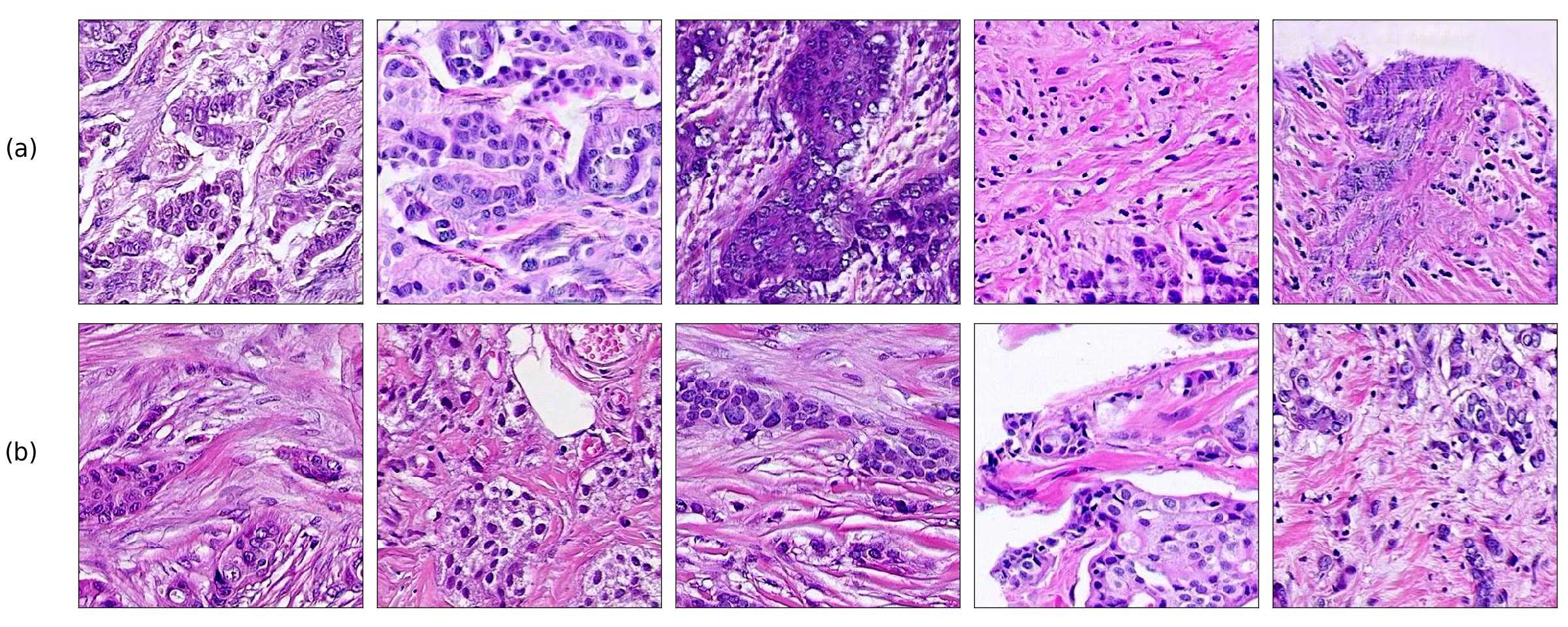}
        \includegraphics[scale=0.15]{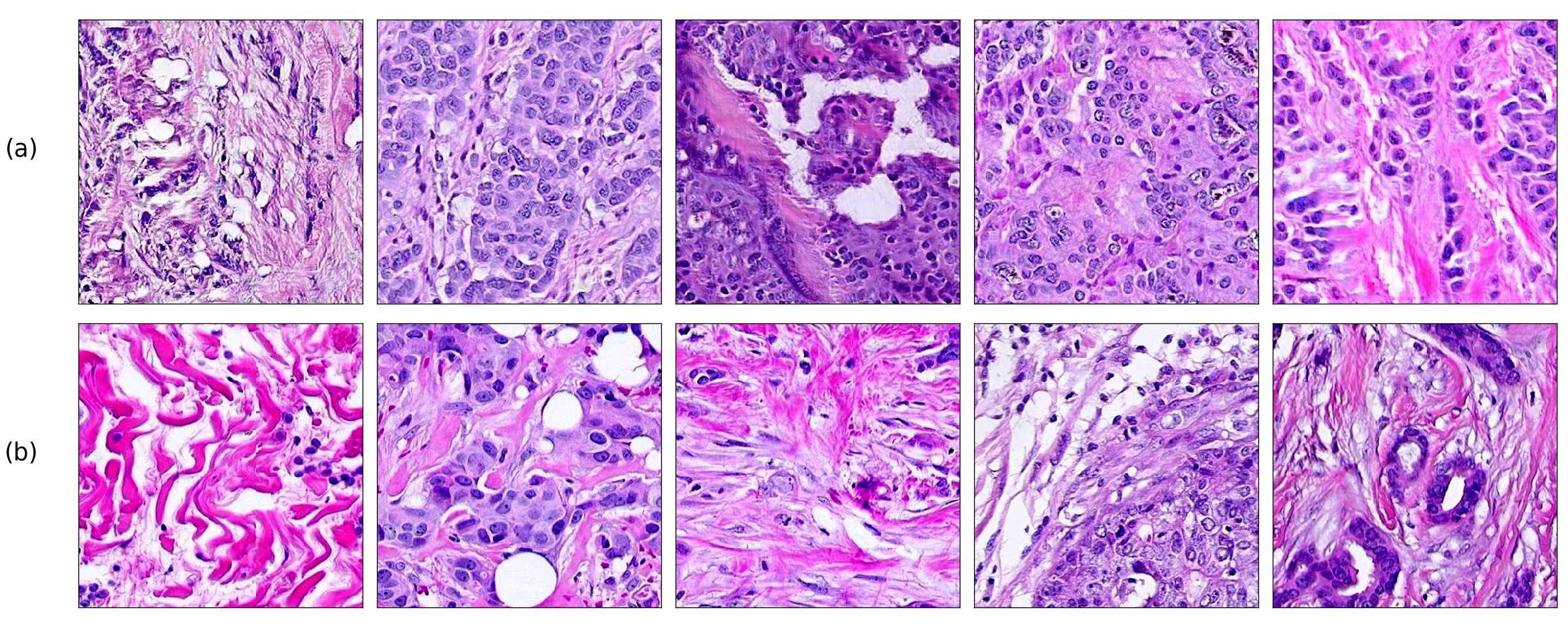}
        \includegraphics[scale=0.15]{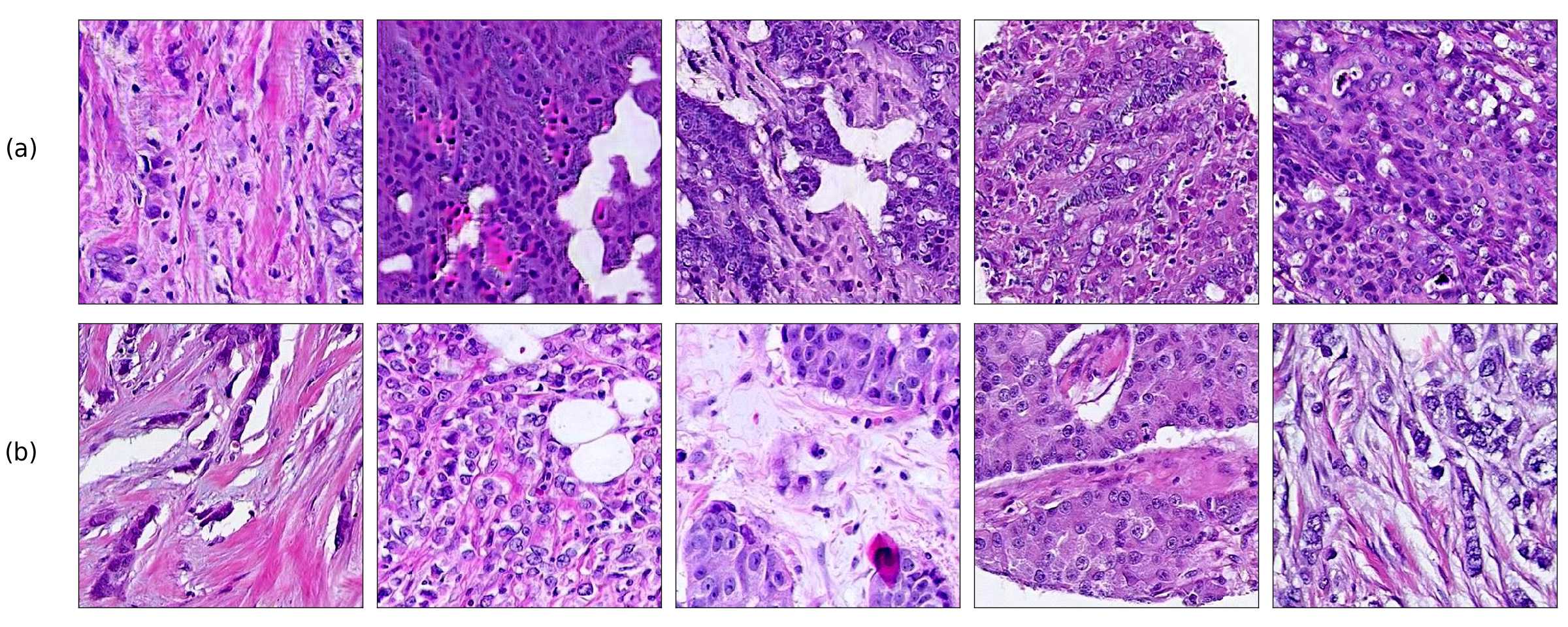}
        \caption{(a): Images ($448\times448$) from PathologyGAN trained on H\&E breast cancer tissue. (b): Real images.}
        \label{fig:hand_picked_448_samples}
    \end{figure}

\begin{figure}[H]
        \centering
        \includegraphics[scale=0.145]{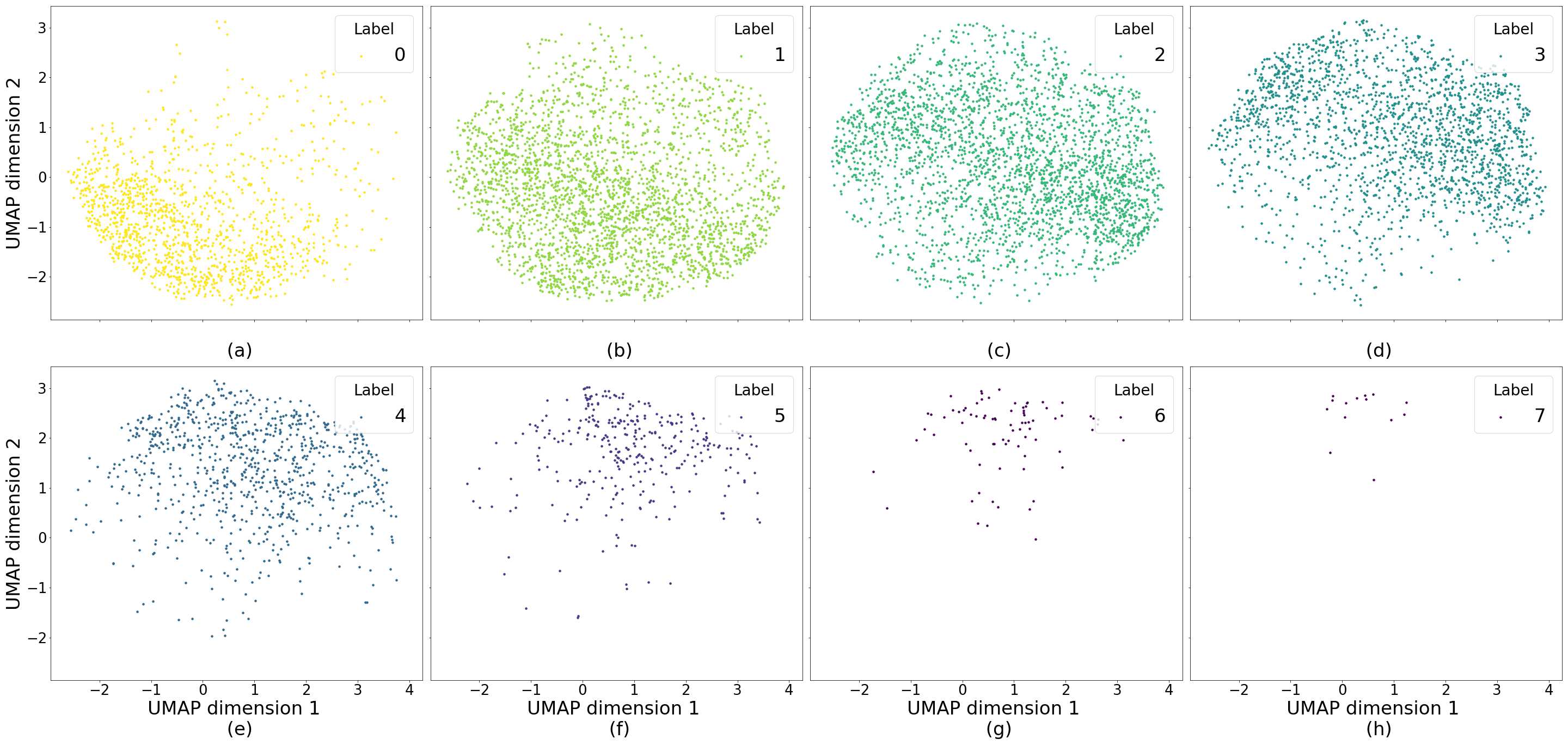}
        \includegraphics[scale=0.20]{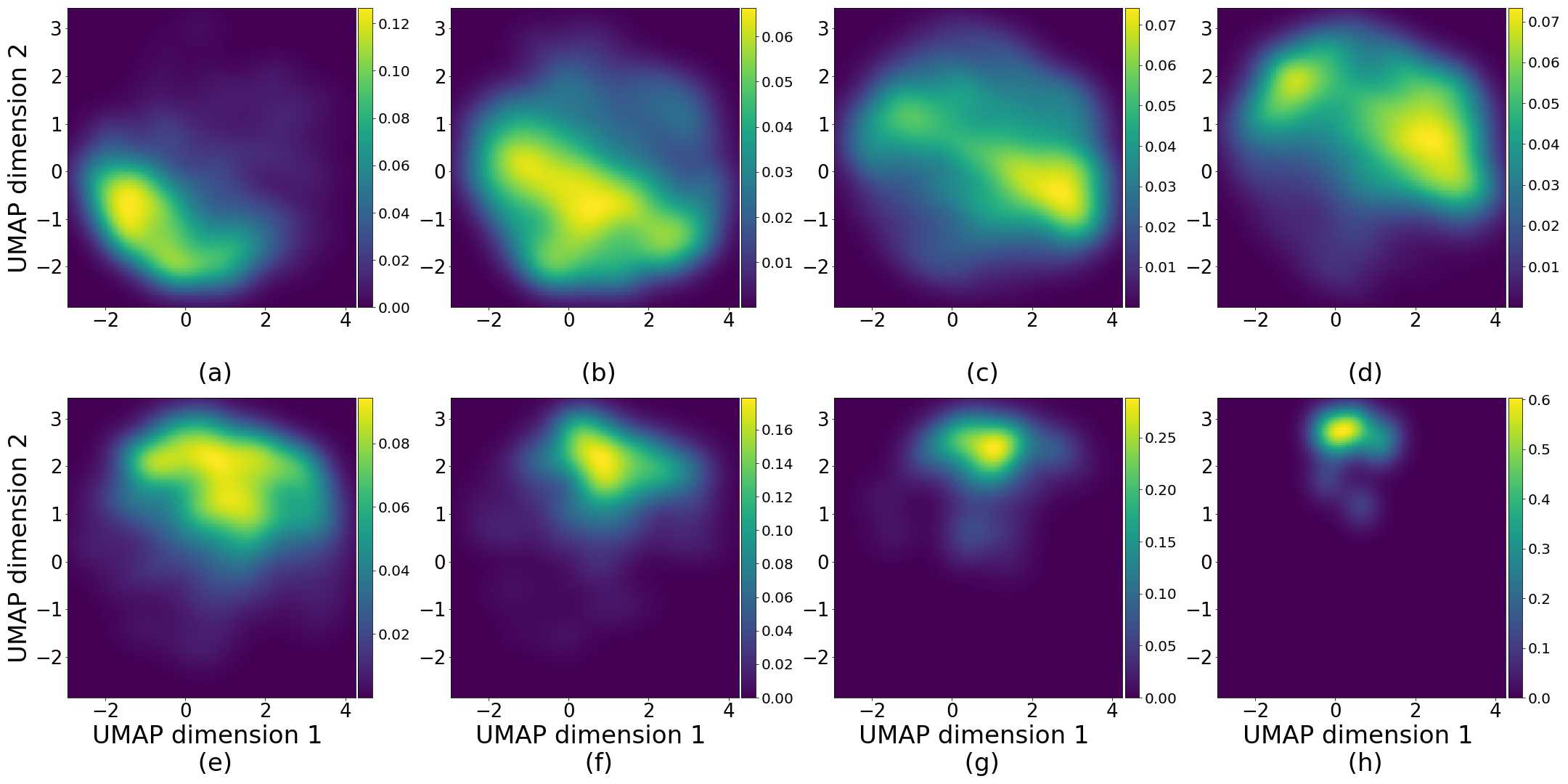}
        \caption{Scatter and density plots of $448\times448$ samples on the UMAP reduced representation of the latent space $w$. Each subfigure (a-h) belongs to samples of only one class, where each class represents a range of counts of cancer cells in the tissue image. (a) accounts for images with the lowest number of cancer cells and (h) corresponds to images with the largest count, subfigures from (a) to (h) belong to increasing number of cancer cells. As previously shown in Figure \ref{fig:latent_space_comp_point_label_breast} and \ref{fig:latent_space_comp_density_label_breast} for the $224\times224$ resolution, representation learning properties are still held at $448\times448$.}
        \label{fig:latent_space_448}
    \end{figure}

    Figure \ref{fig:hand_picked_448_samples} show three examples of comparisons between (a) PathologyGAN images and (b) real images. Additionally, the representation learning properties are still preserved in the latent space. Figure \ref{fig:latent_space_448} captures the density of cancer cells in the $448\times448$ tissue images as previously presented for the $224\times224$ case in \textbf{Appendix C.}
    
\section{Visualization of linear interpolations and vector operations in the latent space.}
\label{appendix:latent_li_vp}
    
    \begin{figure}[H]
		\centering
	    \includegraphics[scale=0.19]{./images/crc/linear_interpolation/PathologyGAN_linear_interpolation_crc_stroma_tumor.jpg}
        \includegraphics[scale=0.19]{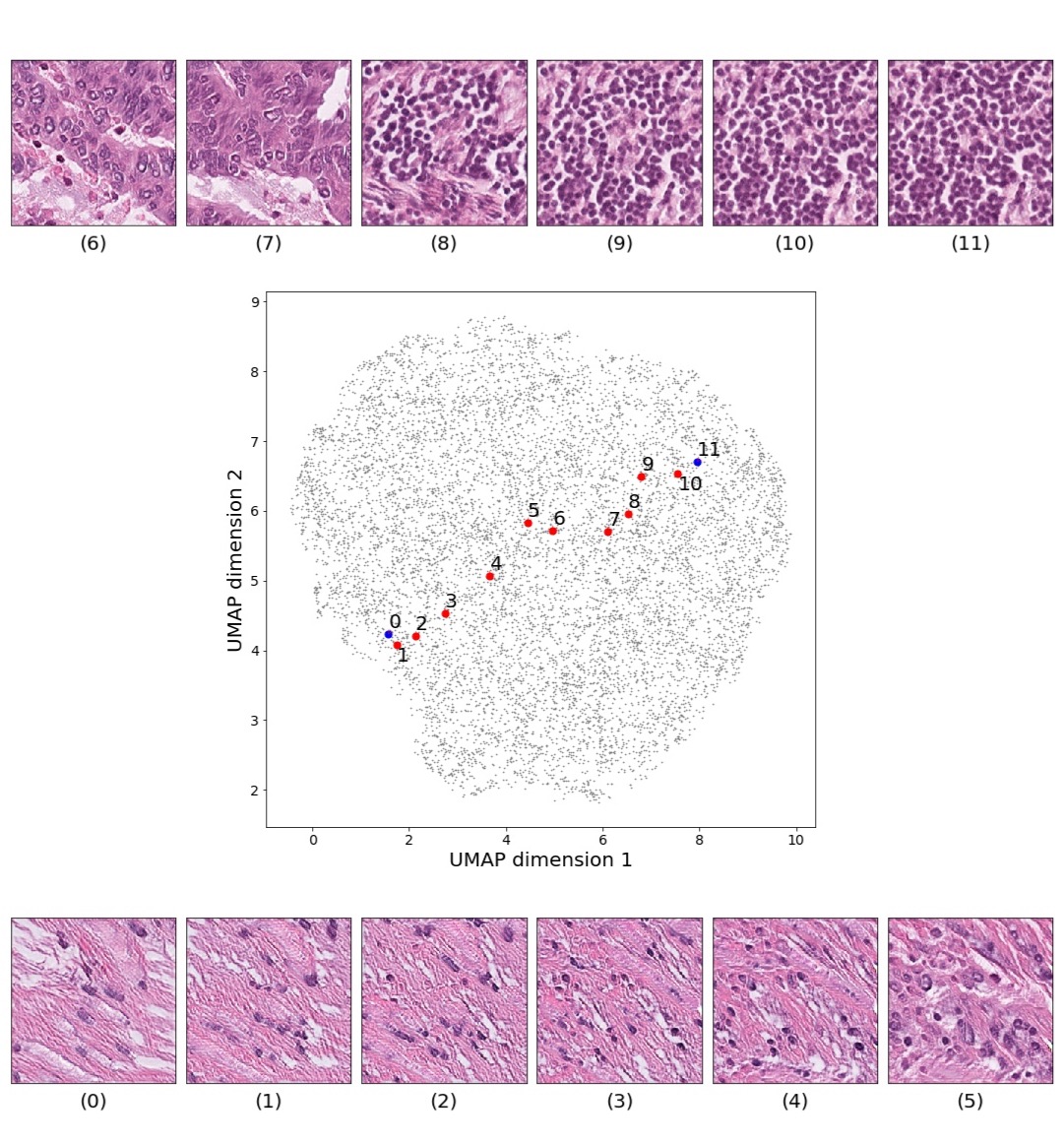}
		\includegraphics[scale=0.19]{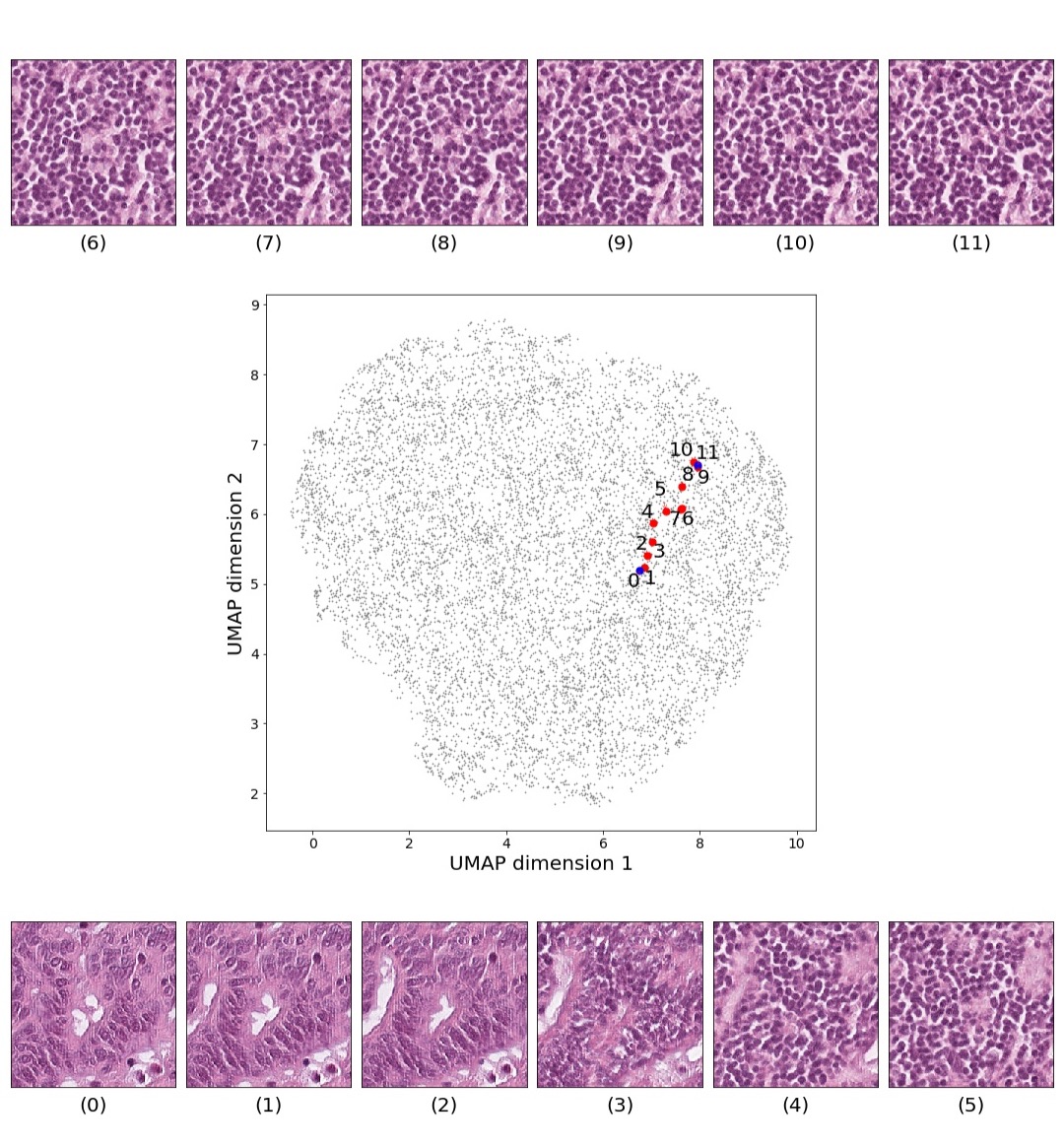}
        \vspace{-0.5cm}
		 \caption{Uniform Manifold Approximation and Projection (UMAP) representations of generated tissue samples where linear interpolations in the latent space are highlighted. We present colorectal cancer transition from stroma to tumor, stroma to lymphocytes, and tumor to lymphocytes. Starting vectors are colored as blue while intermediate points of the interpolations are colored as red. Through the intermediate vectors we show that gradual transitions in the latent space translate into smooth feature transformations, increase/decrease of tumorous cells or increase of lymphocyte counts.}
		 \label{fig:linear_interpolation_latent_space_appendix_crc}
	\end{figure}  
    
     \begin{figure}[H]
		\centering
	    \includegraphics[scale=0.19]{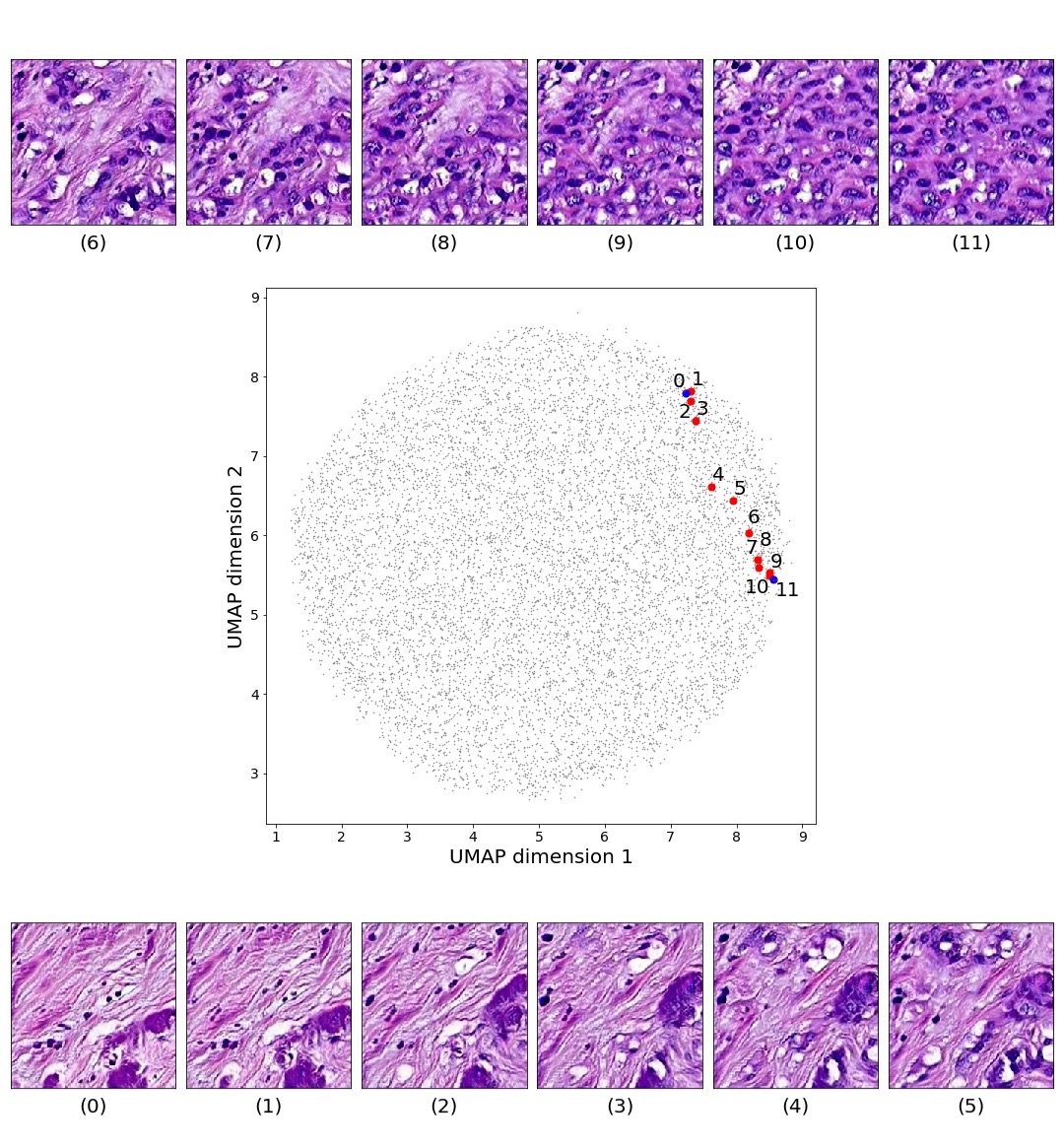}
        \includegraphics[scale=0.19]{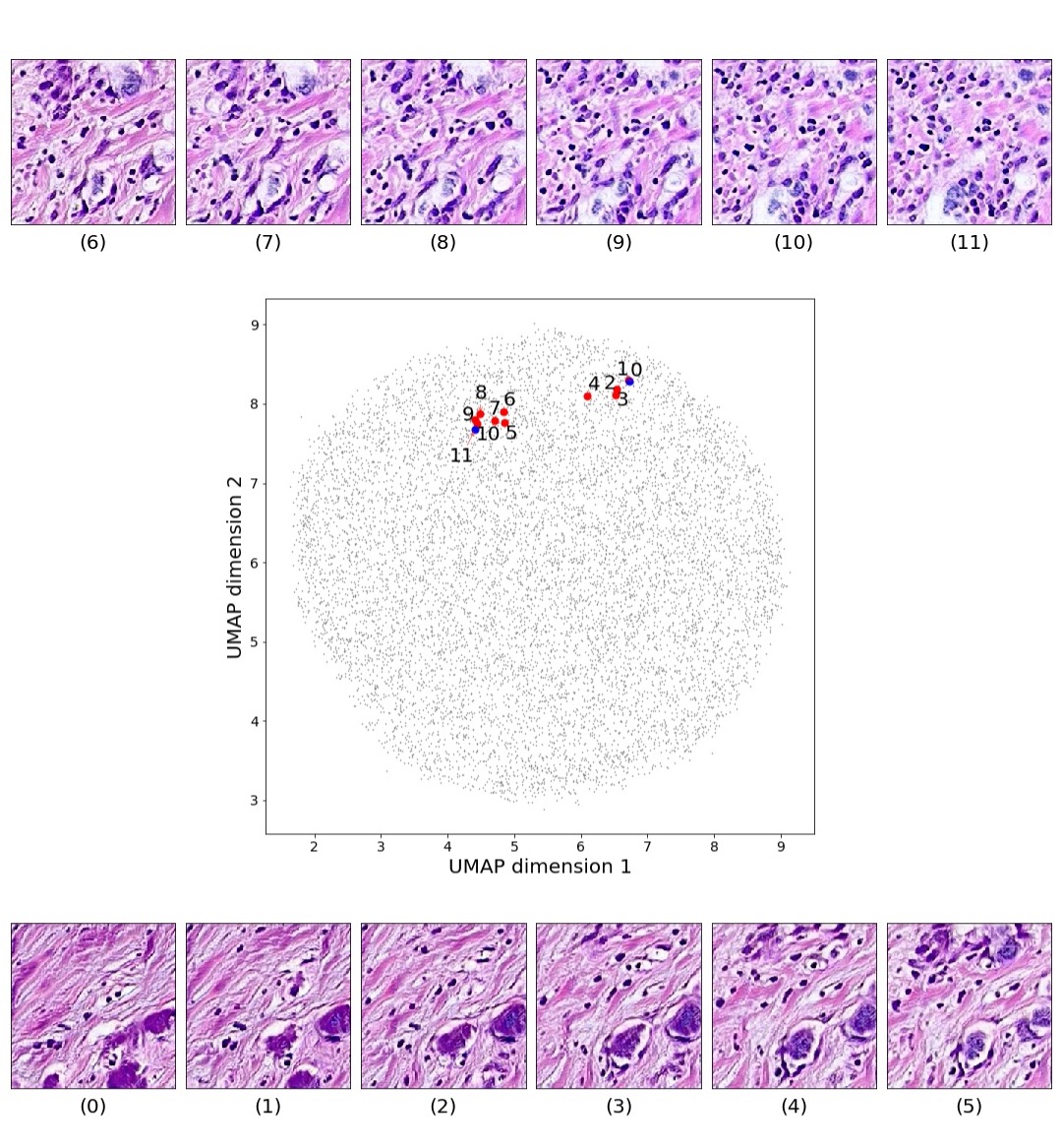}
		\includegraphics[scale=0.19]{./images/vgh_nki/linear_interpolation/PathologyGAN_linear_interpolation_vgh_nki_tumor_lymp.jpg}
        \vspace{-0.25cm}
        \caption{Uniform Manifold Approximation and Projection (UMAP) representations of generated tissue samples where linear interpolations in the latent space are highlighted. We present breast cancer transition from stroma to tumor, stroma to lymphocytes, and tumor to lymphocytes. Starting vectors are colored as blue while intermediate points of the interpolations are colored as red. Through the intermediate vectors we show that gradual transitions in the latent space translate into smooth feature transformations, increase/decrease of tumorous cells or increase of lymphocyte counts.}
		 \label{fig:linear_interpolation_latent_space_appendix_vgh}
	\end{figure}  
	
	\begin{figure}[H]
		\centering
		\minipage{0.5\textwidth}
		  \includegraphics[scale=0.25]{./images/crc/vector_op/PathologyGAN_vector_op_tumor_stroma.jpg}
		\endminipage\hfill
		\minipage{0.5\textwidth}
		  \includegraphics[scale=0.25]{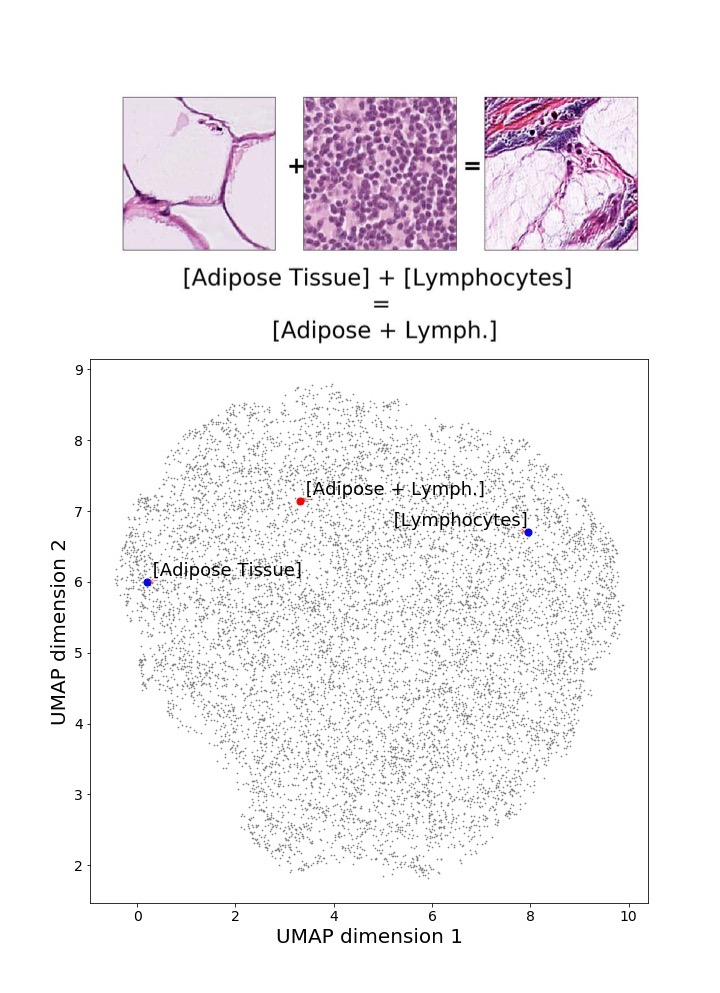}
		\endminipage
		\vfill 
		\includegraphics[scale=0.25]{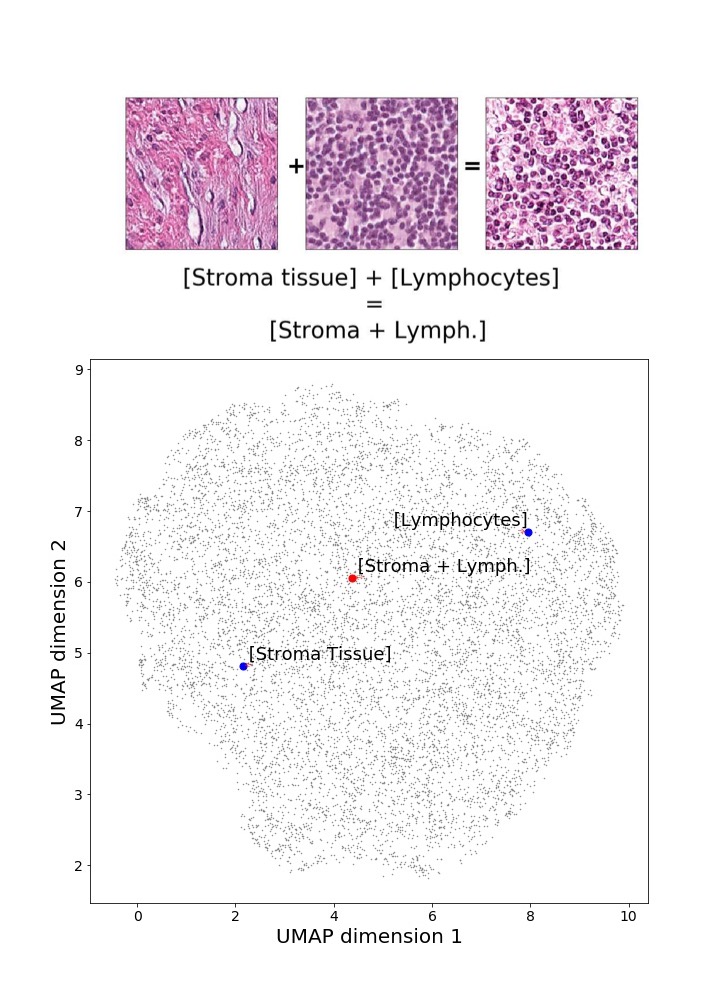}
        \vspace{-0.5cm}
		 \caption{Uniform Manifold Approximation and Projection (UMAP) representations of generated tissue samples where vectors involved in the linear vector operations are highlighted. Original vectors are colored in blue while the results are colored in red. We show colorectal cancer examples where after vector operations  the results fall into regions of the latent space that correspond to the tissue type.}
		 \label{fig:vector_op_latent_space_appendix_crc}
	\end{figure}  
	
	\begin{figure}[H]
		\centering
		\minipage{0.5\textwidth}
		  \includegraphics[scale=0.25]{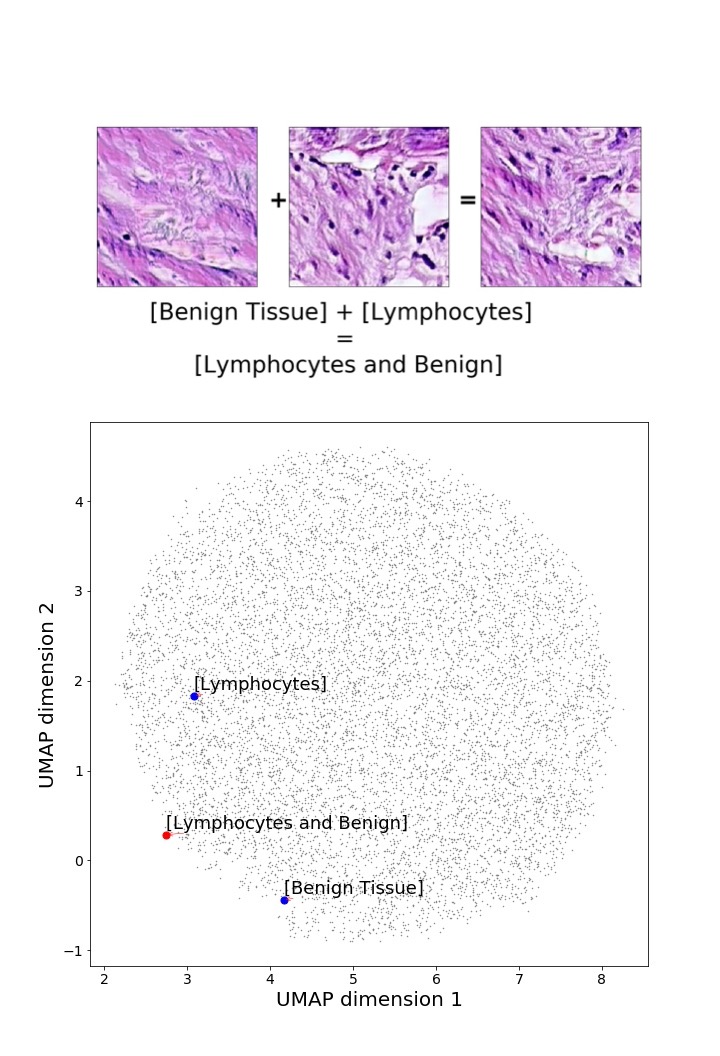}
		\endminipage\hfill
		\minipage{0.5\textwidth}
		  \includegraphics[scale=0.25]{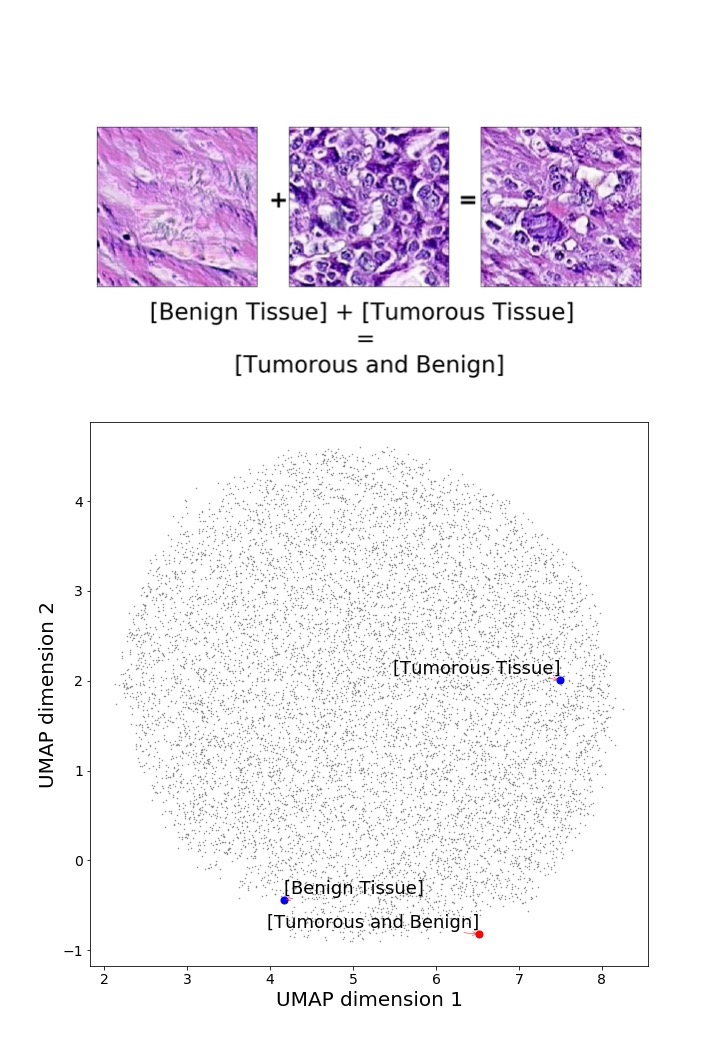}
		\endminipage
		\vfill 
		\includegraphics[scale=0.25]{./images/vgh_nki/vector_op/PathologyGAN_vector_op_tumor_lymph.jpg}
        \vspace{-0.5cm}
		 \caption{Uniform Manifold Approximation and Projection (UMAP) representations of generated tissue samples where vectors involved in the linear vector operations are highlighted. Original vectors are colored in blue while the results are colored in red. We show  breast cancer examples where after vector operations the results combination of different features such as tumor, benign tissue, or lymphocytes.}
		 \label{fig:vector_op_latent_space_appendix_vgh}
	\end{figure}      

\section{GAN evaluation metrics for digital pathology}
\label{GANevalDP}

    In this section, we investigate how relevant GAN evaluation metrics perform on distinguishing differences in cancer tissue distributions. We center our attention on metrics that are model agnostic and work with a set of generated images. We focus on Fr\'echet Inception distance (FID), Kernel Inception Distance (KID), and 1-Nearest Neighbor classifier (1-NN) as common metrics to evaluate GANs. We do not include Inception Score and Mode Score because they do not compare to real data directly, they require a classification network on survival times and estrogen-receptor (ER), and they have also showed lower performance when evaluating GANs \citep{Barratt2018, Xu2018}.
    
    \cite{Xu2018} reported that the choice of feature space is critical for evaluation metrics, so we follow these results by using the 'pool\_3' layer from an ImageNet trained Inception-V1 as a convolutional feature space. 

    We set up two experiments to test how the evaluation metrics capture:
    \begin{itemize}
        \item Artificial contamination from different staining markers and cancer types.
        \item Consistency when two sample distributions of the same database are compared.
    \end{itemize}
    
    \subsection{Detecting changes in markers and cancer tissue features}
\begin{figure}[htbp]
        \centering
        \includegraphics[scale=0.21]{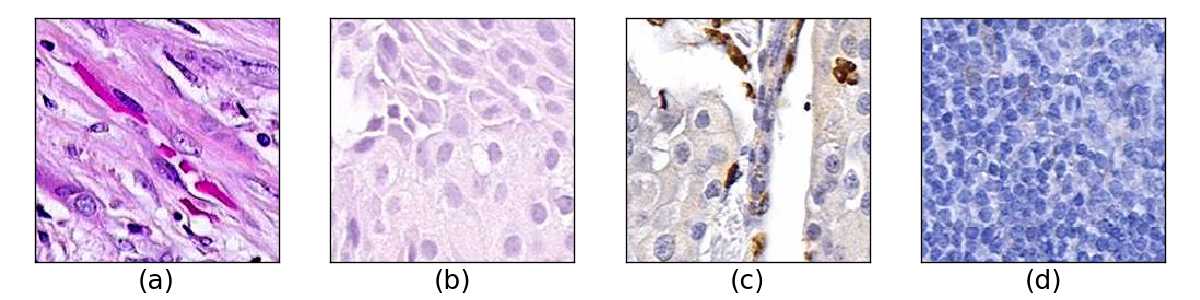}
        \caption{Different cancer types and markers. (a) H\&E Breast cancer, (b) H\&E Bladder cancer, (c) Capthepsin-L Breast cancer, and (d) CD137 Bone marrow cancer. We can see different coloring per marker, and tissue architecture per cancer type.}
        \label{fig:cancer_types_markers}
    \end{figure}

    We used multiple cancer types and markers to account for alterations of color and shapes in the tissue. Markers highlight parts of the tissue with different colors, and cancer types have distinct tissues structures. Examples of these changes are displayed in Figure~\ref{fig:cancer_types_markers}.

    We constructed one reference image set with 5000 H\&E breast cancer images from our data sets of NKI and VGH, and compared it against another set of 5000 H\&E breast cancer images contaminated with other markers and cancer types. We used three types of marker-cancer combinations for contamination, all from the Stanford TMA Database \citep{stanford_tma}: H\&E - Bladder cancer, Cathepsin-L - Breast cancer, and CD137 - Lymph/Colon/Liver cancer.
    
    Each set of images was constructed by randomly sampling from the respective marker-cancer type data set, which is done to minimize the overlap between the clean and contaminated sets.
    
\begin{figure}[ht]
        \centering
        \includegraphics[scale=0.275]{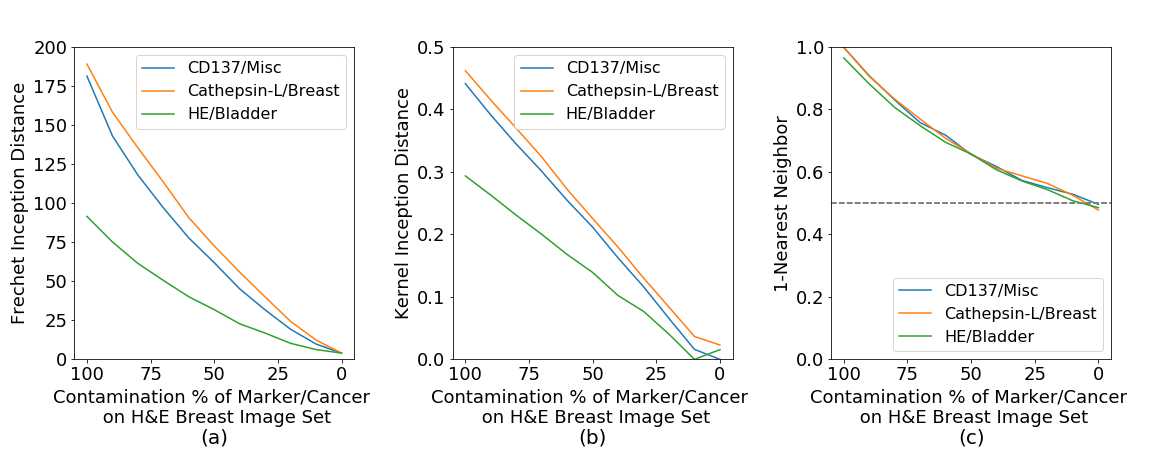}
        \caption{Distinguishing a set H\&E Breast cancer images against different contamination of markers and cancer types. For a metric to be optimnal, the value should decreas along with the contamination. (a) corresponds to FID, (b) KID, (c) 1-NN. FID and KID gradually define changes in marker and tissue type, 1-NN does not provide a clear measure of the changes.}
        \label{fig:contamination_example}
    \end{figure}

    Figure~\ref{fig:contamination_example} shows how (a) FID, (b) KID, (c) 1-NN behave when the reference H\&E breast cancer set is measured against multiple percentage of contaminated H\&E breast cancer sets. Marker types have a large impact due to color change and all metrics capture this except for 1-NN. Cathepsin-L highlights parts of the tissue with brown colors and CD137 has similar color to necrotic tissue on H\&E breast cancer, but still far from the characteristic pink color of H\&E. Accordingly, H\&E-Bladder has a better score in all metrics due to the color stain, again expect for 1-NN. Cancer tissue type differences are captured by all the metrics, which shows a marker predominance, but we can see that on the H\&E marker the differences between breast and bladder types are still captured. 
    
    In this experiment, we find that FID and KID have a gradual response distinguishing between markers and cancer tissue types, however 1-NN is not able to give a measure that clearly defines these changes.
    
    \subsection{Reliability on evaluation metrics}
    Another evaluation we performed was to study which metrics are consistent when two independent sample distributions with the same contamination percentage are compared. To construct this test, for each contamination percentage, we create two independent sample sets of 5000 images and compare them against each other. Again, we constructed these image sets by randomly selecting images for each of the marker-cancer databases. We do this to ensure there are no overlapping images between the distributions.
    
\begin{figure}[!htbp]
        \centering
        \includegraphics[scale=0.275]{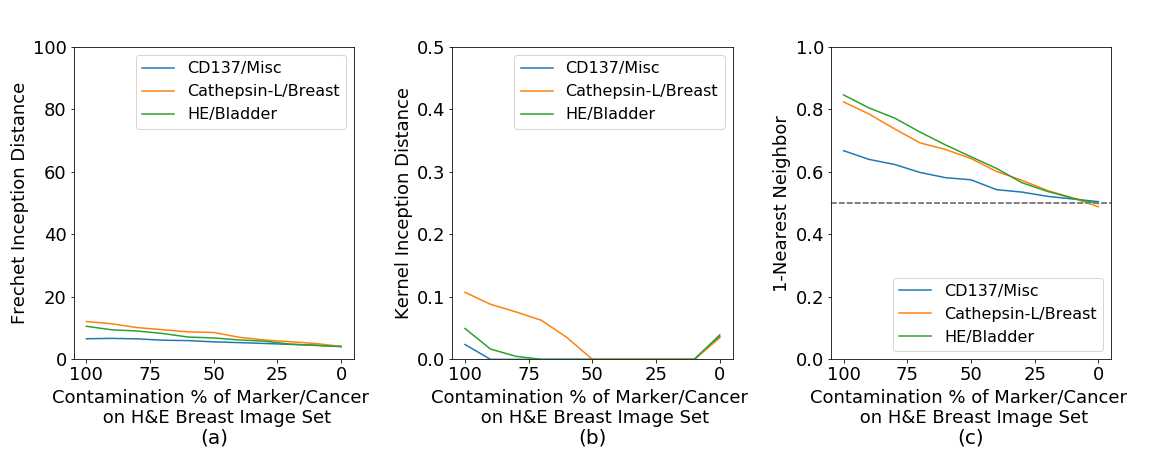}
        \caption{Consistency of metrics when two independent sets of images with the same contamination are compared. Consistent metrics should be close to zero for each of the contamination rates. (a) FID, (b) KID, and (c) 1-NN, we can see that FID is the metric that shows a close to zero constant measure.}
        \label{fig:test_train_example}
    \end{figure}

    In Figure~\ref{fig:test_train_example} we show that (a) FID has a stable performance,  compared to (b) KID, and especially (c) 1-NN. The metrics should show a close to zero distance for each of the contamination rates since we are comparing two sample-distributions from the same data set. This shows that only FID has a close to zero constant behavior across different data sets when comparing the same tissue image distributions.
    
    Based on these two experiments, we argue that 1-NN does not clearly represent changes in the cancer types and marker, and both KID and 1-NN do not give a constant reliable measure across different markers and cancer types. Therefore we focused on FID as the most promising evaluation metrics.
    
\section{Pathologists Tests}
    We provide in here examples of the tests taken by the pathologists:
    \begin{figure}[H]
            \centering
            \includegraphics[scale=0.15]{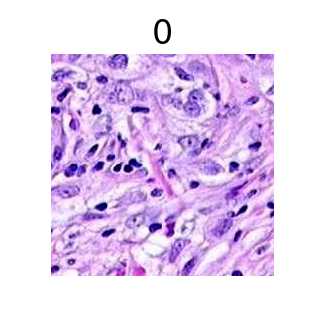}
            \includegraphics[scale=0.15]{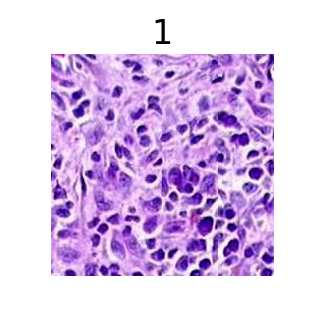}
            \includegraphics[scale=0.15]{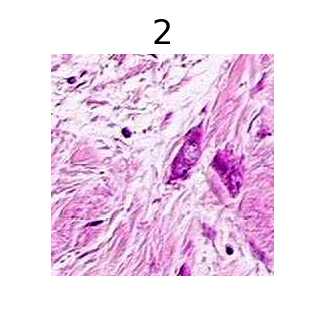}
            \includegraphics[scale=0.15]{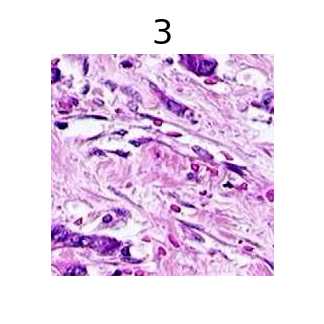}
            \includegraphics[scale=0.15]{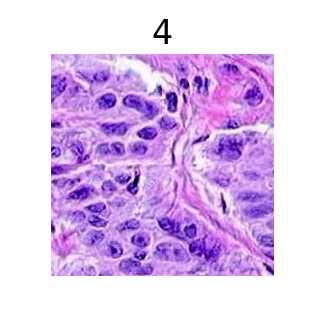}
            \includegraphics[scale=0.15]{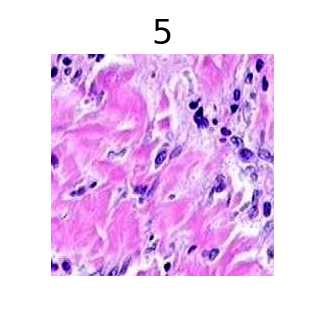}
            \includegraphics[scale=0.15]{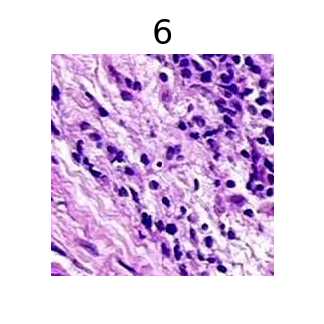}
            \includegraphics[scale=0.15]{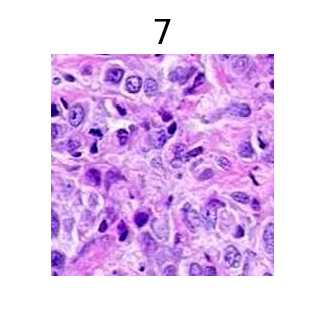}
            \includegraphics[scale=0.15]{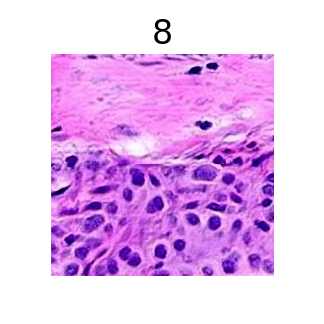}
            \includegraphics[scale=0.15]{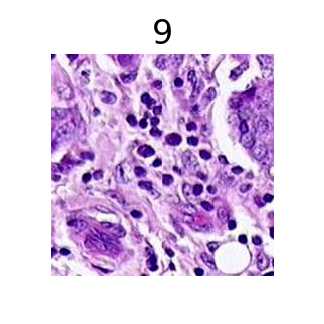}
            \caption{Individual images for breast cancer tissue.}
            \label{fig:test_breast}
    \end{figure}
    
    \begin{figure}[H]
            \centering
            \includegraphics[scale=0.15]{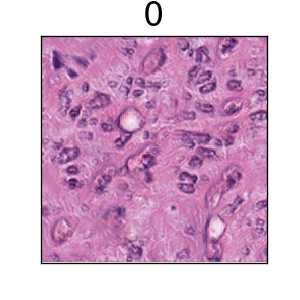}
            \includegraphics[scale=0.15]{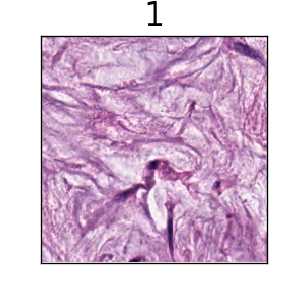}
            \includegraphics[scale=0.15]{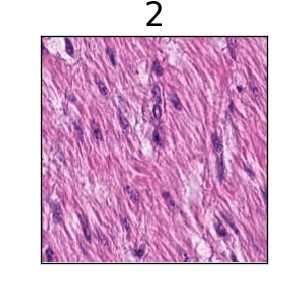}
            \includegraphics[scale=0.15]{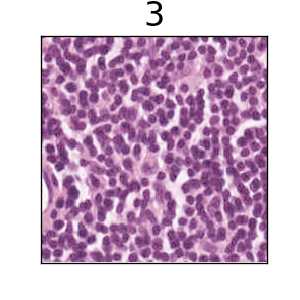}
            \includegraphics[scale=0.15]{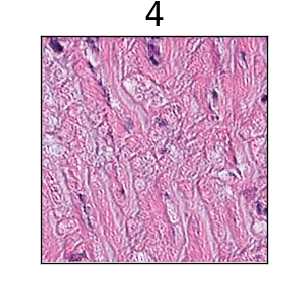}
            \includegraphics[scale=0.15]{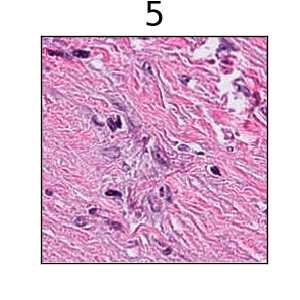}
            \includegraphics[scale=0.15]{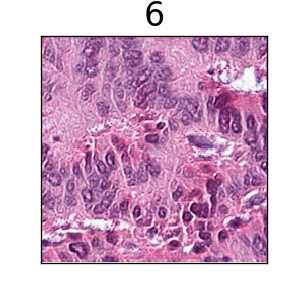}
            \includegraphics[scale=0.15]{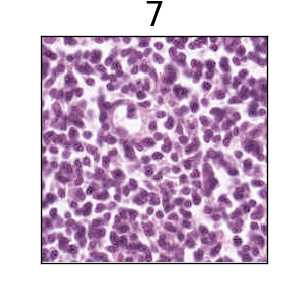}
            \includegraphics[scale=0.15]{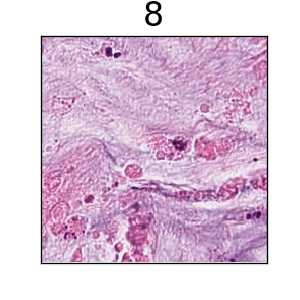}
            \includegraphics[scale=0.15]{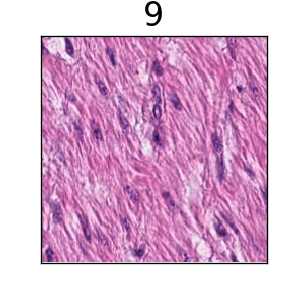}
            \caption{Individual images for colorectal cancer tissue.}
            \label{fig:test_crc}
    \end{figure}

\section{Model Architecture}
	\label{appendix:model_arch}
	     \begin{table}[H]
	        \centering
	        \begin{tabular}{c}
	        Generator Network $G:w \rightarrow x$ \\
	        \toprule
	        \midrule
	        Dense Layer, adaptive instance normalization (AdaIN), and leakyReLU \\
	        $200 \rightarrow 1024$ \\
	        \midrule
	        Dense Layer, AdaIN, and leakyReLU \\
	        $1024 \rightarrow 12544$ \\
	        \midrule
	        Reshape $7\times7\times256$ \\
	        \midrule
	        ResNet Conv2D Layer, 3x3, stride 1, pad same, AdaIN, and leakyReLU $0.2$ \\
	        $ 7\times7\times256 \rightarrow 7\times7\times256 $ \\
	        \midrule
	        ConvTranspose2D Layer, 2x2, stride 2, pad upscale, AdaIN, and leakyReLU $0.2$ \\
	        $ 7\times7\times256 \rightarrow 14\times14\times512 $ \\
	        \midrule
	        ResNet Conv2D Layer, 3x3, stride 1, pad same, AdaIN, and leakyReLU $0.2$ \\
	        $ 14\times14\times512 \rightarrow 14\times14\times512 $ \\
	        \midrule
	        ConvTranspose2D Layer, 2x2, stride 2, pad upscale, AdaIN, and leakyReLU $0.2$ \\
	        $ 14\times14\times512 \rightarrow 28\times28\times256 $ \\
	        \midrule
	        ResNet Conv2D Layer, 3x3, stride 1, pad same, AdaIN, and leakyReLU $0.2$ \\
	        $ 28\times28\times256 \rightarrow 28\times28\times256 $ \\
	        \midrule
	        Attention Layer at $28\times28\times256$ \\
	        \midrule
	        ConvTranspose2D Layer, 2x2, stride 2, pad upscale, AdaIN, and leakyReLU $0.2$ \\
	        $ 28\times28\times256 \rightarrow 56\times56\times128 $ \\
	        \midrule
	        ResNet Conv2D Layer, 3x3, stride 1, pad same, AdaIN, and leakyReLU $0.2$ \\
	        $ 56\times56\times128 \rightarrow 56\times56\times128 $\\
	        \midrule
	        ConvTranspose2D Layer, 2x2, stride 2, pad upscale, AdaIN, and leakyReLU $0.2$ \\
	        $ 56\times56\times128 \rightarrow 112\times112\times64 $ \\
	        \midrule
	        ResNet Conv2D Layer, 3x3, stride 1, pad same, AdaIN, and leakyReLU $0.2$ \\
	        $ 112\times112\times64 \rightarrow 112\times112\times64 $ \\
	        \midrule
	        ConvTranspose2D Layer, 2x2, stride 2, pad upscale, AdaIN, and leakyReLU $0.2$ \\
	        $ 112\times112\times64 \rightarrow 224\times224\times32 $ \\
	        \midrule
	        Conv2D Layer, 3x3, stride 1, pad same, $ 32 \rightarrow 3 $ \\
	        $ 224\times224\times32 \rightarrow 224\times224\times3 $ \\
	        \midrule
	        Sigmoid \\
	        \bottomrule
	        \bottomrule
	        \end{tabular}
	        \caption{Generator Network Architecture details of PathologyGAN model.}
	        \label{generator_arch}
	    \end{table}
	    
	    \begin{table}[H]
	        \centering
	        \begin{tabular}{c}
	        Discriminator Network $C:x \rightarrow d$ \\
	        \toprule
	        \midrule
	        $x \in  \mathbb{R}^{224\times224\times3}$ \\
	        \midrule
	        ResNet Conv2D Layer, 3x3, stride 1, pad same, and leakyReLU $0.2$ \\
	        $ 224\times224\times3 \rightarrow 224\times224\times3 $ \\
	        \midrule
	        Conv2D Layer, 2x2, stride 2, pad downscale, and leakyReLU $0.2$ \\
	        $ 224\times224\times3 \rightarrow 122\times122\times32 $ \\
	        \midrule
	        ResNet Conv2D Layer, 3x3, stride 1, pad same, and leakyReLU $0.2$ \\
	        $ 122\times122\times32 \rightarrow 122\times122\times32 $ \\
	        \midrule
	        Conv2D Layer, 2x2, stride 2, pad downscale, and leakyReLU $0.2$ \\
	        $ 122\times122\times32 \rightarrow 56\times56\times64 $ \\
	        \midrule
	        ResNet Conv2D Layer, 3x3, stride 1, pad same, and leakyReLU $0.2$ \\
	        $ 56\times56\times64 \rightarrow 56\times56\times64 $ \\
	        \midrule
	        Conv2D Layer, 2x2, stride 2, pad downscale, and leakyReLU $0.2$ \\
	        $ 56\times56\times64 \rightarrow 28\times28\times128 $ \\
	        \midrule
	        ResNet Conv2D Layer, 3x3, stride 1, pad same, and leakyReLU $0.2$ \\
	        $ 28\times28\times128 \rightarrow 28\times28\times128 $ \\
	        \midrule
	        Attention Layer at $28\times28\times128$ \\
	        \midrule
	        Conv2D Layer, 2x2, stride 2, pad downscale, and leakyReLU $0.2$ \\
	        $ 28\times28\times128 \rightarrow 14\times14\times256 $ \\
	        \midrule
	        ResNet Conv2D Layer, 3x3, stride 1, pad same, and leakyReLU $0.2$ \\
	        $ 14\times14\times256 \rightarrow 14\times14\times256 $ \\
	        \midrule
	        Conv2D Layer, 2x2, stride 2, pad downscale, and leakyReLU $0.2$ \\
	        $ 14\times14\times256 \rightarrow 7\times7\times512 $ \\
	        \midrule
	        Flatten $ 7\times7\times512 \rightarrow 25088 $ \\
	        \midrule
	        Dense Layer and leakyReLU, $25088 \rightarrow 1024$ \\
	        \midrule
	        Dense Layer and leakyReLU, $1024 \rightarrow 1$ \\
	        \bottomrule
	        \bottomrule
	        \end{tabular}
	        \caption{Discriminator Network Architecture details of PathologyGAN model.}
	        \label{Discriminator_arch}
	    \end{table}
	    
	    \begin{table}[H]
	        \centering
	        \begin{tabular}{c}
	        Mapping Network $M:z \rightarrow w$ \\
	        \toprule
	        \midrule
	        $z \in  \sim \mathbb{R}^{200} \sim \mathcal{N}(0, I)$ \\
	        \midrule
	        ResNet Dense Layer and ReLU, $200 \rightarrow 200$ \\
	        \midrule
	        ResNet Dense Layer and ReLU, $200 \rightarrow 200$ \\
	        \midrule
	        ResNet Dense Layer and ReLU, $200 \rightarrow 200$ \\
	        \midrule
	        ResNet Dense Layer and ReLU, $200 \rightarrow 200$ \\
	        \midrule
	        Dense Layer, $200 \rightarrow 200$ \\
	        \bottomrule
	        \bottomrule
	        \end{tabular}
	        \caption{Mapping Network Architecture details of PathologyGAN model.}
	        \label{mapping_network_arch}
	    \end{table}
	    
\vskip 0.2in

\end{document}